\documentclass[12pt]{article}
\usepackage{natbib}       
\usepackage[utf8]{inputenc}
\usepackage{authblk}
\usepackage{booktabs}
\usepackage{geometry}
\usepackage[T1]{fontenc}
\usepackage{graphicx}
\usepackage{url}
\usepackage{amsmath, amssymb}      
\usepackage{float}
\usepackage{setspace}
\usepackage{placeins}
\usepackage{subcaption} 
\usepackage{multirow}
\usepackage{array}
\usepackage{appendix}
\usepackage{chngcntr}   
\usepackage[normalem]{ulem} 
\usepackage{pdflscape}
\usepackage[T1]{fontenc}

\usepackage{pdflscape}
\usepackage{makecell}
\usepackage{xcolor}
\newcommand{\review}[1]{\textcolor{black}{#1}} 
\usepackage{hyperref} 
\usepackage{xcolor}
\usepackage{array}
\usepackage{tikz}
\usetikzlibrary{arrows.meta}
\usepackage[export]{adjustbox}
\usepackage[svgnames]{xcolor}
\definecolor{steelblueR}{HTML}{4682B4}
\definecolor{tomatoR}{HTML}{FF6347}
\definecolor{darkorchidR}{HTML}{9932CC}
\usepackage[percent]{overpic}
\usepackage{myfiguremacros}
\usepackage{algorithm}
\usepackage{algpseudocode}
\usepackage{caption}
\title{
Robust normality transformation \review{and} outlier detection in diverse distributions,
with application to volume censoring in functional MRI
}

\author{
Fatma Parlak\textsuperscript{1*},
Saranjeet Singh Saluja\textsuperscript{2*},
Amanda Mejia\textsuperscript{3}
\\
\textsuperscript{1}Department of Mathematics,
Ankara Yıldırım Beyazıt University, Ankara,  T\"urkiye
\\
\textsuperscript{2}Department of Statistics,
Indiana University Bloomington, Bloomington, IN, USA
\\
\textsuperscript{3}Department of Statistics,
North Carolina State University, Raleigh, NC, USA
\\
\textsuperscript{*}These authors contributed equally to this work.
}

\date{}
\begin{document}

\maketitle 

\doublespacing

\begin{abstract}

\review{Normality transformations improve conformity with parametric assumptions and, for outlier detection, facilitate parametric thresholds. However, transformation parameters must be estimated from data that may contain contamination, which can distort the transformation of non-outlying observations and lead to masking and swamping.} Recently developed robust power transformations \review{address skew but not} more general departures from normality, and they remain vulnerable to outliers \review{due to reliance on Huber M-estimators of location and scale, which are heavily biased under contamination}. Here, we propose a novel robust transformation based on the flexible sinh--arcsinh (SHASH) family of distributions, which accommodates skewness, non-Gaussian tail weights, and combinations thereof. \review{To robustly estimate the transformation, we consider two methods for initializing outliers:} a conventional distance-based approach and a novel variant of anomaly detection. \review{Through extensive simulations and real data analyses, we find that robust SHASH transformation more successfully transforms non-outlying observations toward normality and better identifies outliers at contamination levels up to 20\%, compared with existing approaches. In an application to fMRI volume censoring using robust MCD distances, which are approximately $F$-distributed under Gaussianity, robust normality transformation improves distinction of artifactual volumes, illustrating its potential as a preprocessing step for automatic, robust multivariate outlier detection in large neuroimaging studies.}
\end{abstract}


\section{Introduction} \label{sec:introduction}

\review{Normality transformations are widely used to improve conformity with distributional assumptions underlying parametric statistical methods \citep{box1964analysis, carroll1980robust} and are increasingly leveraged in machine learning and artificial intelligence, for example to improve the performance of deep neural networks \citep{eftekhari2025gaussianizing} and to efficiently quantize large language models \citep{dettmers2023qlora}. Normality transformations are also useful for outlier detection, where knowledge of the source distribution of non-outlying observations permits the use of distribution-based thresholds to distinguish unusual observations while controlling the false positive rate. In univariate settings, transformation to standard normality facilitates outlier detection by adopting known Gaussian quantiles as thresholds. In multivariate settings, robust variants of Mahalanobis distance, such as those based on minimum covariance determinant (MCD) estimators of location and scatter \citep{rousseeuw1993alternatives}, facilitate identification of multivariate outliers while mitigating their influence. Under multivariate normality, \cite{hardin2005distribution} showed that MCD distances approximately follow a scaled F-distribution, the quantiles of which can be used as an outlier detection threshold. 
}

\review{The fundamental challenge is that the transformation must be estimated from observed data that may contain outliers, which can influence the transformation. A classical framework for studying robustness to such observations is the contamination model \citep{huber1964robust}, in which the observed distribution is represented as
$F_{\epsilon}(x)=(1-\epsilon)F_0(x)+\epsilon G(x)$,
where $F_0$ is the distribution of non-outlying observations, $G$ is an unspecified contamination distribution, and $\epsilon$ is the contamination proportion. In the context of normality transformation, the objective is to utilize observations drawn from $F_{\epsilon}$ to estimate a transformation that maps $F_0$ toward normality, without allowing observations arising from $G$ to distort the transformation. Naively estimating a normality transformation from all observations without accounting for contamination can instead yield a transformation that accommodates $F_\epsilon$ rather than $F_0$, likely failing to transform $F_0$ to normality. This reflects an inherent challenge for transformation-based outlier detection: the outliers we aim to identify can themselves influence estimation of the transformation used to identify them. With knowledge of the outliers, the transformation could be estimated; with knowledge of the correct transformation parameters, the outliers could be identified; yet, in most cases we lack knowledge of both.}

\review{The influence of outliers on the procedures used to identify them is a well-studied problem in robust statistics \citep{rousseeuw2011robust}.} For instance, outliers can influence regression fits or estimates of location and scale used to compute distances, diminishing their apparent outlyingness. This can lead to \textit{masking}, where true outliers go undetected, and/or \textit{swamping}, where regular observations falsely appear to be outliers \citep{hawkins1980identification}. There is a long literature on robust estimation of location and scale in univariate \citep{hampel1974influence,rousseeuw1993alternatives} and multivariate \citep{maronna1976robust,donoho1982breakdown,rousseeuw1984least,maronna2002robust} settings, allowing robust distances to be readily computed. \review{However, flagging outliers based on these distances still requires an appropriate reference distribution from which outlier detection thresholds can be derived.} \review{When the distribution of non-outlying observations is unknown, determining an appropriate threshold is challenging. One potential solution is to transform the data toward normality, after which known distribution-based thresholds can be applied.}

One well-known family of transformations is the Box--Cox power transformation \citep{tukey1957comparative, box1964analysis} for positive-valued data or Yeo--Johnson (YJ) transformation \citep{yeo2000new} for real-valued data. \review{To address the sensitivity of transformation parameter estimation to outliers, \citet{carroll1980robust} developed a robust approach to estimating power transformation parameters, and subsequent work considered robust transformation estimation in regression models \citep{carroll1987diagnostics}. A different approach was taken by \citet{yuan2000robust}, who propose Winsorizing multivariate observations far from the center based on robust distances to reduce skewness and kurtosis; however, the shape of the inlying distribution is not modeled or altered, and the shrinkage of extreme observations obscures the outlyingness, precluding outlier detection.} More recently, \citet{raymaekers2021transforming} developed robust versions of Box-Cox and YJ power transformations to central normality, designed to transform the central portion of the data toward normality while allowing outliers to remain outlying. Their reweighted maximum likelihood (RewML) procedure first obtains a robust initial estimate of the transformation parameter using a bounded objective and rectified transformation. The data are then transformed using this initial estimate, and robust z-scores using Huber M-estimators of center and scale are used to identify and exclude potential outliers from subsequent reweighted maximum likelihood estimation.

However, two main limitations restrict the broad applicability of robust Box–Cox and YJ transformations. First, these power transformations primarily address skewness and have limited flexibility to accommodate non-Gaussian tail weights. \review{Second, the robustness of the RewML procedure depends critically on both the initial estimate of the transformation parameter and the subsequent identification of outliers using robust z-scoring. Under substantial contamination, outliers may distort the initial transformation estimate as well as the robust location and scale estimates used for z-scoring, compromising the identification of outliers for reweighting. Consequently, contaminating observations may continue to exert influence on the transformation.} In our simulation studies, we indeed find that the performance of robust Box-Cox or YJ transformation suffers when there are more than a few outliers, \review{in terms of non-normality of inlying observations and poor sensitivity to outliers.}

In this paper, we develop a novel robust transformation to standard \review{n}ormality, which overcomes the two limitations above. First, we consider transformations that can accommodate a wide range of non-Gaussian distributions. In recent years, several transformations to \review{n}ormality have been developed that can handle skew as well as tail weight \citep{jones2009sinh, goerg2015LambertW, Tsai2017HP}. \review{However, these transformations are typically estimated by maximum likelihood or moment-based methods, which are themselves sensitive to outliers, and to our knowledge no contamination-robust estimation procedure has been developed for them.} \review{Alternatively, the semiparametric ordered quantile normalization of \citet{peterson2020}, which maps the ranks of the observations to the corresponding standard normal quantiles, can transform data of virtually any shape toward normality. However, since the transformed value of each observation depends only on its rank, the most extreme observations are mapped to the largest normal quantiles regardless of their magnitude, so outliers are no longer distinguishable from regular observations after transformation, making this approach unsuitable for outlier detection.}
Here, we adopt the highly flexible sinh-arcsinh (SHASH) family of distributions and associated transformation, which can address non-Gaussian tail weight, skew and combinations of both \citep{jones2009sinh}. \review{Specifically, we adopt the 4-parameter SHASH family, which incorporates location and scale parameters in addition to skew and tail weight parameters, in contrast to power transformations which require post-transformation centering and scaling via robust z-scoring to achieve standard normality.} To address the second limitation, namely initialization and iterative exclusion of outliers during transformation, we consider alternatives to conventional robust z-scoring, which fails in many settings, as we will show.  Specifically, we propose a novel variant of nonparametric anomaly detection, bridging the statistical and machine learning outlier detection literatures.  

The remainder of the paper proceeds as follows. In section \ref{sec:Method}, we introduce the SHASH family of distributions and associated transformation, and we describe the robust transformation and outlier initialization procedures.  In section \ref{sec:simulation}, we present extensive simulation studies using a wide array of distribution shapes and outlier contamination levels to assess the performance of robust SHASH transformation for achieving normality and for outlier detection, compared with existing techniques, \review{including non-robust SHASH transformation} and robust \review{RewML Box-Cox/YJ} transformation. In section \ref{sec:toydatasets}, we analyze several real world \review{outlier detection} datasets to compare method performance in different scenarios.  Finally, in section \ref{sec:neuroimaging}, \review{we illustrate the value of the proposed methods in the context of censoring artifact-contaminated volumes in functional neuroimaging data. We apply robust SHASH normality transformation as a preprocessing step for multivariate robust outlier detection to improve conformity with parametric assumptions. We observe improved identification of volumes containing subtle artifacts and fewer apparent false positives.}  We conclude with a discussion of our findings, as well as limitations and directions for future research.

\section{Methods} \label{sec:Method}

\begin{figure}
    \centering
    \Huge 
    \resizebox{\textwidth}{!}{
    \begin{tabular}{c|c|c|c|c|c}

  & \Huge Heavy‑tailed & \Huge Light‑tailed & \Huge Normal
  & \Huge Left‑skewed  & \Huge Right‑skewed \\[10pt]

  & \Huge ($\mu = 0,\ \sigma = 1$, & \Huge ($\mu = 0,\ \sigma = 1$,
  & \Huge ($\mu = 0,\ \sigma = 1$, & \Huge ($\mu = 0,\ \sigma = 1$,
  & \Huge ($\mu = 0,\ \sigma = 1$, \\[10pt]

  & \Huge \ $\nu = 0,\ \tau = 0.5$)  & \Huge \ $\nu = 0,\ \tau = 10$)
  & \Huge \ $\nu = 0,\ \tau = 1$)   & \Huge \ $\nu = -2,\ \tau = 1$)
  & \Huge \ $\nu = 2,\ \tau = 1$) \\[10pt]\hline
      \begin{picture}(20,220)\put(-10,100){\rotatebox[origin=c]{90}{%
      \fontsize{22.5}{16}\selectfont Before Transformations}}\end{picture} &
      \includegraphics[width = 0.45\textwidth]{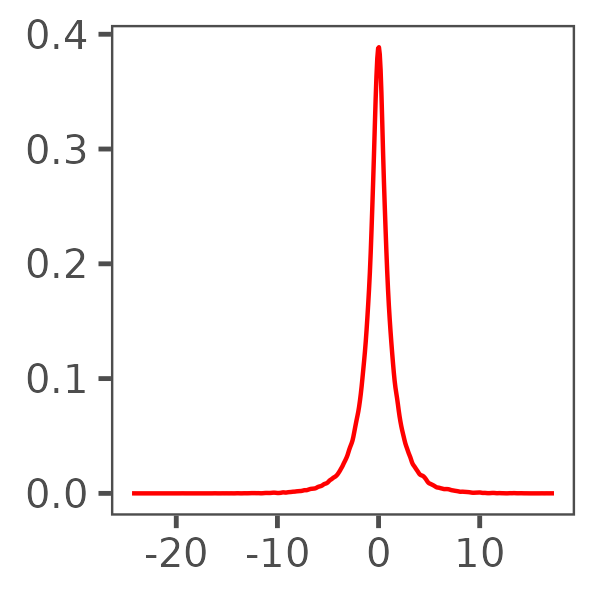} &
      \includegraphics[width = 0.45\textwidth]{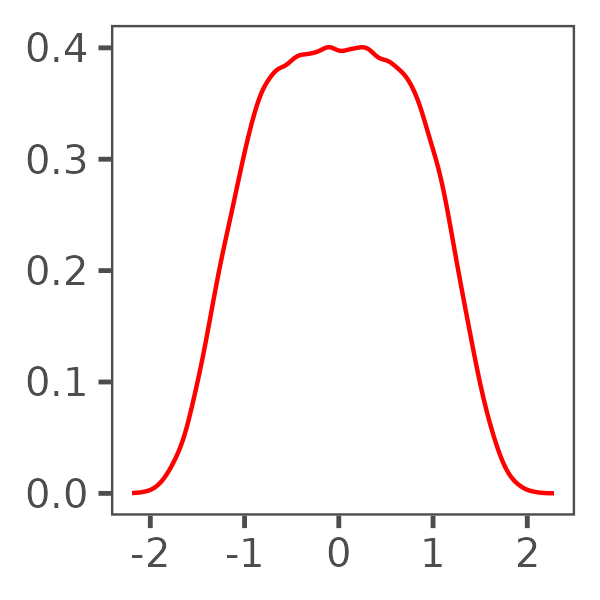} &
      \includegraphics[width = 0.45\textwidth]{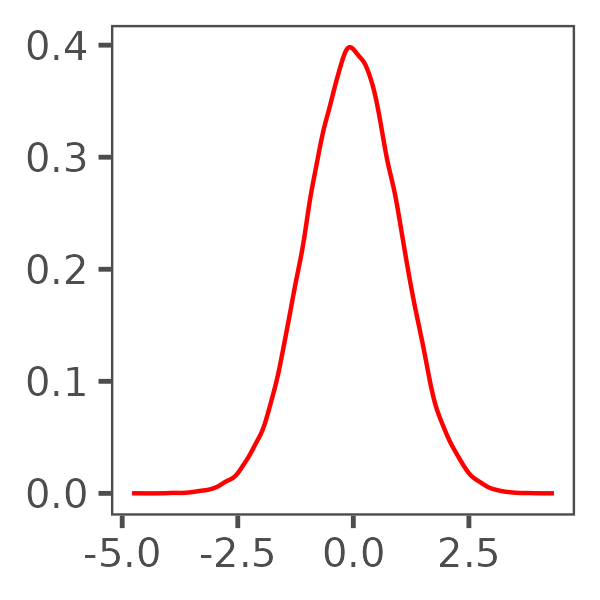} &
      \includegraphics[width = 0.45\textwidth]{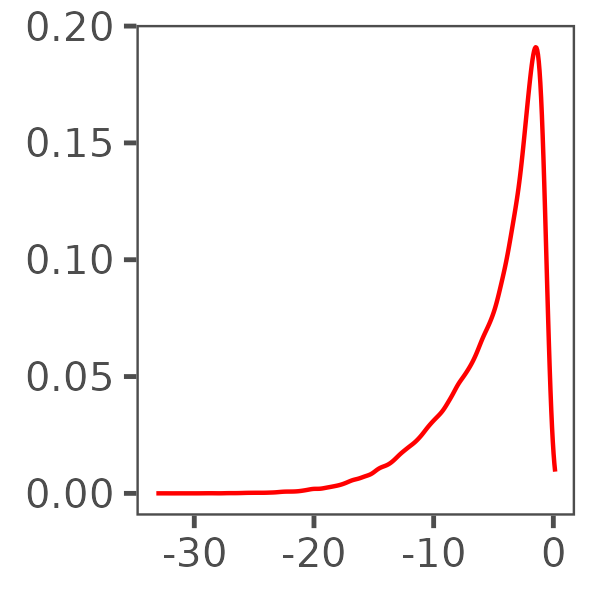} &
      \includegraphics[width = 0.45\textwidth]{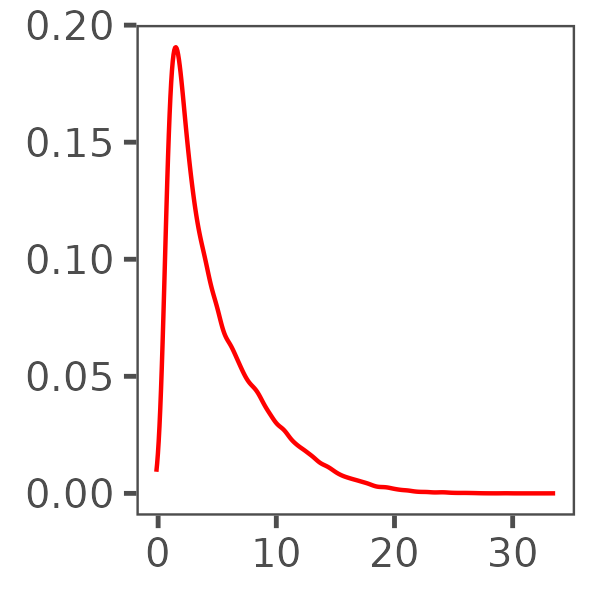}
      \\
      \hline
      \begin{picture}(20,220)\put(-10,100){\rotatebox[origin=c]{90}{%
      \fontsize{22.5}{16}\selectfont After Transformations}}\end{picture} 
      &
      \includegraphics[width = 0.45\textwidth]{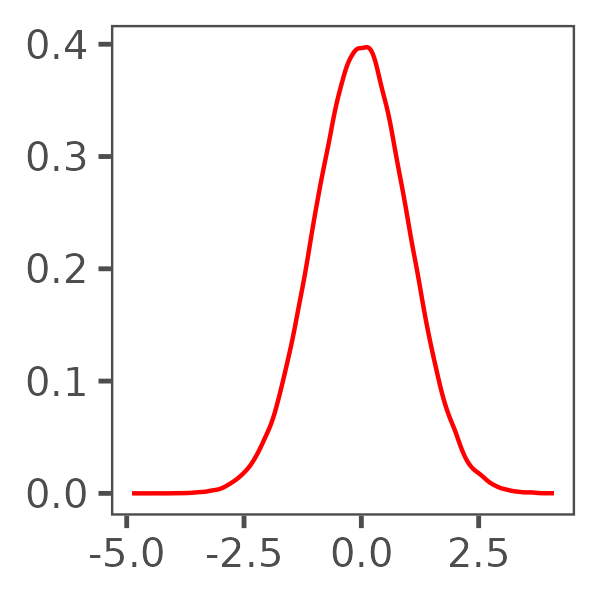} &
      \includegraphics[width = 0.45\textwidth]{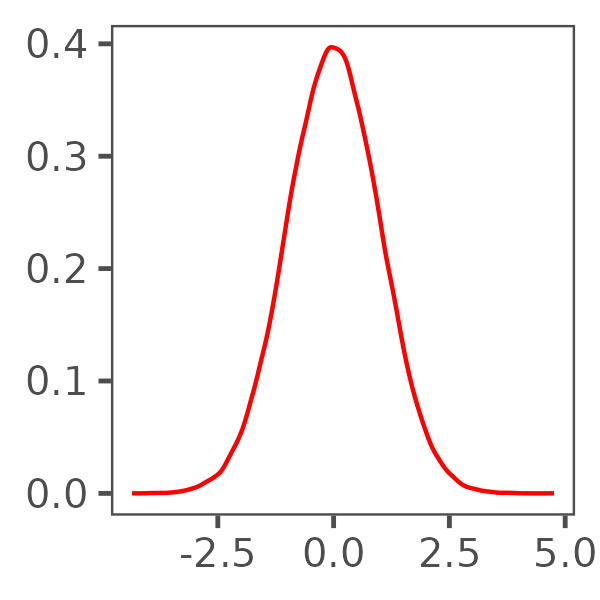} &
      \includegraphics[width = 0.45\textwidth]{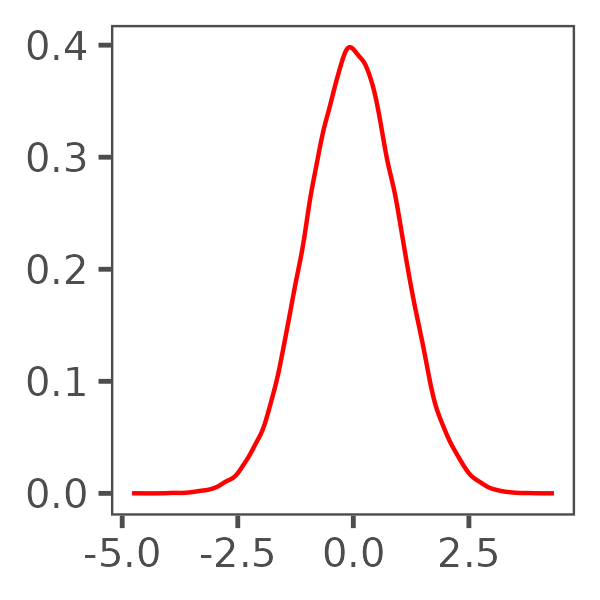} &
      \includegraphics[width = 0.45\textwidth]{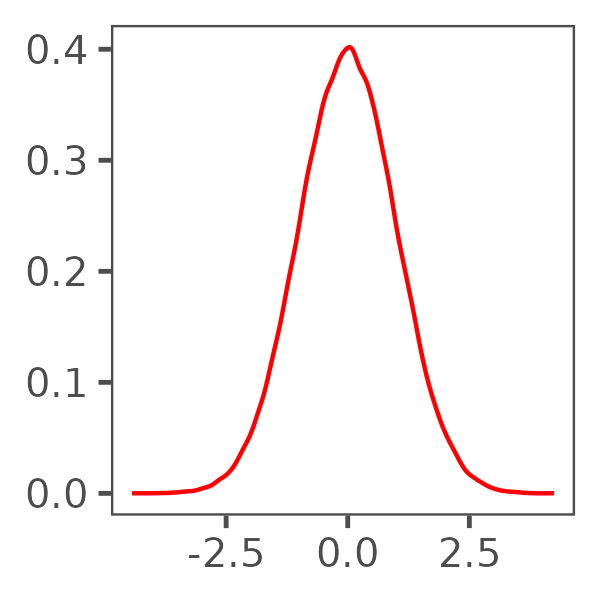} &
      \includegraphics[width = 0.45\textwidth]{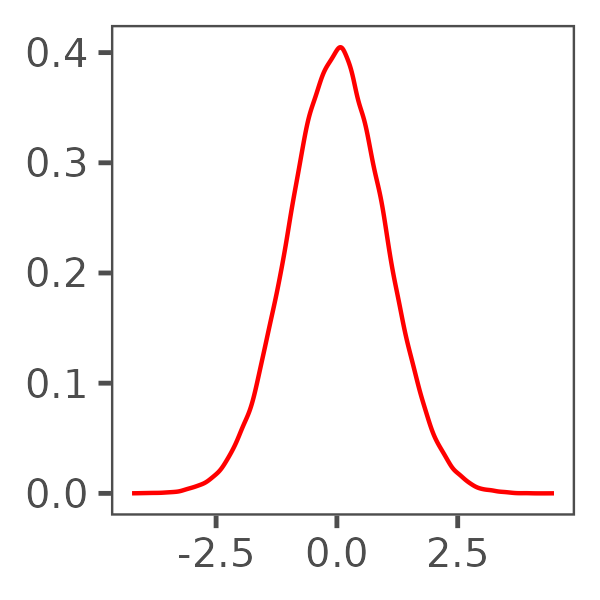} \\
      \hline
      \begin{picture}(20,220)\put(-10,100){\rotatebox[origin=c]{90}{%
      \fontsize{22.5}{16}\selectfont Before Transformations}}\end{picture} 
      &
      \includegraphics[width = 0.45\textwidth]{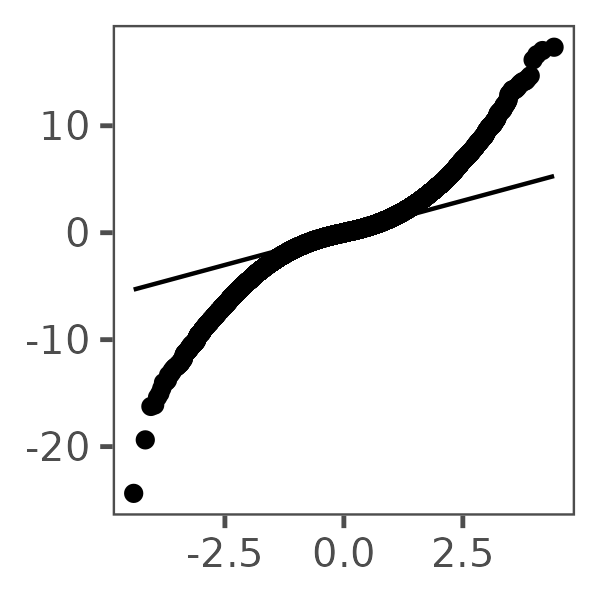} &
      \includegraphics[width = 0.45\textwidth]{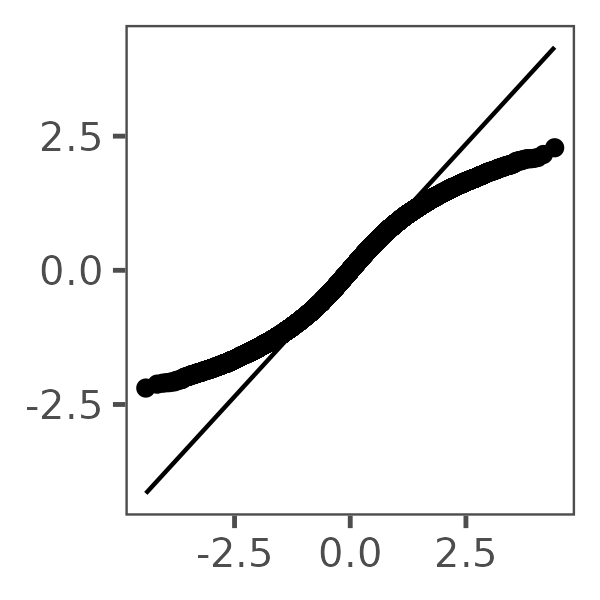} &
      \includegraphics[width = 0.45\textwidth]{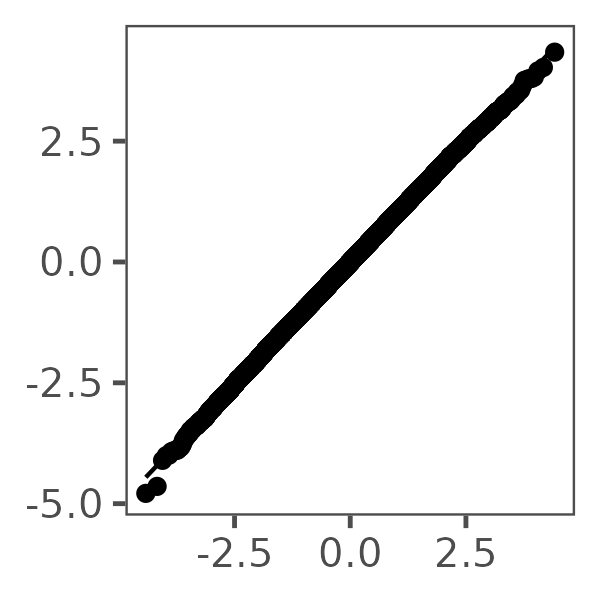} &
      \includegraphics[width = 0.45\textwidth]{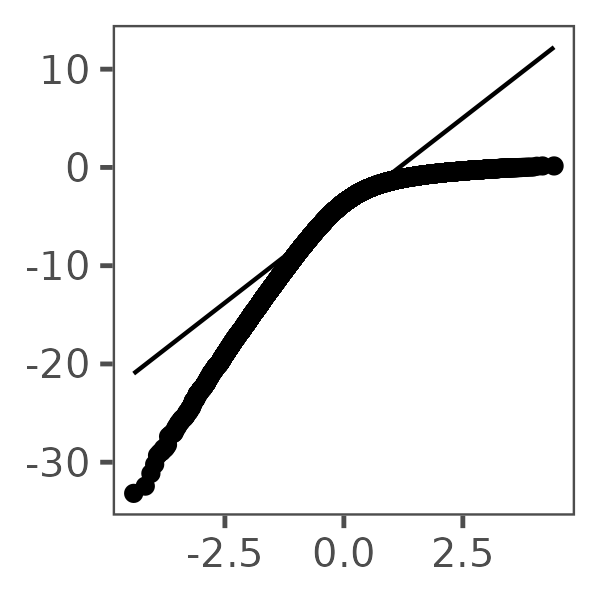} &
      \includegraphics[width = 0.45\textwidth]{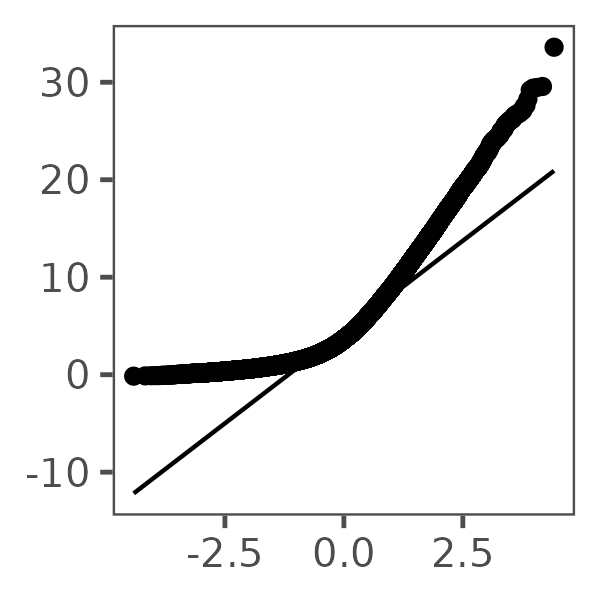}\\
      \hline
      \begin{picture}(20,220)\put(-10,100){\rotatebox[origin=c]{90}{%
      \fontsize{22.5}{16}\selectfont After Transformations}}\end{picture} 
      &
      \includegraphics[width = 0.45\textwidth]{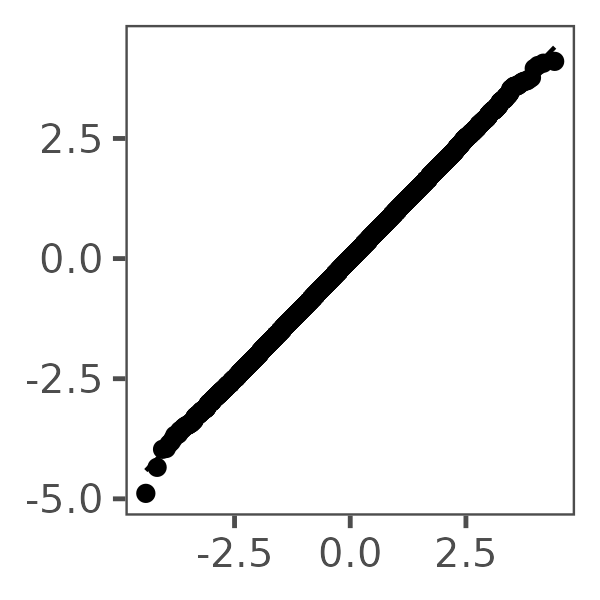} &
      \includegraphics[width = 0.45\textwidth]{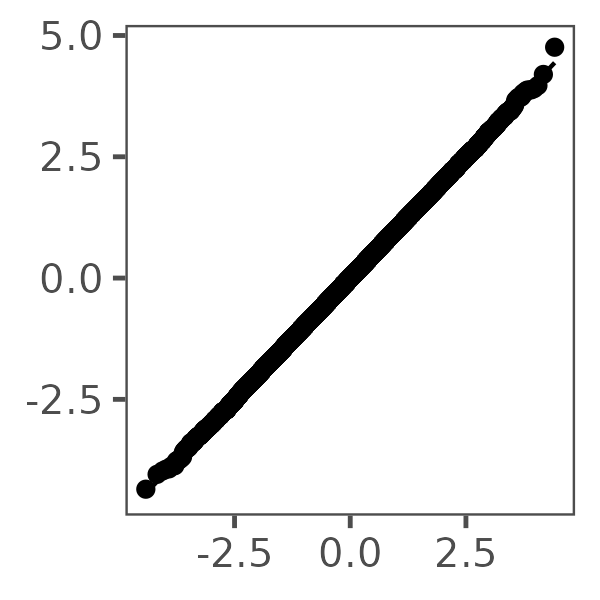} &
      \includegraphics[width = 0.45\textwidth]{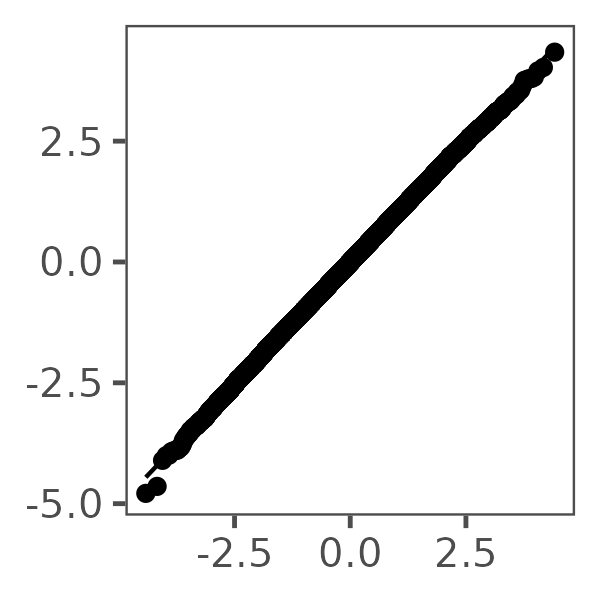} &
      \includegraphics[width = 0.45\textwidth]{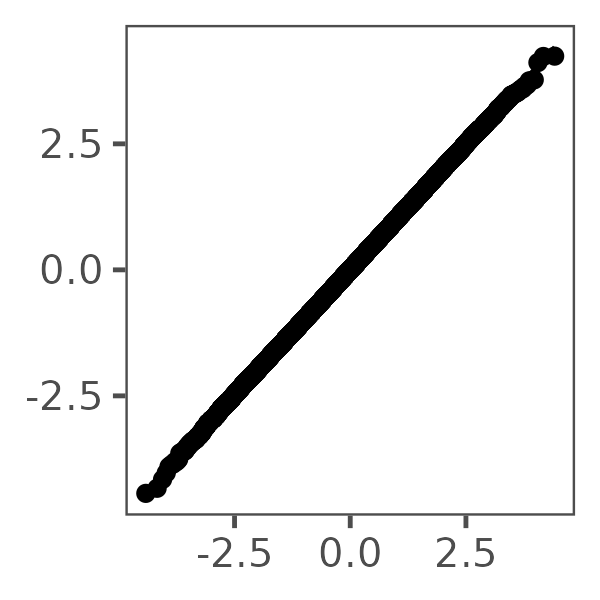} &
      \includegraphics[width = 0.45\textwidth]{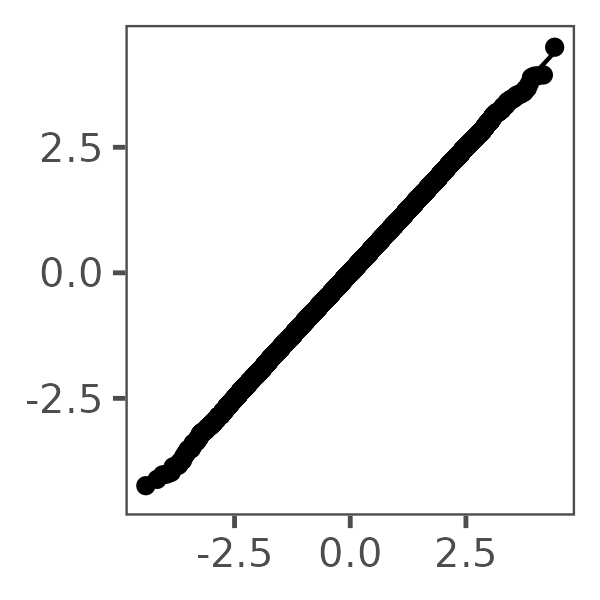}
      \end{tabular}}
      
    \caption{\small\textbf{Example SHASH random variables and their transformation to \review{n}ormality.} Each data set contains $100{,}000$ randomly sampled draws from a SHASH distribution with given parameter values. These represent a variety of distributional shapes: heavy-tailed, light-tailed, \review{n}ormal, left-skewed, and right-skewed, respectively. For each random variable, an estimated density curve and \review{n}ormal QQ plot before and after transformation to \review{n}ormality is shown, illustrating how SHASH random variables can be transformed to standard \review{n}ormal random variables.}
    \label{fig:SHASHo2}
\end{figure}

\subsection{The SHASH distribution family and transformation}

The sinh-arcsinh (SHASH) family of distributions is a versatile family characterized by four parameters controlling the location ($\mu$), scale ($\sigma$), skewness ($\nu$), and tailweight ($\tau$) \citep{jones2009sinh}. Its most basic form has only two parameters, $\nu$ and $\tau$, and arises by applying an inverse SHASH transformation to a standard \review{n}ormal random variable. Letting $Z$ be a standard \review{n}ormal random variable, $X = S_{\nu, \tau}^{-1}(Z)$ is said to follow a SHASH distribution with parameters $\nu$ and $\tau$. The SHASH transformation $S_{\nu, \tau}$ and its inverse $S^{-1}_{\nu, \tau}$ are defined as
\begin{align}
    Z = S_{\nu, \tau}(X)  &= \sinh{[\tau \sinh^{-1}{(X)} - \nu]} \label{eq:SHASHtr} \\
    X = S^{-1}_{\nu, \tau}(Z)  &= \sinh{[ \tau^{-1}(\sinh^{-1}{(Z)} + \nu)]}.
    \label{eq:invSHASHtr}
\end{align}
By a change of variables, the density function of a 2-parameter SHASH distribution is 
\begin{align}
    f_X (x | \nu, \tau) &= \tau \left(\frac{1 + (S^2_{\nu, \tau}(x))}{2 \pi (1+x^2)}\right)^{\frac{1}{2}} \exp{\left(\frac{-S^2_{\nu, \tau}(x)}{2} \right)}.
\end{align}
The four-parameter family arises by incorporating location and scale parameters $\mu$ and $\sigma$. Under regularity conditions of the maximum likelihood theory, \cite{jones2009sinh} demonstrated that the estimation of the location parameter is asymptotically independent of the scale and tailweight parameters, but the scale and tailweight parameters are asymptotically dependent. They alleviate this issue through a reparameterization, replacing $\sigma$ with $\sigma \tau$. The four-parameter {SHASH} transformation and its inverse are given by 
\begin{align}
    \label{eq:4paramtrans}
    Z =& S_{\mu, \sigma, \nu, \tau}(X) 
    = \sinh{\big[\tau \sinh^{-1}{\Big(\frac{X-\mu}{\sigma \tau}\Big)} - \nu \big]} \\
    X =& S^{-1}_{\mu, \sigma, \nu, \tau} (Z) = \sigma \tau \sinh{\left[ \frac{\sinh^{-1}(Z) + \nu}{\tau} \right] + \mu }.
\end{align}

This leads to what is known as the {SHASHo2} distribution (hereafter ``SHASH'') implemented in the \texttt{gamlss.dist} R package \citep{rigby2019distributions}. Its probability density function, using the identities $\text{sinh}(x) = (e^x - e^{-x})/2$ and $\text{cosh}(x) = (e^x + e^{-x})/2$, is (see Appendix \ref{app:derivation} for detailed derivation): 
\begin{align}
    f_X (x | \mu, \sigma, \nu, \tau) =  \frac{c(x)}{\sigma \sqrt{2 \pi \big[1+ \big(\frac{x - \mu}{\sigma \tau} \big)^2\big]}}\exp\left(-\frac{r^2(x)}{2}\right) 
    \label{eq:shash-density}
\end{align}

\noindent where $r(x)$ and $c(x)$ are given by
\begin{align}
    r(x) &= \frac{1}{2} \left\{ 
    \exp \left[ \tau \operatorname{asinh}\!\left(\frac{x - \mu}{\sigma \tau}\right) - \nu \right]
    - 
    \exp\!\left[-\tau \operatorname{asinh}\!\left(\frac{x - \mu}{\sigma \tau} \right) + \nu \right]
\right\} \\
    c(x) &= \frac{1}{2} 
\left\{
    \exp\!\left[ \tau \operatorname{asinh}\!\left(\frac{x - \mu}{\sigma \tau}\right) -\nu \right]
    +
    \exp\!\left[-\tau \operatorname{asinh}\!\left(\frac{x - \mu}{\sigma \tau}\right) +\nu \right]
\right\}.
\end{align}


\begin{table}
    \begin{center}
    \begin{tabular}{c c c c }
    \hline
    Parameter & & Range & Function \\
    \hline
       $\mu$     &  &  $-\infty < \mu < \infty$, & modulates the location       \\
       $\sigma$  &  & $0 < \sigma < \infty$,     & modulates the scale    \\ 
       $\nu$     &  & $-\infty < \nu < \infty$,  & modulates the skewness    \\ 
       $\tau$    &  & $0 < \tau < \infty$,      & modulates the tailweight \\
       \hline
    \end{tabular}
    \end{center}
    \caption{Range and role of each parameter in the \texttt{SHASH}($\mu$, $\sigma$, $\nu$, $\tau$) family of distributions.}
    \label{shash}
\end{table}

The range and role of each parameter are given in \textbf{Table \ref{shash}}. 
To illustrate the influence of each parameter on the distribution shape, \textbf{Figure \ref{fig:SHASHo2}} shows five different distributions from the SHASH family, including heavy-tailed, light-tailed, \review{n}ormal, left-skewed, and right-skewed examples. \review{Comparing the heavy-tailed and light-tailed examples, $\tau$ is the only parameter that changes, from $0.5$ to $10$. Under the SHASHo2 parameterization used in this manuscript and implemented in \texttt{gamlss.dist}, smaller values of \(\tau\) produce heavier tails, and larger values produce lighter tails; \(\tau=1\) corresponds to the \review{n}ormal tailweight when  \(\nu=0\). Comparing the left-skewed and right-skewed examples, \(\nu\) is the only parameter that changes, from \(-2\) to \(2\). Negative values of \(\nu\) produce a left-skewed distribution, while positive values produce a right-skewed distribution.}

Recall that SHASH random variables can be transformed to standard \review{n}ormality by applying the 4-parameter SHASH transformation, given in Equation \ref{eq:4paramtrans}. The second row of \textbf{Figure \ref{fig:SHASHo2}} displays the density curves after applying the SHASH transformation with the corresponding parameters.  The last two rows show \review{n}ormal quantile-quantile (QQ) plots before and after transformation. After applying the SHASH transformation with correct parameter values, each distribution exactly follows a standard \review{n}ormal distribution. 

\begin{figure}
    \centering
    \Huge
    \resizebox{\textwidth}{!}{
    \begin{tabular}{c|c|c|c|c|c|c}
    & \Huge Normal
     & \Huge T
     & \Huge Laplace
     & \Huge Gamma
     & \Huge Chi-Square
     & \Huge Weibull \\[10pt]

  & \Huge \review{($\mu = 10,  \sigma = 3$)} & \Huge ($df = 4$)
  & \Huge ($\mu = 0, \beta = 3$) & \Huge ($\alpha = 2,  \beta = 1$)
  & \Huge ($k = 3$) & \review{\Huge($\lambda = 3, k = 1$)} \\[10pt]
    \hline
    \begin{picture}(20, 220)\put(-10,97){\rotatebox[origin=c]{90}{\Huge Before transformation}}\end{picture} & 
    \includegraphics[width=0.45\textwidth]{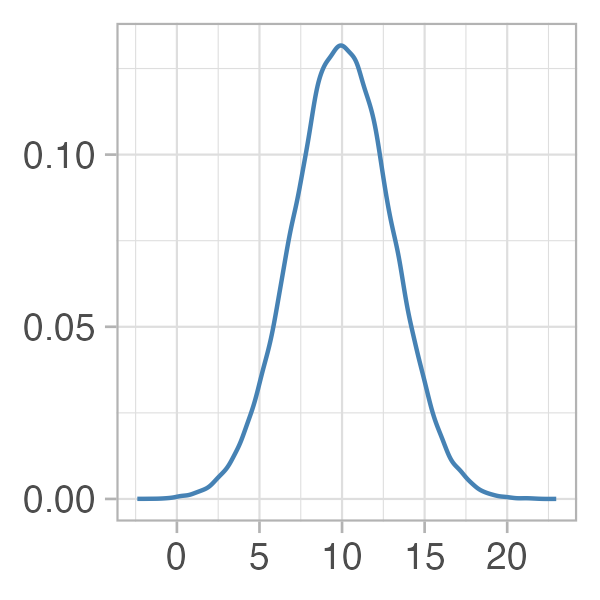} &
    \includegraphics[width=0.45\textwidth]{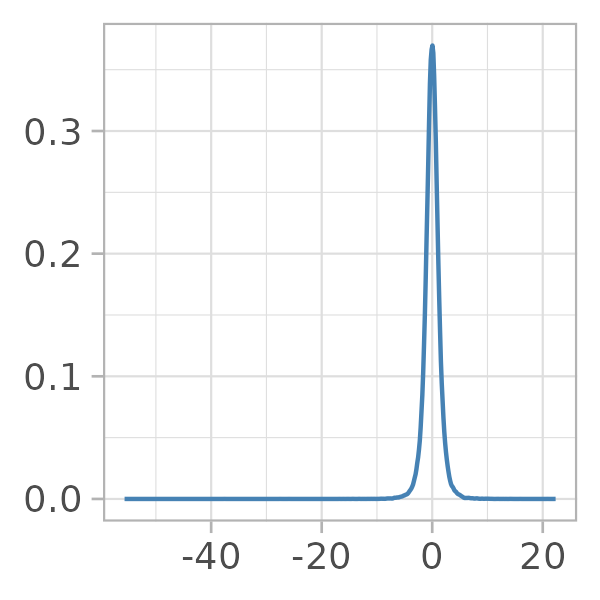} &
    \includegraphics[width=0.45\textwidth]{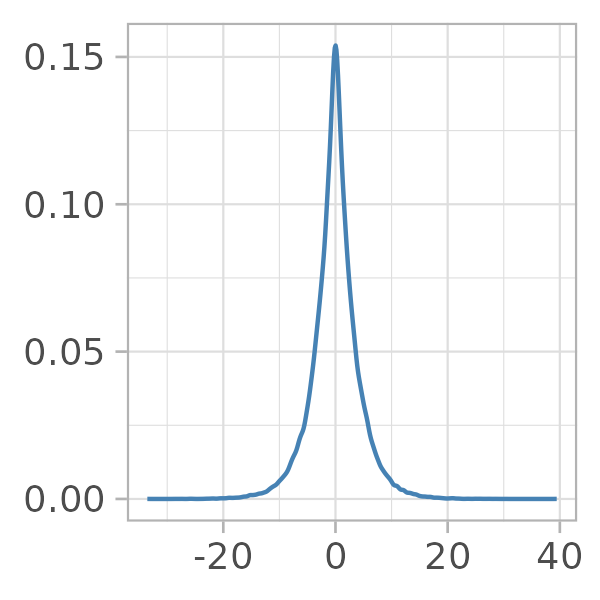} & 
    \includegraphics[width=0.45\textwidth]{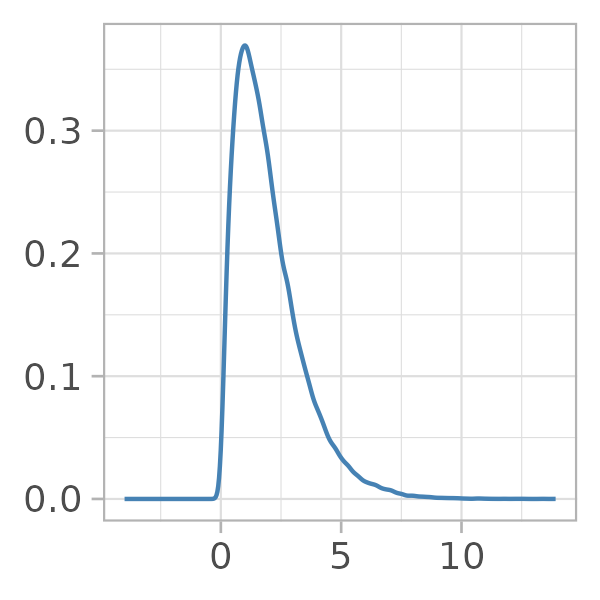}  & 
    \includegraphics[width=0.45\textwidth]{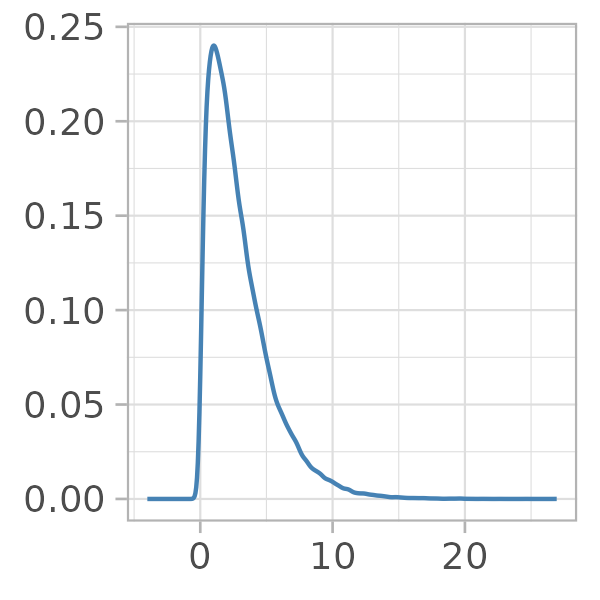} &
    \includegraphics[width=0.45\textwidth]{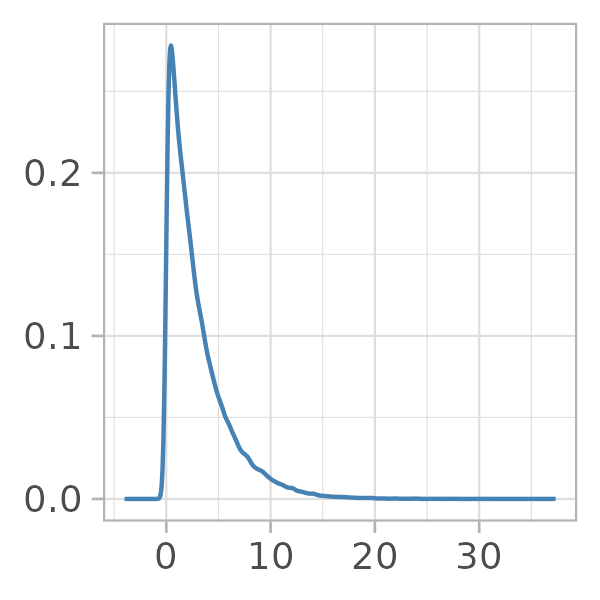} \\
    \hline
    \begin{picture}(20, 220)\put(-10,100){\rotatebox[origin=c]{90}{\Huge After transformation}}\end{picture} &
    \includegraphics[width=0.48\textwidth]{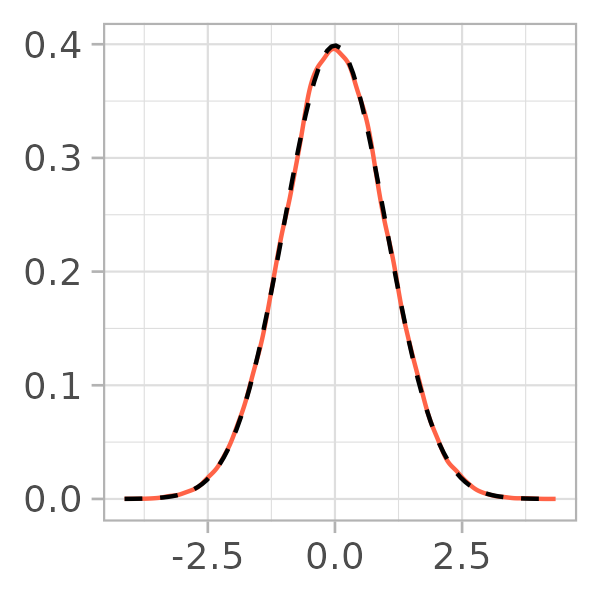} &
    \includegraphics[width=0.48\textwidth]{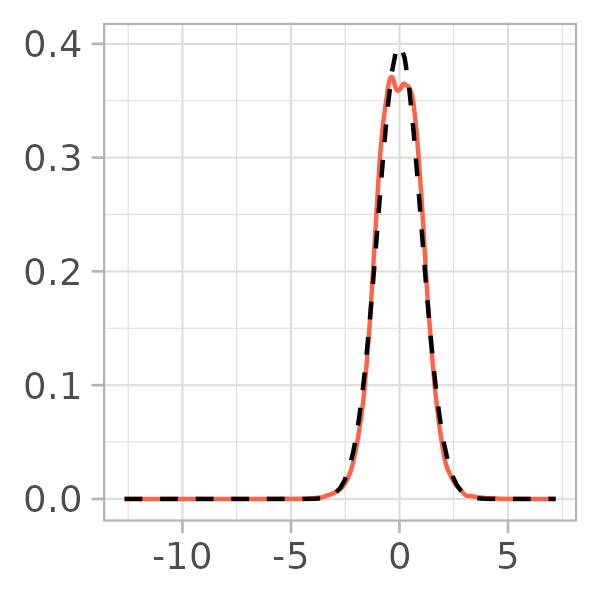} &
    \includegraphics[width=0.48\textwidth]{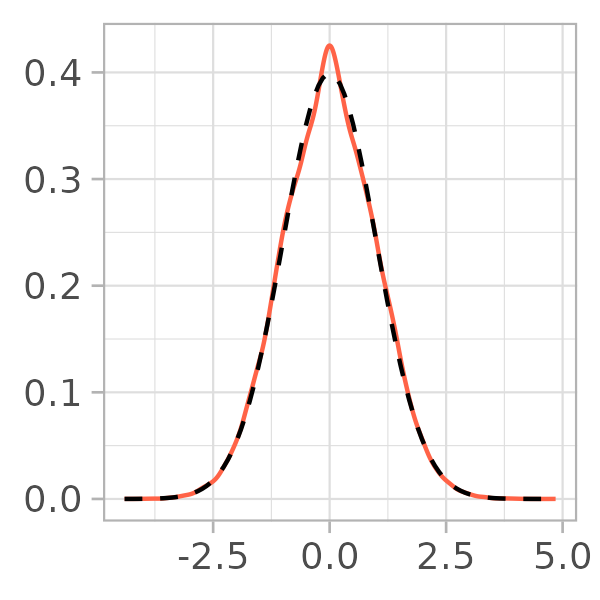} & 
    \includegraphics[width=0.48\textwidth]{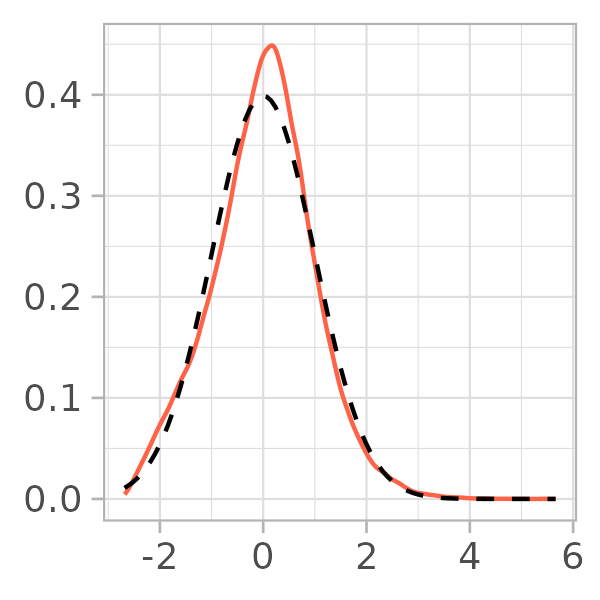}  & 
    \includegraphics[width=0.48\textwidth]{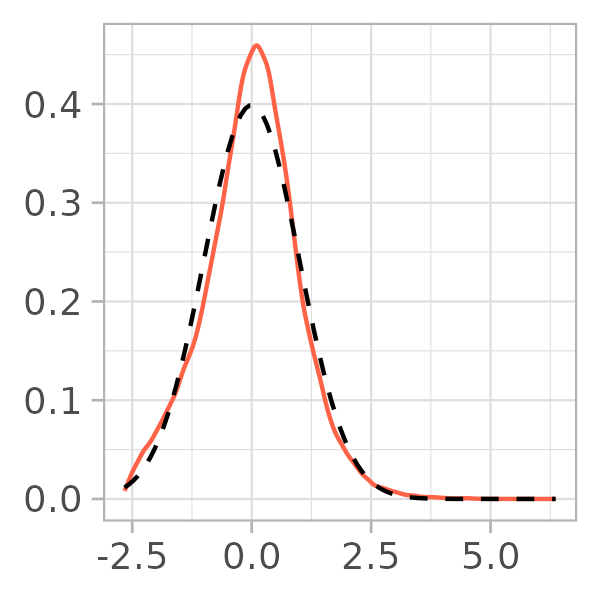} &
    \includegraphics[width=0.48\textwidth]{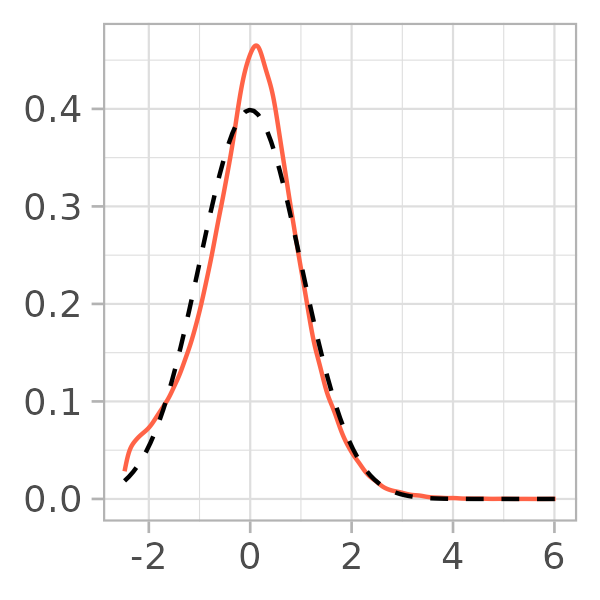}
    \\
    \hline
    \begin{picture}(20, 220)\put(-10,97)
    {\rotatebox[origin=c]{90}{\Huge QQ Plots}}\end{picture} & 
    \includegraphics[width=0.45\textwidth, height=0.55\textwidth, keepaspectratio, trim = 0 6mm 0 0, clip]{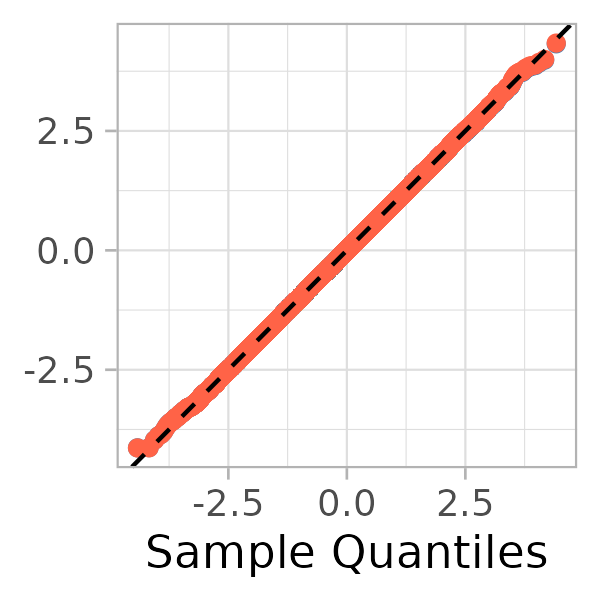} &
    \includegraphics[width=0.45\textwidth, height=0.55\textwidth, keepaspectratio, trim = 0 6mm 0 0, clip]{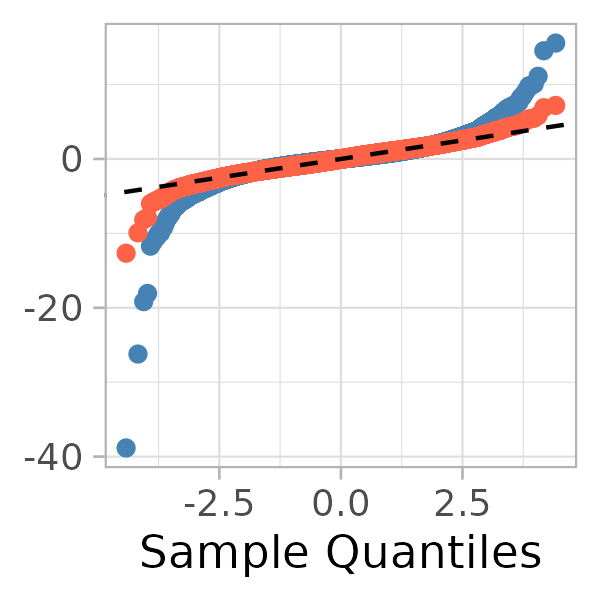} & 
    \includegraphics[width=0.45\textwidth, height=0.55\textwidth, keepaspectratio, trim = 0 6mm 0 0, clip]{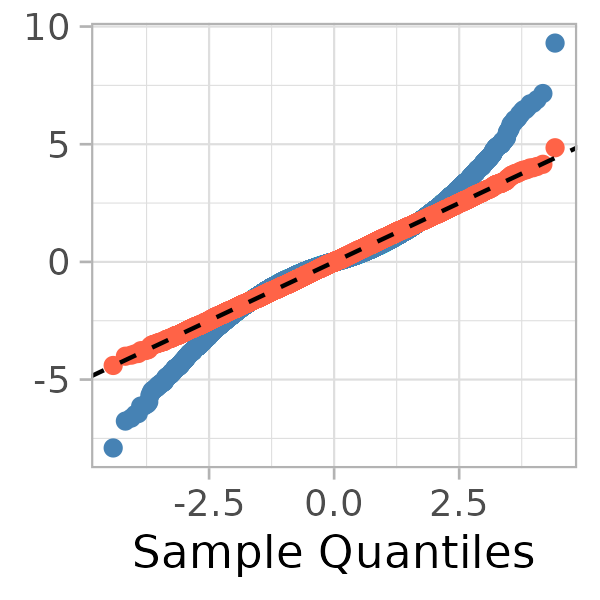} &
    \includegraphics[width=0.45\textwidth, height=0.55\textwidth, keepaspectratio, trim = 0 6mm 0 0, clip]{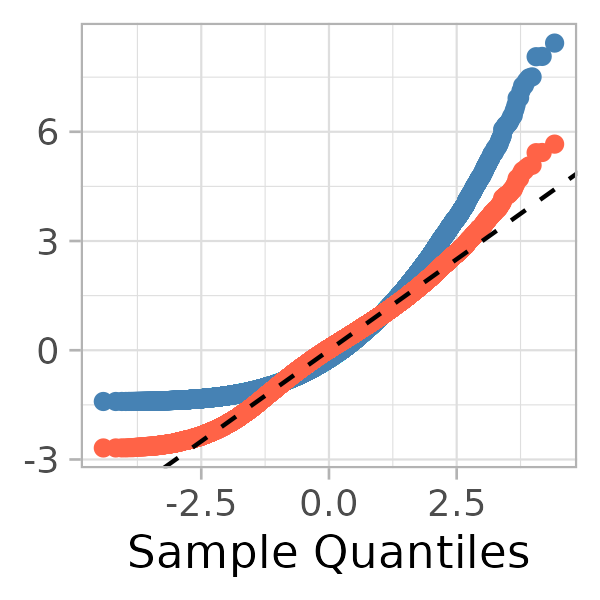}  & 
    \includegraphics[width=0.45\textwidth, height=0.55\textwidth, keepaspectratio, trim = 0 6mm 0 0, clip]
    {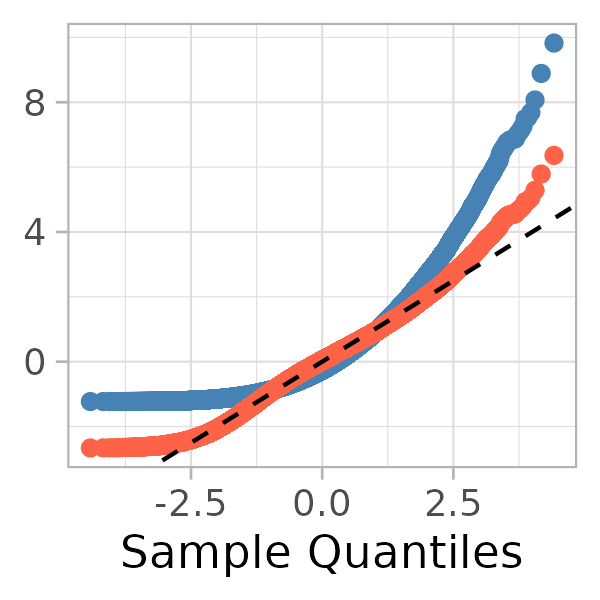} &
    \includegraphics[width=0.45\textwidth, height=0.55\textwidth, keepaspectratio, trim = 0 6mm 0 0, clip]{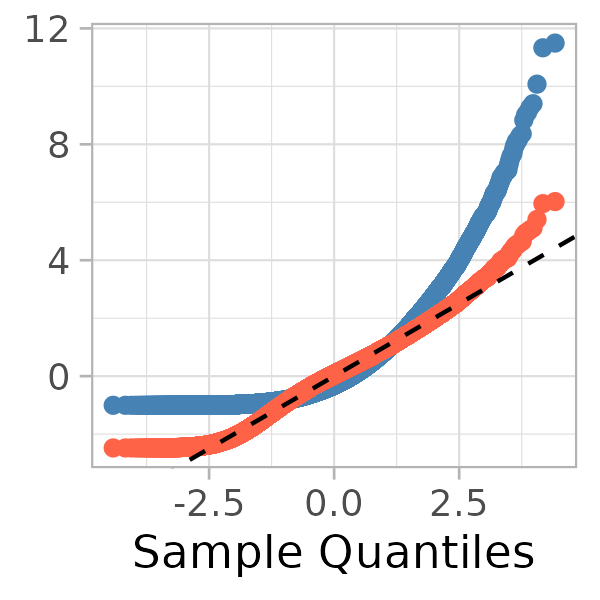} \\
    \hline
    \rule{0pt}{3em}
    & \review{\Huge \shortstack{$\hat{\mu} = 10.0$, $\hat{\sigma} = 3.0$ \\
                               $\hat{\nu} = 0.0$, $\hat{\tau} = 1.0$}}
    & \review{\Huge \shortstack{$\hat{\mu} = 0.0$, $\hat{\sigma} = 1.0$ \\
                               $\hat{\nu} = 0.0$, $\hat{\tau} = 0.6$}}
    & \review{\Huge \shortstack{$\hat{\mu} = 0.0$, $\hat{\sigma} = 2.7$ \\
                               $\hat{\nu} = 0.0$, $\hat{\tau} = 0.6$}}
    & \review{\Huge \shortstack{$\hat{\mu} = 0.5$, $\hat{\sigma} = 0.8$ \\
                               $\hat{\nu} = 1.2$, $\hat{\tau} = 1.0$}}
    & \review{\Huge \shortstack{$\hat{\mu} = -0.1$, $\hat{\sigma} = 0.9$ \\
                               $\hat{\nu} = 1.8$, $\hat{\tau} = 1.1$}}
    & \review{\Huge \shortstack{$\hat{\mu} = -0.4$, $\hat{\sigma} < 0.1$ \\
                               $\hat{\nu} = 14.5$, $\hat{\tau} = 0.9$}} \\[10pt]
    \end{tabular}}
    \includegraphics[width=0.7\textwidth, height=0.5\textwidth, keepaspectratio]{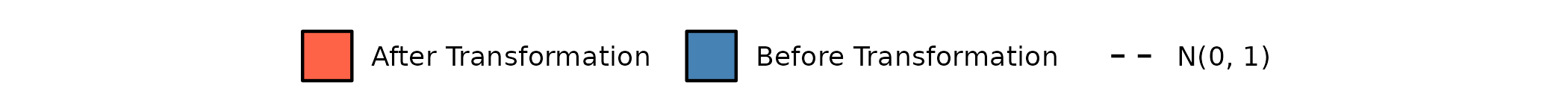}
    \caption{\small\textbf{Illustration of applying SHASH transformation to non-SHASH distributions \review{to approximately achieve standard normality}.} Each distribution was modeled as a SHASH random variable, although most distributions shown here do not fall within the SHASH family of distributions. The second and third rows shows density and QQ plots after applying SHASH transformation using maximum likelihood estimates of the SHASH parameters. While only the \review{n}ormal distribution is perfectly transformed, the other example distributions become much closer to \review{n}ormal than before transformation. \review{The bottom row gives the four SHASH parameter estimates, obtained by maximum likelihood from a sample of size $n = 100{,}000$, with $\sigma$ and $\tau$ on the natural scale.}}
    \label{fig:SHASHo2dist}
\end{figure}

Our proposed transformation assumes that the underlying data can be modeled using a SHASH distribution. Even when the data does not actually arise from a SHASH distribution, this may be a reasonable approximation in many settings given the flexibility of the SHASH family to represent widely diverse distributional shapes. \textbf{Figure~\ref{fig:SHASHo2dist}} illustrates the use of the SHASH transformation on several common parametric distributions. Since the SHASH transformation parameters are unknown here, we estimate them by maximum likelihood using \texttt{gamlssML()} from the \texttt{gamlss} R package, with the \texttt{SHASHo2} family provided by \texttt{gamlss.dist} \citep{rigby2019distributions}. With the exception of the \review{n}ormal distribution, none of these distributions arises from the SHASH family. Yet, applying the SHASH transformation using estimated parameters brings all of them much closer to \review{n}ormality.  The transformation is strikingly accurate for the Laplace distribution, which resembles the SHASH heavy tailed distribution shown in \textbf{Figure \ref{fig:SHASHo2}}.  The $t$ distribution retains some heaviness in the tails after transformation, but much less than prior to transformation.  The three skewed distributions all retain some degree of skew, but it is strongly mitigated by transformation, especially in the right tail where outliers are more likely to be present. 

\subsection{Robust SHASH transformation}
\label{sec:robust_SHASH}

Robust SHASH transformation 
involves the following steps, implemented in the open-source
\texttt{rrobot} package (version 0.2), which is available on
GitHub (\url{https://github.com/mandymejia/rrobot}) and CRAN.

\begin{algorithm}[htbp]
\captionsetup{labelformat=empty}
\caption{Robust SHASH Transformation to Standard Normality}
\label{alg:shash_outlier_detection}
\begin{algorithmic}
    \State {Step 1:} Initialize outlier labels using robust techniques.
    \Statex \textbf{Repeat}
    \State {Step 2: }Estimate SHASH parameters using weighted MLE, excluding outliers.
    \State {Step 3:} Apply the SHASH transformation to standard normality (Equation~\ref{eq:4paramtrans}).
    \State {Step 4:} Update outlier labels using $N(0,1)$ quantiles ($\pm 3$, or $+ 3$ for right tail only)
    \Statex \textbf{Until} the identified set of outliers remains unchanged.
\end{algorithmic}
\end{algorithm}


\review{For outlier detection following the transformation to standard normality,} we apply a threshold of $\pm 3$, or $+ 3$ if the original data is strictly positive. If the transformation is successful at achieving standard \review{n}ormality, a threshold of $\pm 3$ should result in a false positive rate of $0.27\%$, or $0.14\%$ for strictly positive data.  Depending on the context, a more aggressive threshold or a more lenient threshold can be adopted in order to prioritize maximizing power to detect outliers or minimizing false positives.


 This procedure follows \review{a similar} strategy as the \review{robust RewML power} transformation approach \citep{raymaekers2021transforming}, \review{but with several key differences: we use the more flexible SHASH transformation; we do not rely on robust z-scoring during iteration, since the SHASH transformation adjusts for location and scale as well as skew and tail weight;} and we consider alternative outlier initialization methods. Outlier initialization is a crucial step for any transformation \review{to avoid the influence of outliers on transformation}, especially given the high degree of flexibility of the SHASH transformation. If outliers are allowed to influence the SHASH parameter estimates, they are likely to have a strong effect and cause masking of true outliers.  
 
 Specifically, we consider two outlier initialization methods: robust z-scoring, and a novel variant of an anomaly detection technique.  
 For robust z-scoring, a z-score is computed by subtracting a robust estimate of location and dividing by a robust estimate of scale. We use the median, $m_x$, and the median absolute deviation, $MAD_x = 1.4826\ \text{med}(|x_i - m_x|)$, which is consistent for the standard deviation under Gaussianity \citep[p.~118]{ruppert2010statistics}. Observations exceeding the standard \review{n}ormal 0.995 quantile or below the 0.005 quantile, equal to $\pm2.58$ (or simply exceeding $2.58$ if the original data is strictly positive), are initialized as outliers, as in \cite{raymaekers2021transforming}. \review{The median and MAD are more consistent under contamination than Huber M-estimators of location and scale \citep{martin1993bias}, on which RewML and some other existing methods rely \citep{carroll1980robust, yuan2000robust}, and hence are adopted to maximize robustness to outliers at initialization (see \textbf{Figure \ref{fig:CenterScaleSideBySide}}).  Yet, even using these more conservative estimators,} robust z-scoring breaks down in skewed distributions and \review{under substantial outlier contamination}, as our simulation studies will show \review{(see \textbf{Figure \ref{fig:MedianMADNormalAndGamma}})}.
 
Thus, we consider an alternative approach for initializing outliers. Isolation forest (iForest) is a tree-based anomaly detection algorithm that identifies outliers by recursively isolating data points \citep{liu2008isolation}. The key
idea is that outliers are more susceptible to isolation and therefore require fewer splits. Each data point is assigned an anomaly score between 0 and 1, where values closer to 1 indicate a higher likelihood of being an outlier. While no principled method of selecting an anomaly score threshold exists, thresholds between 0.5 and 0.6 are commonly considered in the literature \citep{liu2008isolation, chabchoub2022}. Here, we propose a variant of iForest tailored for univariate settings, wherein outliers assume extreme values. We \review{first apply iForest based on a threshold of 0.6 \citep{liu2008isolation}, then identify the anomaly closest to the median, and finally use that distance from the median to define the outlier threshold(s). All observations more extreme than the smallest anomaly, in terms of distance from the median, are labeled as outliers. In the case of real-line support, this is applied across both tails symmetrically. If we only expect outliers in one tail, such as in strictly positive, right-skewed data, we only consider iForest anomalies in that tail. This process, which we call iForest+, is illustrated in \textbf{Figure ~\ref{fig:iForestInit}}.}


We call the complete SHASH pipeline using robust z-scoring for outlier initialization \textit{SHASH-z}, and using our iForest variation for outlier initialization \textit{SHASH-i}.


\review{\begin{table}[t]
    \centering
    \begin{tabular}{llll}
    \hline
    Distribution & Parameters & Outlier tails \\
    \hline
    Normal        & $\nu = 0$, $\tau = 1$     & Both \\
    Heavy-tailed  & $\nu = 0$, $\tau = 0.5$   & Both \\
    Light-tailed  & $\nu = 0$, $\tau = 2$     & Both \\
    Right-skewed  & $\nu = 2$, $\tau = 1$     & Upper \\
    \hline
    \end{tabular}
    \caption{Summary of SHASH distributions used in the first simulation study. All distributions have $\mu = 0$ and $\sigma = 1$.}
    \label{table:shash_summary}
\end{table}}

\section{Simulation Studies} \label{sec:simulation}

To evaluate the performance of SHASH \review{transformation and subsequent} outlier detection across a wide range of scenarios, we perform extensive simulation studies wherein the true outliers are known. For comparison, we apply several existing methods: \review{non-robust SHASH transformation using MLE, RewML robust power transformations \citep{raymaekers2021transforming}, and (for outlier detection only) robust z-scoring using the median and MAD, as defined in Section \ref{sec:robust_SHASH}.} For all methods, we use a final post-transformation outlier detection threshold of $\pm 3$ (or $+3$ for positively valued data). 

\begin{figure}
    \centering
    \newcommand{\simpanel}[1]{%
        \raisebox{-0.5\height}{\includegraphics[width=0.23\textwidth, trim={0cm 0cm 0cm 0cm}, clip]{#1}}%
    }
    \newcommand{\panelarrow}{%
        \tikz[baseline=0pt]\draw[-{Stealth[length=6pt,width=9pt]}, line width=2pt] (0,0) -- (0.4,0);%
    }
    \newcommand{\rowlabel}[1]{\rotatebox[origin=c]{90}{#1}}
    \setlength{\tabcolsep}{0.5pt}
    \resizebox{\textwidth}{!}{%
    \begin{tabular}{c c c c c c c c}
        & Original Data & & Transformed & & Outliers Added & & Untransformed \\[2pt]
        \rowlabel{Symmetric} &
        \simpanel{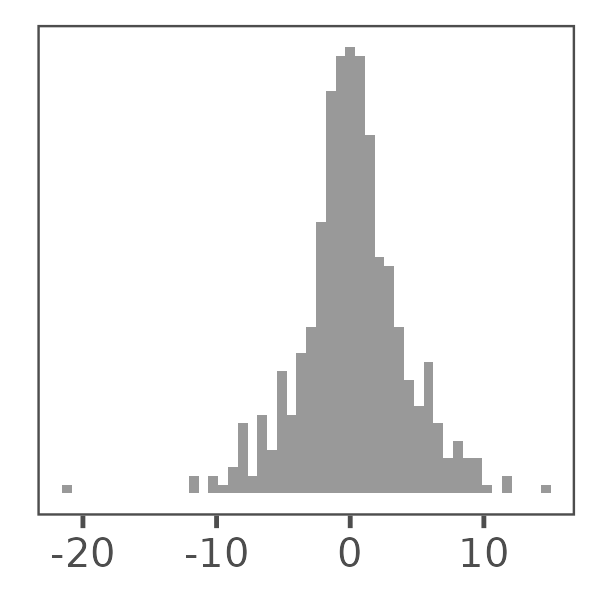} & \panelarrow &
        \simpanel{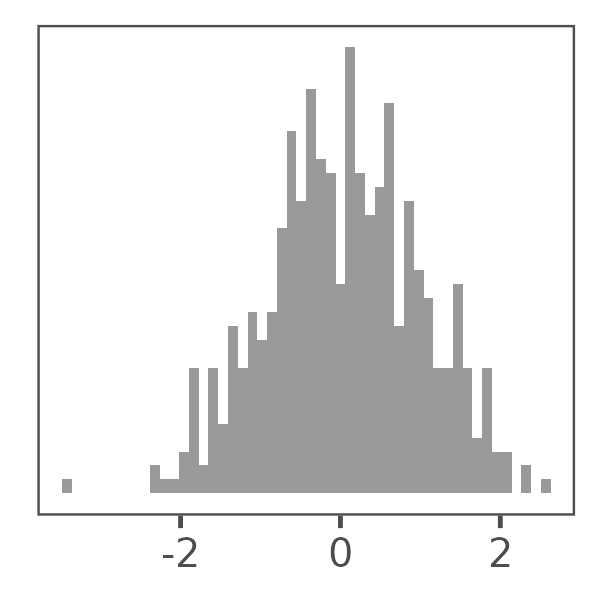} & \panelarrow &
        \simpanel{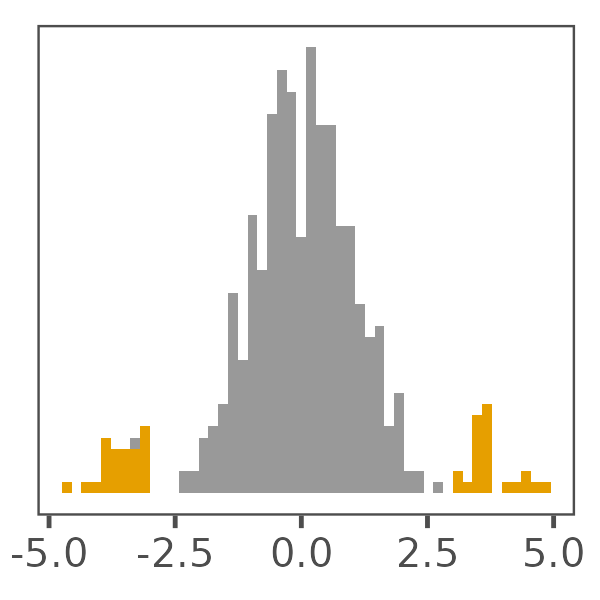} & \panelarrow &
        \simpanel{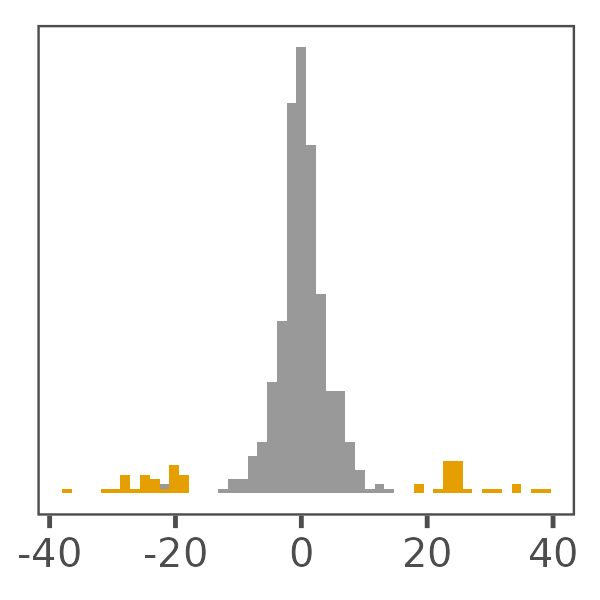} \\[4pt]
        \rowlabel{Right Skewed} &
        \simpanel{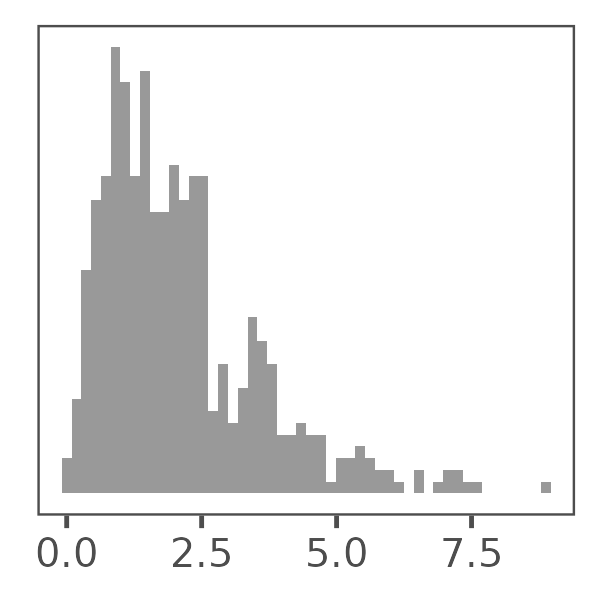} & \panelarrow &
        \simpanel{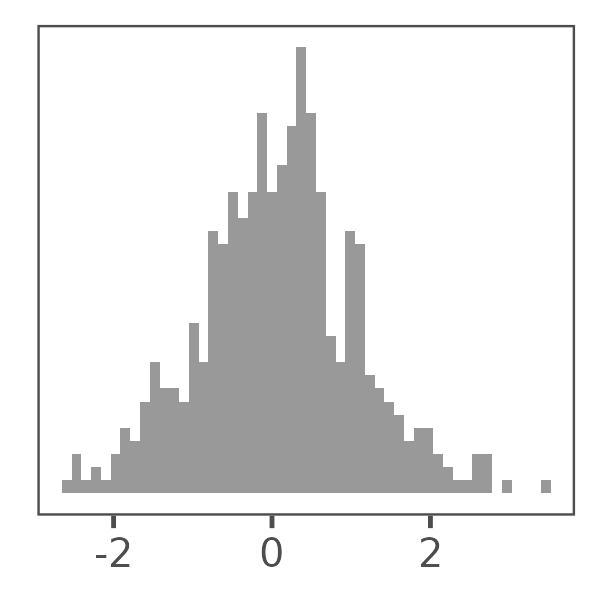} & \panelarrow &
        \simpanel{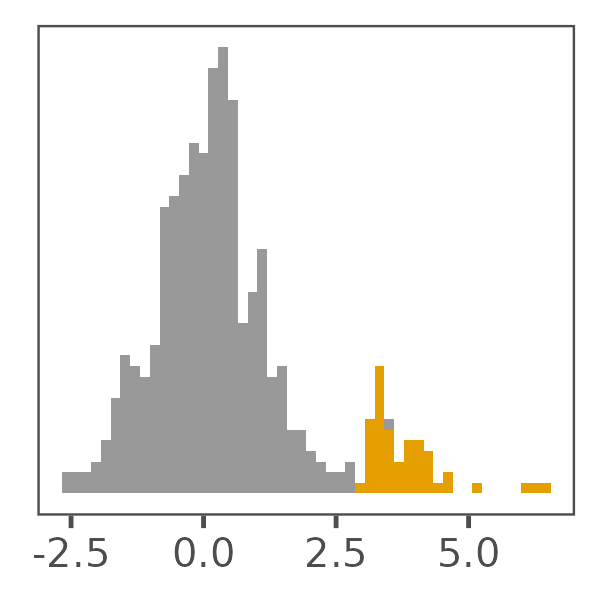} & \panelarrow &
        \simpanel{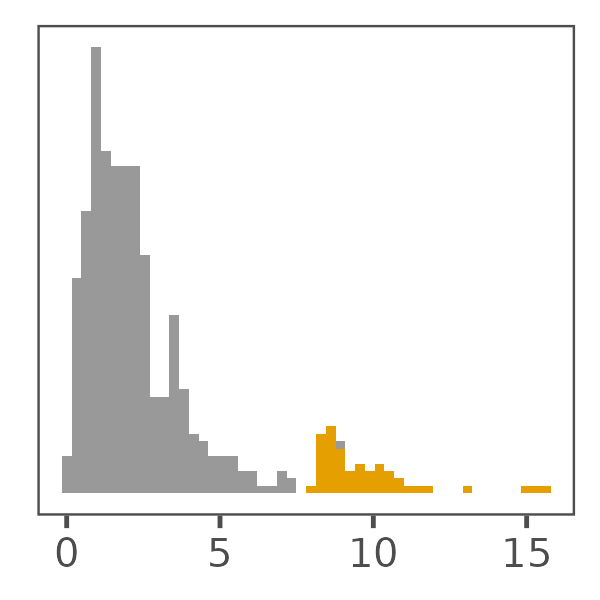} \\
    \end{tabular}}
    \caption{\small\textbf{The outlier contamination process for simulation studies.} The process is illustrated for one symmetric distribution (Laplace$(0,3)$; top row), and one right-skewed distribution (Gamma$(2,1)$; bottom row), with arrows indicating the order of the steps. The first column shows an original outlier-free sample, which generally follows a non-Gaussian distribution. The second column shows the same data after applying a SHASH transformation to (approximately) achieve standard \review{n}ormality.  If the original data arises from a SHASH distribution (as in the first simulation study), the true SHASH parameters are used for transformation; if the original data arises from a non-SHASH distribution (as in the second simulation study), SHASH parameters estimated via MLE from the large sample of $n=100{,}000$ shown in \textbf{Figure \ref{fig:SHASHo2dist}}. In the third column, outliers are introduced by replacing $x\%$ of the data (here $10\%$ for illustration) with values randomly sampled as $\chi^2(4)/5 + 3$, shown in orange. For distributions defined on the real line, outliers are randomly assigned positive or negative signs. Finally, the fourth column shows the contaminated data inversely transformed back to the original scale and distribution.}
    \label{fig:shash_transform_real_line}
\end{figure}

\review{
First, we perform a set of simulations to evaluate how well each transformation achieves normality. This requires data for which an exact transformation to normality exists. We therefore generate the non-outlying observations directly from four SHASH family distributions, representing examples of Gaussianity, heavy tails, light tails, and skew (Table \ref{table:shash_summary}). As illustrated in \textbf{Figure \ref{fig:shash_transform_real_line}}, outliers are generated beyond $\pm 3$ (as $\chi^2(4)/5 + 3$) on the normal scale into the tail(s) indicated in Table \ref{table:shash_summary}, at contamination levels ranging from $0\%$ to $20\%$; the outlier values are then transformed to the SHASH distribution by applying the inverse SHASH transformation with known parameter values. For each scenario, we perform $100$ independent simulation repetitions with sample sizes ranging from $n=100$ to $n=5{,}000$.
}

\review{Second, we perform a set of simulation studies using data arising from several parametric distributions, varying the degree of outlier contamination from $0\%$ to $20\%$.} 
We consider two main categories of distributions: (1) symmetric distributions with support on  $(-\infty, \infty)$ and (2) right-skewed distributions with support on $(0, \infty)$. Table \ref{table:distribution_summary} lists the distributions, their parameter values, and the rationale for their inclusion in this study. Again, outliers are generated beyond $\pm 3$ on the (approximately) normal scale, before applying the inverse transformation back to the SHASH distribution scale, returning the non-outlying observations back to their original values (see \textbf{Figure \ref{fig:shash_transform_real_line}}). Unlike the first simulation study, here the true SHASH parameters are unknown; we hence estimate them using MLE from the large sample of $n=100{,}000$ shown in \textbf{Figure \ref{fig:SHASHo2dist}}. For each scenario, we perform \review{$1{,}000$} independent simulation repetitions, each with a fixed sample size of $n = 500$.

\begin{table}[t]
    \centering
    \begin{tabular}{llll}
    \hline
    Support & Distribution & Parameters & Purpose \\
    \hline
    \multirow{3}{*}{$\mathbb{R}$} 
        & Normal($\mu$, $\sigma$)       & \review{$\mu = 10$, $\sigma = 3$ }     & Optimal case for all methods  \\
        & Student's $t$($\nu$)          & $\nu = 4$                     & Heavy-tailed \\
        & Laplace($\mu$, $\beta$)       & $\mu = 0$, $\beta = 3$        & Exponential shape \\
    \hline
    \multirow{3}{*}{$\mathbb{R}_+$} 
        & Gamma($\alpha$, $\beta$)      & $\alpha = 2$, $\beta = 1$     & Skewed with heavy tails \\
        & Chi-square($k$)               & $k = 3$                       & Skewed with heavier tails \\
        & Weibull($\lambda$, $k$)       & \review{$\lambda = 3$, $k = 1$} & Skewed with exponential shape \\
    \hline
    \end{tabular}
    \caption{Summary of distributions used in the second simulation study, along with selected parameters and evaluation purpose.}
    \label{table:distribution_summary}
\end{table}

\newcommand{\myfig}[1]{\includegraphics[width=0.17\textwidth,trim=2mm 0 0 0,clip]{#1}}
\newcommand{\rottextcell}[2]{%
  \rotatebox[origin=c]{90}{\shortstack{#1\\#2}}%
}


\newcommand{\vertlabel}[1]{%
  \begin{picture}(0,70)%
    \put(0,42){\rotatebox[origin=c]{90}{\small #1}}%
  \end{picture}%
}

\subsection{Evaluation Metrics} \label{sec:evaluation_metrics}

\review{We evaluate the performance of robust SHASH transformation and existing methods based on two metrics: ability to transform the data towards standard \review{n}ormality, and ability to detect outliers after transformation while minimizing false positives. Since we employ $N(0,1)$ quantiles for outlier detection, the accuracy of outlier detection provides an indirect measure of success for normality transformation, in addition to being an objective in its own right. }

\review{\textbf{Transformation toward standard normality.} For the first simulation study, we assess how close the transformed inlying observations are to the standard \review{n}ormal distribution, based on visual inspection of transformed distributions and a Cram\'er--von Mises goodness-of-fit test. Transformation parameters are estimated from each simulated dataset using three methods: robust SHASH transformation using SHASH-i, non-robust SHASH transformation via MLE, and RewML robust power transformation.  The resulting transformations are applied, with the objective of transforming the truly inlying observations to standard normality.  Kernel density estimates and QQ plots are used to visualize the distributions of the transformed inliers.  As a formal test, we compute the Cram\'er--von Mises test statistic, $W^2 = n\int \{F_n(z)-\Phi(z)\}^2\,d\Phi(z)$, where $F_n$ is the empirical distribution function of the transformed inlying observations, and $\Phi$ is the standard normal distribution function. Larger values indicate deviations from standard normality and thus reflect error in the estimated transformations.}
 
\review{\textbf{Outlier detection.}} \review{For the second simulation study, the primary focus is on the accuracy of outlier detection, since the non-Gaussian parametric distributions considered cannot be exactly transformed to normality. However, we do use visual inspection to qualitatively assess the ability of each method to approximately transform the inlying data observations to standard normality.} The accuracy of outlier detection is evaluated by comparing the flagged outliers against the true outliers in each simulated dataset. Two performance metrics are used in conjunction: the true positive rate (TPR) or sensitivity, and the false positive rate (FPR) or 1 - specificity. True positives (TP) are truly outlying observations that are correctly identified as outliers; false positives (FP) are truly non-outlying observations that have been incorrectly flagged as outliers; true negatives (TN) are truly non-outlying observations correctly classified as non-outliers; false negatives (FN) are true outliers not identified as outliers. The true positive rate $\text{TPR} = {\text{TP}}/({\text{TP} + \text{FN}})$ quantifies a method’s sensitivity to actual outliers. A TPR close to 1.0 indicates that the method detects most of the true outliers. The false positive rate $\text{FPR} = {\text{FP}}/({\text{FP} + \text{TN}})$ quantifies the proportion of truly non-outlying observations incorrectly identified as outliers. A low FPR \review{indicates} minimal misclassification of non-outlying data observations as outliers, i.e. high specificity. If the data is perfectly transformed to standard \review{n}ormality, applying an outlier detection threshold of $\pm 3$ should yield a false positive rate (FPR) of $0.003$.


\subsection{Simulation Results} \label{sec:results_SHASH}

\review{Here, we present results of both simulation studies and assess the ability of robust SHASH transformation and existing methods to achieve standard normality and accurately identify outliers in a variety of settings. } 

\review{Occasionally, SHASH transformation leads to a degenerate fit, wherein extreme parameter estimates lead to large values after transformation. We flag a fit as degenerate if more that 1\% of observations have absolute transformed values greater than $20$. Degenerate fits tend to arise when the data from which the SHASH parameters are estimated exhibits sharply light tails, which can occur due to overly aggressive outlier initialization. In practice, we recommend adjusting the iForest+ anomaly score threshold from 0.6 to a slightly higher value when this occurs. In the results presented below, we exclude any degenerate fits and report the percentage of iterations affected and how their exclusion impacts performance metrics.}

\begin{figure}
    \centering
    \includegraphics[width=\textwidth]{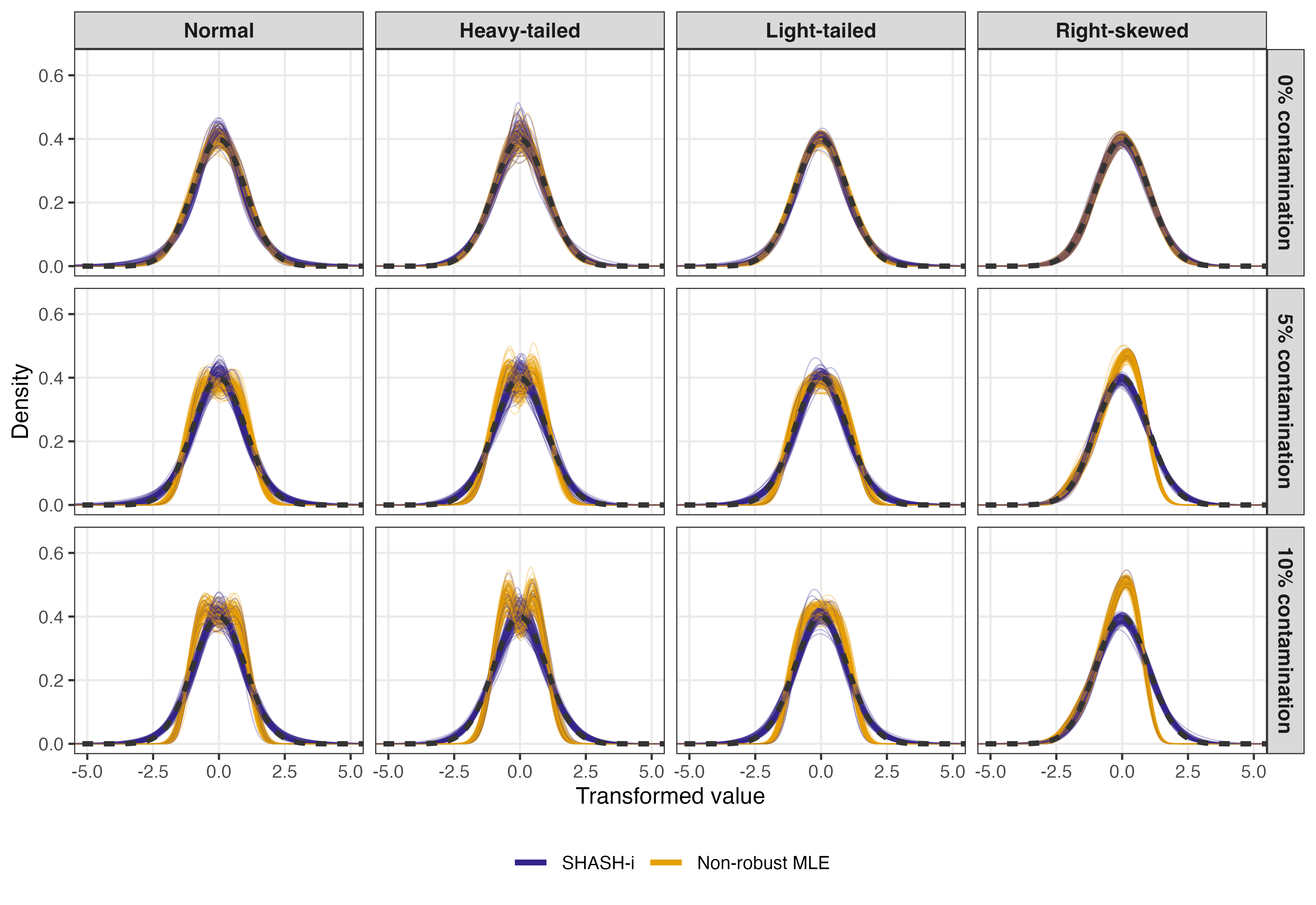}
    \caption{\review{\small\textbf{Densities of transformed data under robust and non-robust SHASH transformation.} For robust SHASH transformation, iForest+ outlier initialization was used. Densities represent the distribution among truly inlying observations, excluding any outliers, after transformation. Sample size is fixed to $n = 500$, with 100 independent iterations for each scenario. Dashed lines show the standard \review{n}ormal density, which represents the transformation with no error.}}
    \label{fig:density_n500}
\end{figure}

\review{\textbf{Simulation study 1.}}
\review{\textbf{Figures \ref{fig:density_n500}} and \textbf{\ref{fig:qq_n500}} display kernel density estimates and normal QQ plots, respectively, of the truly inlying observations arising from SHASH distributions, transformed using estimated transformation parameters ($n=500$). Outlier contamination ranges from $0\%$ to $10\%$. Here we focus on robust versus non-robust SHASH transformation, while the formal tests below also include robust RewML power transformations. With no outliers, both robust and non-robust SHASH transformation are fairly successful at transforming all distributions to standard normality. Robust SHASH transformation results in somewhat heavier tails for the symmetric distributions, due to the incorrect exclusion of some non-outliers in both tails for transformation parameter estimation. Still, robust SHASH transformation performs reasonably well when no true outliers are present. However, non-robust SHASH transformation quickly breaks down in the presence of outliers, as those observations exert a strong influence on the transformation. For normally distributed, heavy tailed, and light tailed data, the non-robustly transformed distributions of inliers appear markedly light-tailed, as the non-robust transformation accommodates the outliers and non-outlying observations are compressed towards the center. For right-skewed data, the non-robustly transformed distribution of inliers appears left-skewed, again due to the non-robust transformation accommodating the outliers in the right tail.  These effects intensify as contamination increases. These distortions would likely cause frequent outlier masking. By contrast, robust SHASH transformation consistently transforms distributions close to normality under outlier contamination, though with somewhat heavier-than-Gaussian tail weight.  This would likely result in slightly elevated false positive outlier detection rates, but would  avoid the masking of outliers. }

\review{\textbf{Figure~\ref{fig:density_sample_sizes}} shows the same comparison at $10\%$ contamination for sample sizes from $n=100$ to $n=5{,}000$. As $n$ increases, robust SHASH transformation becomes less variable and converges close to standard normality, with slightly heavier-than-Gaussian tails, while non-robust transformation also becomes less variable but remains biased and clearly non-Gaussian due to the influence of outliers. In samples as small as $n=100$, robust SHASH transformation is noisier and shows more frequent susceptibility to outliers, but still approximately achieves normality in most cases.}

\begin{figure}
    \centering
    \includegraphics[width=\textwidth]{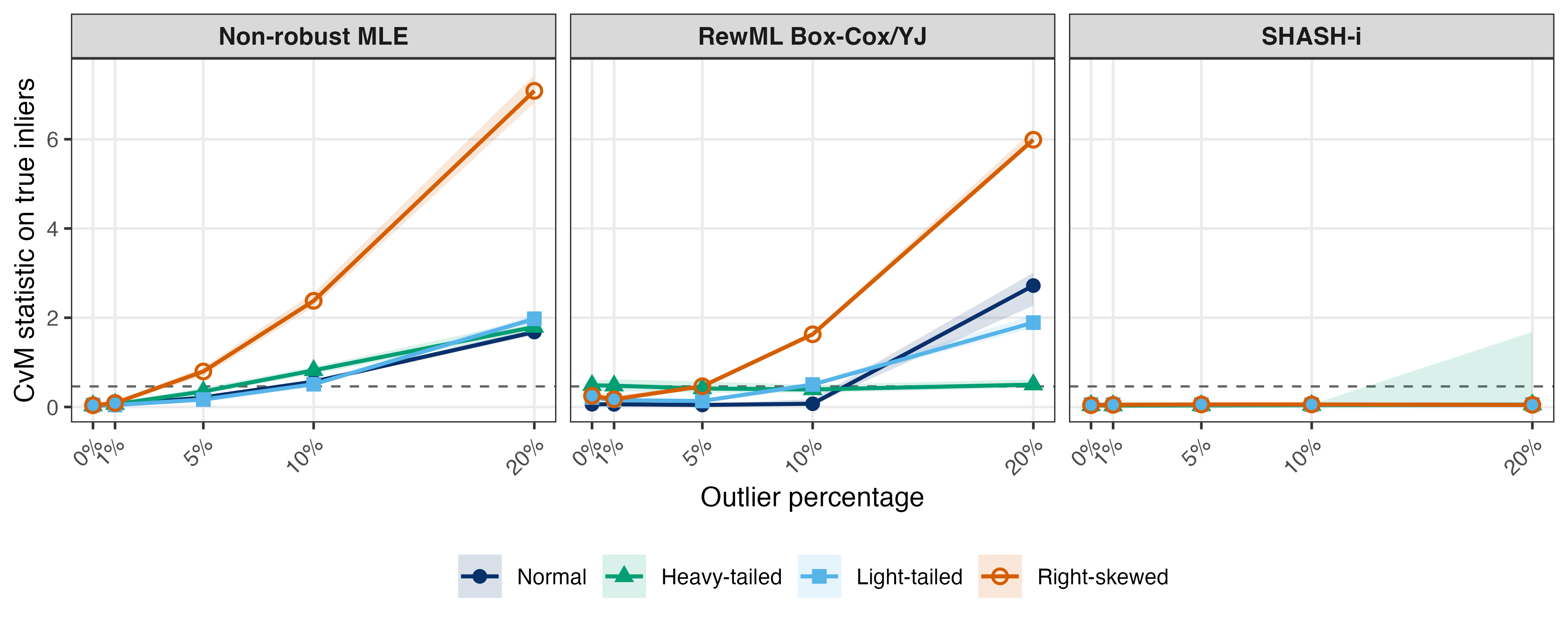}
    \caption{\review{\small\textbf{Hypothesis tests of departure from \review{n}ormality after transformation.} The y-axis shows the Cram\'er--von Mises $W^2$ test statistic between the transformed non-outlying observations and $N(0,1)$, at $n = 500$; larger values indicate departures from normality. Robust SHASH transformation uses iForest+ for outlier initialization. Lines are medians over 100 iterations; bands are IQRs. Dashed line shows the critical value at $\alpha=0.05$ for the null distribution of $W^2$, equal to 0.461. Several degenerate SHASH-i fits were excluded at 20\% contamination: 1 of 100 for \review{n}ormal data, and 11 of 100 for right-skewed data.}}
    \label{fig:cvm_n500}
\end{figure}

\review{\textbf{Figure~\ref{fig:cvm_n500}} shows the Cram\'er--von Mises statistic ($W^2$) for robust SHASH transformation, non-robust SHASH transformation, and RewML robust power transformation ($n=500$). Outlier contamination ranges from $0\%$ to $20\%$. Lines show the median value across $100$ simulations, and bands show the inter-quartile range (IQR). Consider first the case with up to $1\%$ outliers. For robust and non-robust SHASH transformation, $W^2$ is close to zero, indicating no statistically significant deviation from normality, while for RewML robust power transformation, $W^2$ is slightly elevated but generally not significant. As outlier contamination increases, however, non-robust SHASH transformation and RewML robust power transformations exhibit increasingly significant deviations from normality. Surprisingly, RewML performs worst for right-skewed data, given that power transformations target skew. This is consistent with patterns we observe in the second simulation study reported below, wherein outliers sometimes strongly influence the RewML power transformations for skewed distributions, resulting in masking of outliers (see \textbf{Figure~\ref{fig:method_comparison_examples}}).  For robust SHASH transformation, the medians show no significant deviations from normality for any distributions or contamination levels. A wider IQR at 20\% contamination for heavy tailed distributions indicates that a minority of robust SHASH transformations fail under heavy outlier contamination.}

\newcommand{\vertlabelB}[1]{%
  \begin{picture}(10,80)%
    \put(0,45){\rotatebox[origin=c]{90}{\small #1 }}%
  \end{picture}%
}

\newcommand{\imgcrop}[1]{
    \includegraphics[width=0.9\textwidth, height=1.7\textheight, keepaspectratio, trim=0 0 0 8mm, clip]{ #1 } 
}

\begin{figure}
    \centering
    \includegraphics[width=1\textwidth, trim = 0 0 0 0, clip]{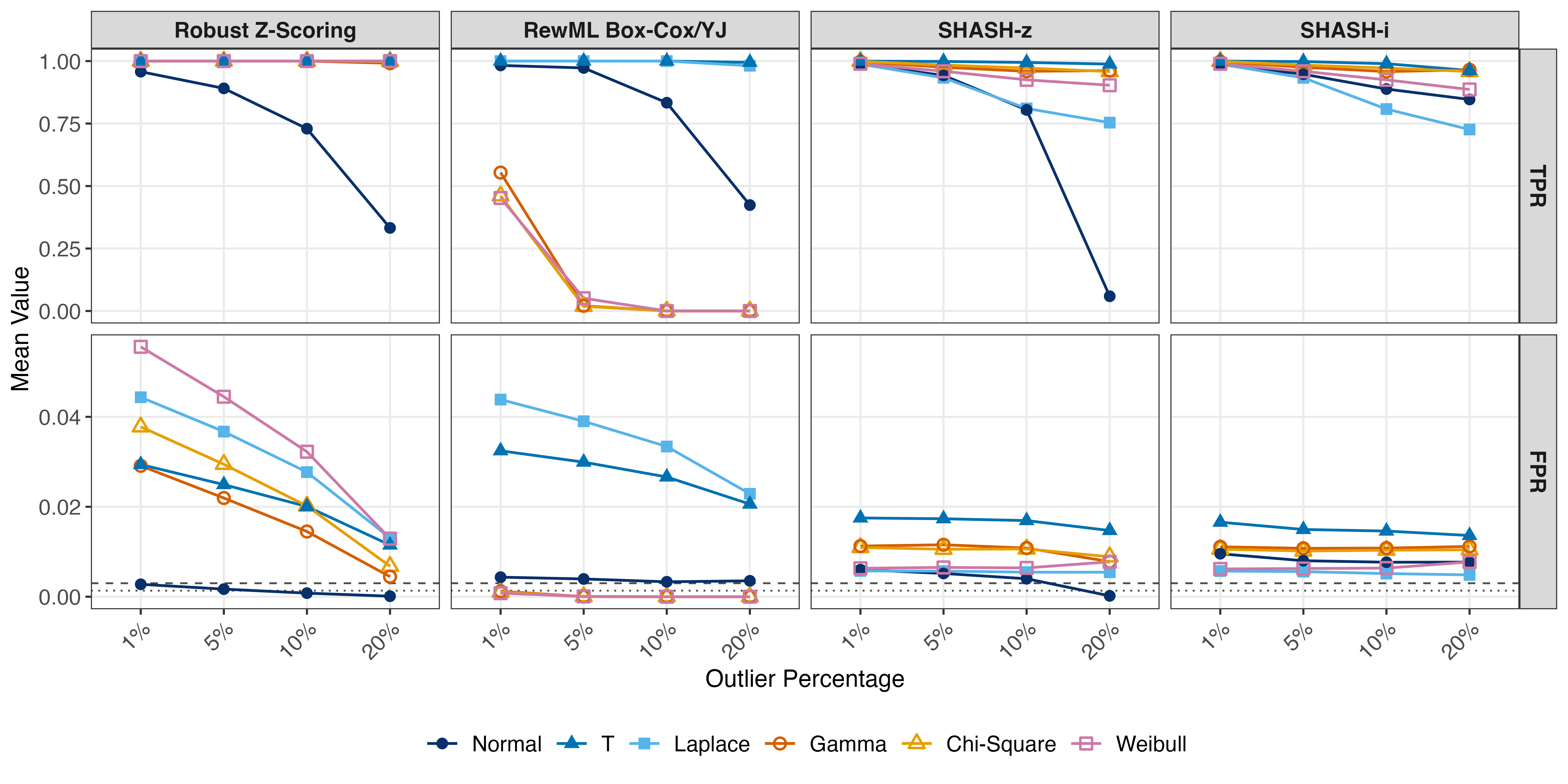}
    \caption{\textbf{Accuracy of outlier detection in simulation study.} True positive rate (TPR) and false positive rate (FPR) (mean over 1000 iterations) across distributions and outlier contamination levels for four robust 
    methods. The dashed and dotted horizontal lines in the FPR panel indicate the nominal false positive rates of 0.003 and 0.0014 in symmetric and skewed distributions, respectively. \review{For SHASH-z and SHASH-i, degenerate fits are excluded (see \textbf{Figure~\ref{fig:degen_rate}}).}}
    \label{fig:FPR_TPR_plot}
\end{figure}

\review{\textbf{Simulation study 2.}}
\textbf{Figure \ref{fig:method_comparison_examples}} shows randomly selected examples for each of the six distributions considered in the simulation study at 10\% outlier contamination; \textbf{Figures \ref{fig:varyingpercentagenormal}} to \textbf{\ref{fig:varyingpercentageweibull}} show examples at \review{contamination levels} from 1\% to \review{20\%}. After a successful transformation, the data should approximately follow a standard \review{n}ormal distribution, and true outliers will fall outside the thresholds, while non-outliers will mainly fall within them. \review{To summarize performance of each method,} \textbf{Figure \ref{fig:FPR_TPR_plot}} shows \review{their} TPR and FPR over $1000$ simulated datasets for each distribution and outlier contamination level.

Consider first the Gaussian case. If the data is known to follow a \review{n}ormal distribution, then no transformation is needed.  However, we may not have this knowledge a priori, so by applying transformation to \review{normally distributed} data, we are evaluating whether there is a detrimental effect of transformation. At 10\% outlier contamination, all methods perform well. However, \textbf{Figure \ref{fig:varyingpercentagenormal}} shows that at higher outlier contamination, robust z-scoring, \review{RewML power} transformation, and SHASH-z transformation all tend to fail.  Since those methods depend on robust z-scoring \review{either for initialization or updating of outliers}, they are likely failing due to the undue influence of outliers with robust z-scoring. By contrast, SHASH-i transformation is seen to work quite well, even at \review{20\%} outlier contamination. \textbf{Figure \ref{fig:FPR_TPR_plot}} likewise shows that the sensitivity of robust z-scoring, RewML power transformation, and SHASH-z to outliers suffers markedly as the outlier contamination level increases, while SHASH-i remains sensitive.

Why would robust z-scoring fail even in Gaussian data under contamination? We examine the Monte Carlo sampling distributions of the median and MAD \review{ across 1{,}000 simulated datasets of size $n=500$} in Gaussian and skewed data (\textbf{Figure \ref{fig:MedianMADNormalAndGamma}}). In the Gaussian case, the median remains unbiased as the proportion of outliers grows. However, the MAD scale estimator grows as the number of outliers increases. \review{This reflects the fact that many robust estimators of location and scale are inconsistent under contamination \citep{martin1993bias, klooster2025inconsistency}.} When outliers are infused disproportionately across both tails, as in our skewed simulation scenarios, the median is also inconsistent. This breakdown of the MAD and even the median under contamination will lead to suboptimal performance of robust z-scoring, and along with it any method that depends on it to identify and exclude outliers. \textbf{Figure \ref{fig:CenterScaleSideBySide}} considers the bias under contamination for several other robust estimators of center and scale, including Huber estimators \citep{huber1981robust} and $Q_n$ and $S_n$ estimators \citep{rousseeuw1993alternatives}.  Across all distributions, the Huber estimators show the greatest bias under contamination. This is likely why \review{RewML power} transformations break down more quickly under contamination than robust z-scoring or SHASH-z. The $Q_n$ and $S_n$ estimators are more biased than the MAD in symmetric distributions, while $Q_n$ shows slightly less bias in skewed distributions. Thus, the existence of randomly occurring, large-magnitude, possibly asymmetric outliers is a problem for many robust estimators of center and scale, and any method that relies on them to identify outliers will likewise deteriorate under contamination.

\begin{figure}
    \centering
    
    \begin{tabular}{@{}c@{\hspace{3pt}}c@{}}

       &{\scriptsize
       \makebox[0.98\textwidth][c]{%
       \parbox[c]{0.19\textwidth}{\centering Outlier-infused values}%
       \parbox[c]{0.19\textwidth}{\centering Robust z-scoring}%
       \parbox[c]{0.19\textwidth}{\centering RewML Box-Cox/YJ}%
       \parbox[c]{0.19\textwidth}{\centering SHASH-z}%
       \parbox[c]{0.19\textwidth}{\centering SHASH-i}%
       }}
       \\[-10pt]
        
        \vertlabelB{\scriptsize Normal $(10,3)$} &
        \includegraphics[
            width=0.98\textwidth,
            trim = {0.4 0.8cm 0 1cm},
            clip,
            keepaspectratio
        ]{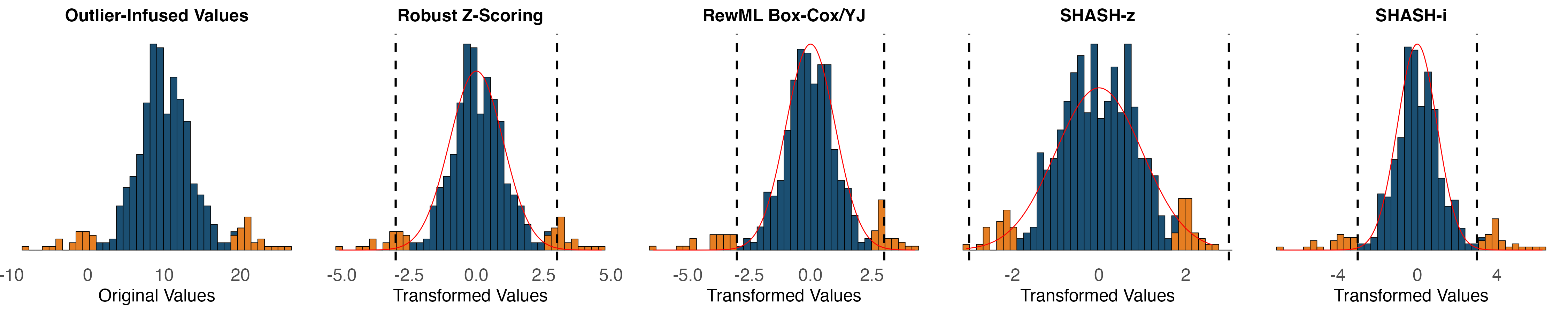}
        \\[-10pt]

        \vertlabelB{\scriptsize $t(4)$} &
        \includegraphics[
            width=0.98\textwidth,
            trim = {0.4 0.8cm 0 1cm},
            clip,
            keepaspectratio
        ]{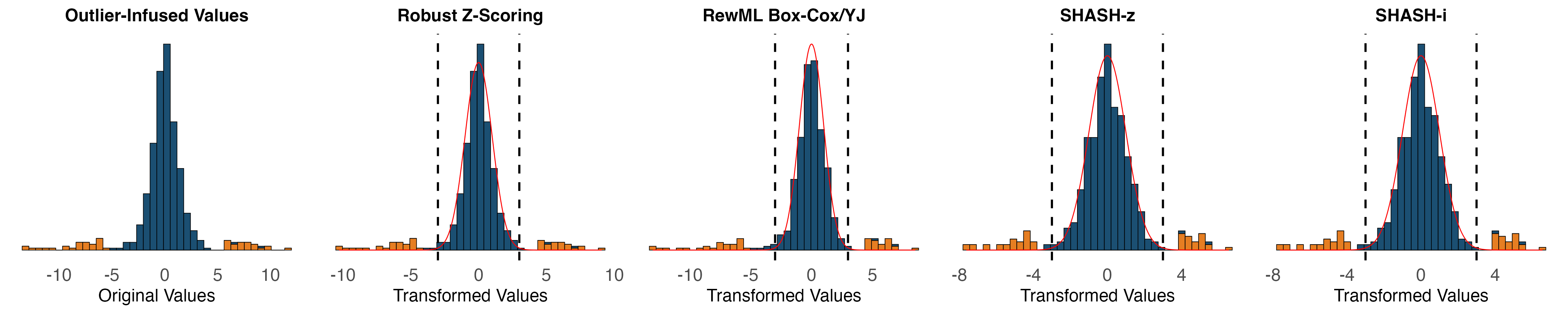}
        \\[-10pt]

        \vertlabelB{\scriptsize Laplace $(0,3)$} &
        \includegraphics[
            width=0.98\textwidth,
            trim = {0.4 0.8cm 0 1cm},
            clip,
            keepaspectratio
        ]{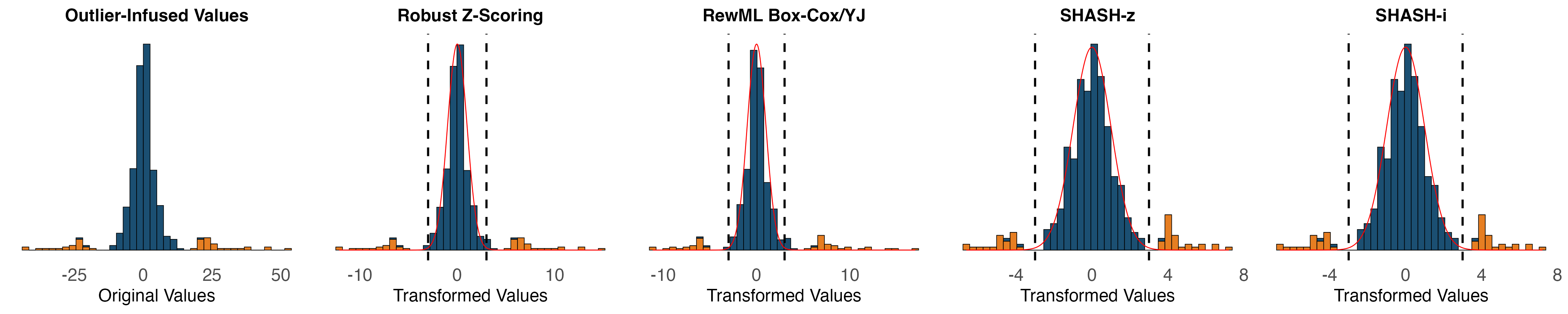}
        \\[-10pt]

        \vertlabelB{\scriptsize Gamma $(2,1)$} &
        \includegraphics[
            width=0.98\textwidth,
            trim = {0.4 0.8cm 0 1cm},
            clip,
            keepaspectratio
        ]{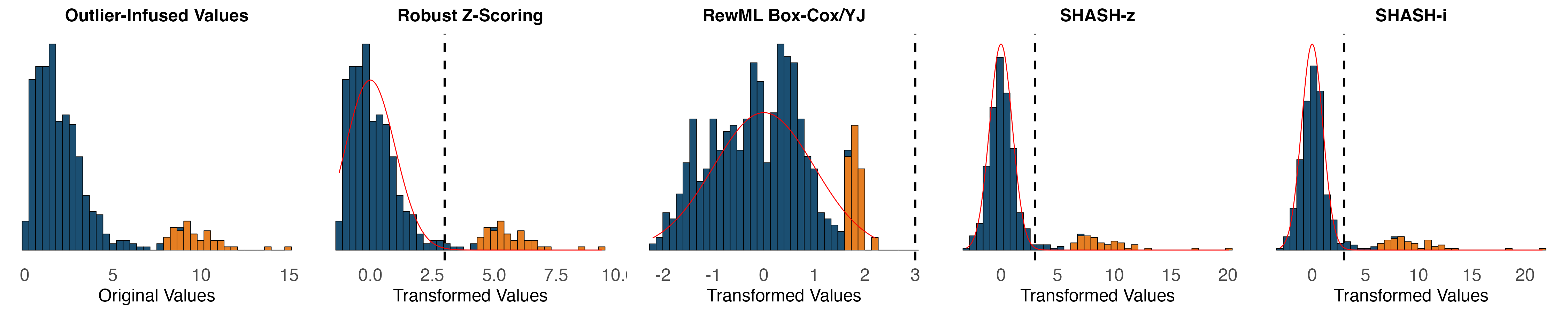}
        \\[-10pt]

        \vertlabelB{\scriptsize Chi-square $(4)$} &
        \includegraphics[
            width=0.98\textwidth,
            trim = {0.4 0.8cm 0 1cm},
            clip,
            keepaspectratio
        ]{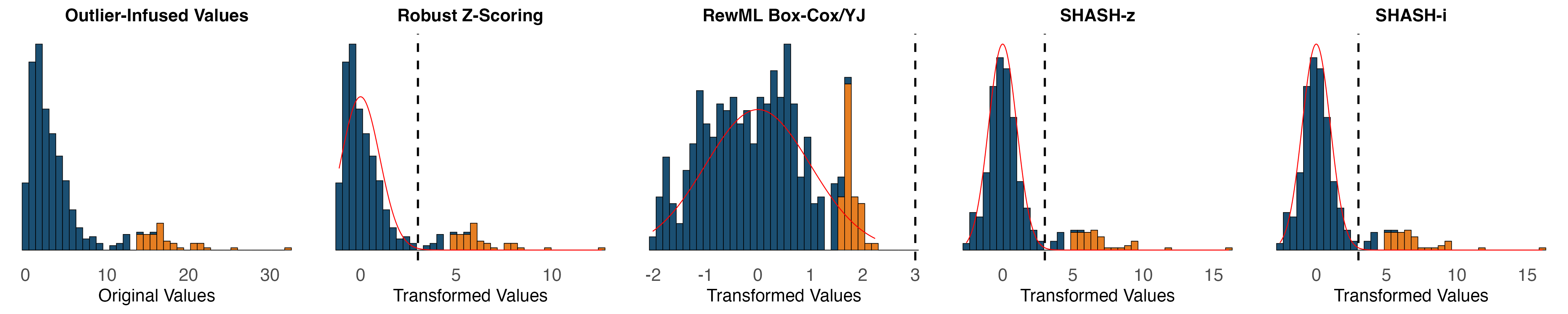}
        \\[-10pt]

        \vertlabelB{\scriptsize Weibull $(3,1)$} &
        \includegraphics[
            width=0.98\textwidth,
            trim = {0.4 0.8cm 0 1cm},
            clip,
            keepaspectratio
        ]{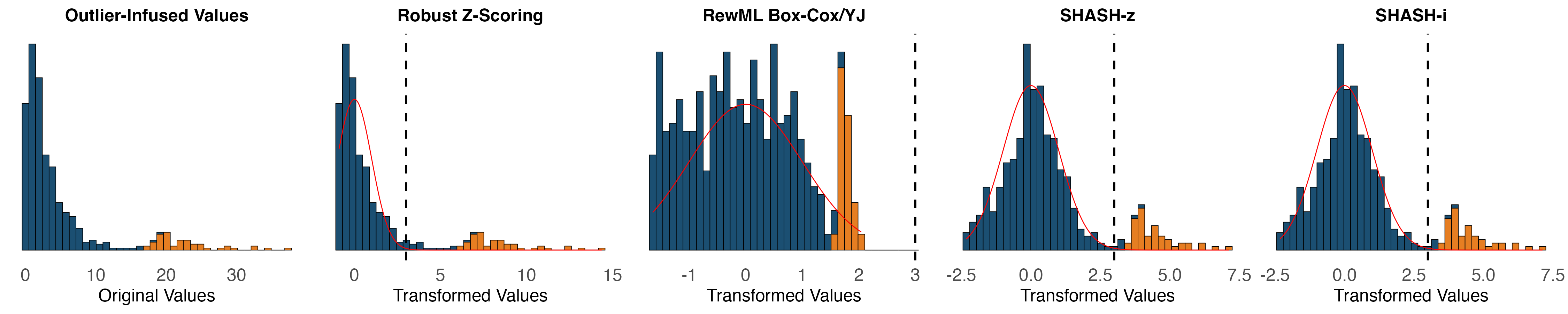}

    \end{tabular}

    \includegraphics[width=4cm]{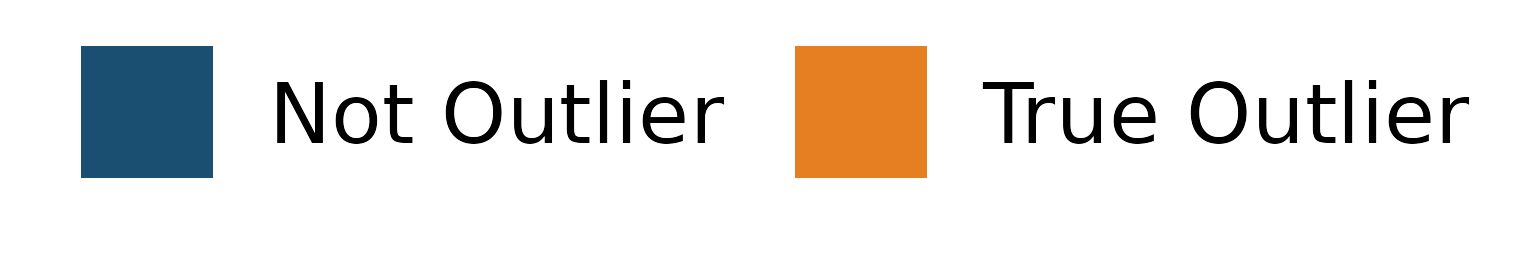}

    \caption{\review{\small \textbf{Example histograms comparing robust z-scoring, RewML Box-Cox/YJ, SHASH-z, and SHASH-i under $10\%$ outlier contamination.} For each distribution, one randomly selected simulated dataset is shown.The first column displays the outlier-infused observations on their original scale, whereas the remaining columns display the values on the corresponding transformed scales. The vertical dashed lines indicate the outlier detection thresholds, the true outliers are highlighted in orange, and the red curve represents the standard \review{n}ormal density for reference. The methods generally identify the true outliers successfully; however, for the right-skewed distributions, RewML Box-Cox/YJ masks all true outliers in the examples shown.}}
    \label{fig:method_comparison_examples}
\end{figure}

Turning to the symmetric heavy tailed (\textit{t} and Laplace) distributions, both robust z-scoring and \review{RewML Box-Cox/YJ} transformation misclassify some non-outliers as outliers due to their heavy tails. The \review{RewML Box-Cox/YJ} transformation is not equipped to transform heavy-tailed distributions to \review{n}ormality, since it is designed \review{primarily to address skew}. In these examples, the SHASH transformations reduce the tailweight and largely avoid misclassification of regular observations as outliers. These differences are easiest to discern at $1\%$ outlier contamination in \textbf{Figures} \textbf{\ref{fig:varyingpercentageT}} and \textbf{\ref{fig:varyingpercentagelaplace}}, where the differential effects of the transformations themselves dominate, rather than the potential influence of outliers on those transformations.

For skewed (Gamma, Chi-square, and Weibull) distributions, both SHASH transformations achieve near symmetry \review{ at 10\% contamination}, with outliers isolated in the right tail. Although some regular observations are misclassified as outliers, they are generally fewer than prior to transformation. Quite surprisingly, \review{RewML Box-Cox/YJ} transformation seems to fail consistently even at 5\% outlier contamination for all three skewed distributions, with true outliers falling well below the outlier detection threshold. This indicates masking of outliers due to their influence on the transformation parameter.

Considering the FPR and TPR over 1000 iterations shown in \textbf{Figure \ref{fig:FPR_TPR_plot}},  SHASH-i exhibits the strongest overall performance, in terms of high TPR while maintaining low FPR even under heavy contamination. While robust z-scoring and RewML power transformation exhibit near-perfect TPR in some scenarios, this is accompanied by very high FPR. Likewise, those methods exhibit nominal FPR in some scenarios, but this is accompanied by low TPR as contamination increases.  As expected based on the slightly heavier-than-Gaussian tails induced by robust SHASH transformation (see \textbf{Figures \ref{fig:density_n500}} and \textbf{\ref{fig:qq_n500}}), SHASH-i and SHASH-z do not achieve the nominal FPR and instead have FPR of 1-2\%, which is consistent across contamination levels and distributions. Yet overall, SHASH-z and especially SHASH-i achieve a better balance between TPR and FPR than the other methods and are more stable under contamination and across distributions. 

\review{Since SHASH-i relies on iForest+ for outlier initialization, we also evaluated iForest+ alone as an outlier detection method (\textbf{Figure~\ref{fig:R_iforest}}). iForest+ achieves near-perfect sensitivity---the goal of our outlier initialization procedure to eliminate the effect of outliers on the transformation---but comes at the cost of a substantially higher false positive rates than SHASH-i across distributions and contamination levels. This indicates that the subsequent robust normality transformation and iterative updating improve specificity beyond the initialization alone. Furthermore, SHASH-i provides transformation towards normality, not just outlier detection alone.}

\review{Degenerate SHASH fits excluded from the TPR and FPR calculations were rare, affecting only 1.3\% of SHASH-i fits and 0.7\% of SHASH-z fits. These were nearly all for the Weibull distribution at 20\% contamination (\textbf{Figure~\ref{fig:degen_rate}}). Excluding them changes the mean TPR and FPR of either method by less than 0.001 overall. In practice, these cases can be avoided by relaxing the outlier initialization step. For SHASH-i, this can be accomplished by increasing the iForest+ anomaly score threshold to avoid overly light tails in the to-be-transformed dataset. To illustrate the importance of anomaly score threshold, we experimented with decreasing it from 0.6 to 0.55 or 0.5. We observed a dramatic impact on the rate of degenerate fits, which grew from 1.3\% to 5.3\% to 11.2\% as the iForest anomaly score threshold was lowered. Another possible remedy in the case of extreme right skew is to apply log transformation as a preprocessing step, which we explore in the next section.}

\section{Real Data Analysis} \label{sec:toydatasets}

\review{We apply the robust SHASH transformation and outlier detection procedure to several real-world outlier detection datasets and compare its performance with RewML Box-Cox/YJ transformation \citep{raymaekers2021transforming} in terms of transformation and outlier detection.} These datasets represent a wide variety of distribution shapes and contexts.  In some cases the outlier identities are known or can be reasoned.  Specifically, we consider three datasets:

\begin{enumerate}
    \item The Top Gear dataset \citep{alfons2019robustHD} includes $297$ production vehicles featured on the BBC’s Top Gear website, with variables such as engine specifications, fuel economy and performance metrics. We consider the MPG and Weight variables, which were previously examined by \citet{raymaekers2021transforming} to evaluate the performance of \review{RewML Box-Cox/YJ} transformation. In addition, we consider Price, given its extreme right skew and the presence of several expensive models that can be considered outliers.  
    \item The HBK dataset is a well-known dataset introduced by
    \cite{hawkins1984robustbase} consisting of $75$ observations on four variables. Fourteen ($18.6\%$) of the observations are known to be outliers \citep{robustbase}\review{: observations 1--14 are outlying in the three
    explanatory variables (X1--X3), while only observations 1--10 are outlying in the response (Y)}. Given the number and size of the outliers, they are potentially influential on the transformation, making this dataset particularly susceptible to the masking effect. Since the outlier identities
    are known, the performance of each method can be assessed.
    \item The Wood dataset \citep{rousseeuw1987robust, robustbase} contains
    $20$ observations on five variables related to wood-specific gravity. A noteworthy feature of this dataset is that the distributions exhibit short tails, which the SHASH transformation is well suited to. These data were
    contaminated by replacing $4$ of the $20$ observations (observations 4, 6, 8 and 19) with outliers, so the outlier identities are known. This data set is analyzed in the principal component space using robust PCA \citep{Hubert2005ROBPCA} since the outliers are multivariate.
\end{enumerate}

\begin{figure}
    \centering
    \includegraphics[width=0.72\textwidth,
                 height=1\textheight,
                 keepaspectratio]{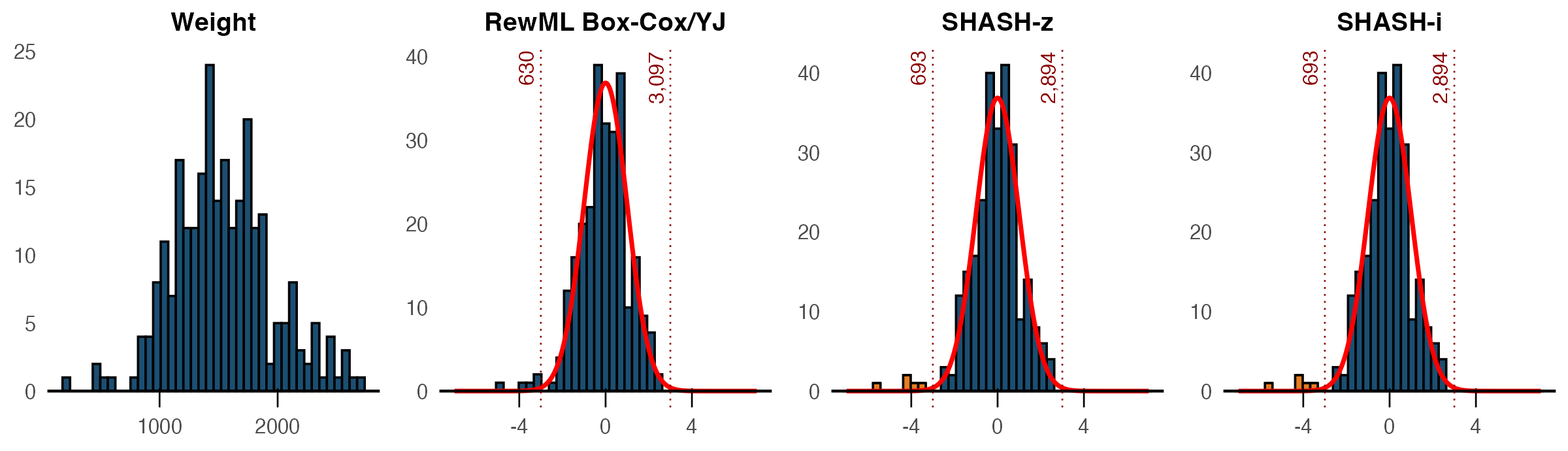} \\[6pt]
    \includegraphics[width=0.72\textwidth,
                 height=1\textheight,
                 keepaspectratio]{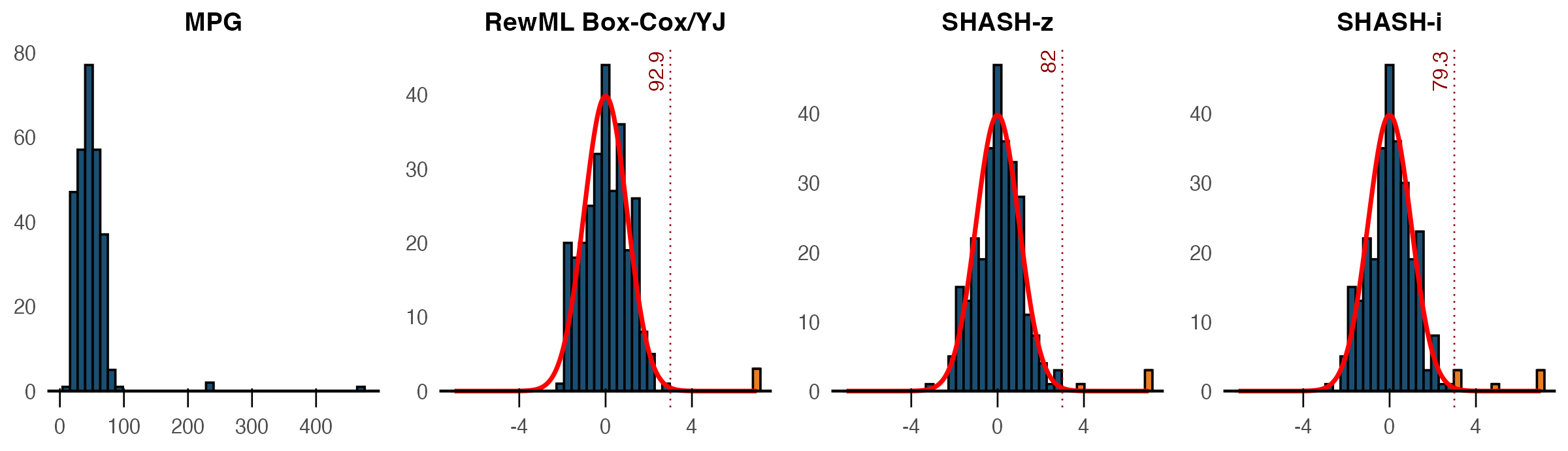} \\[6pt]
    \includegraphics[width=0.72\textwidth,
                 height=1\textheight,
                 keepaspectratio]{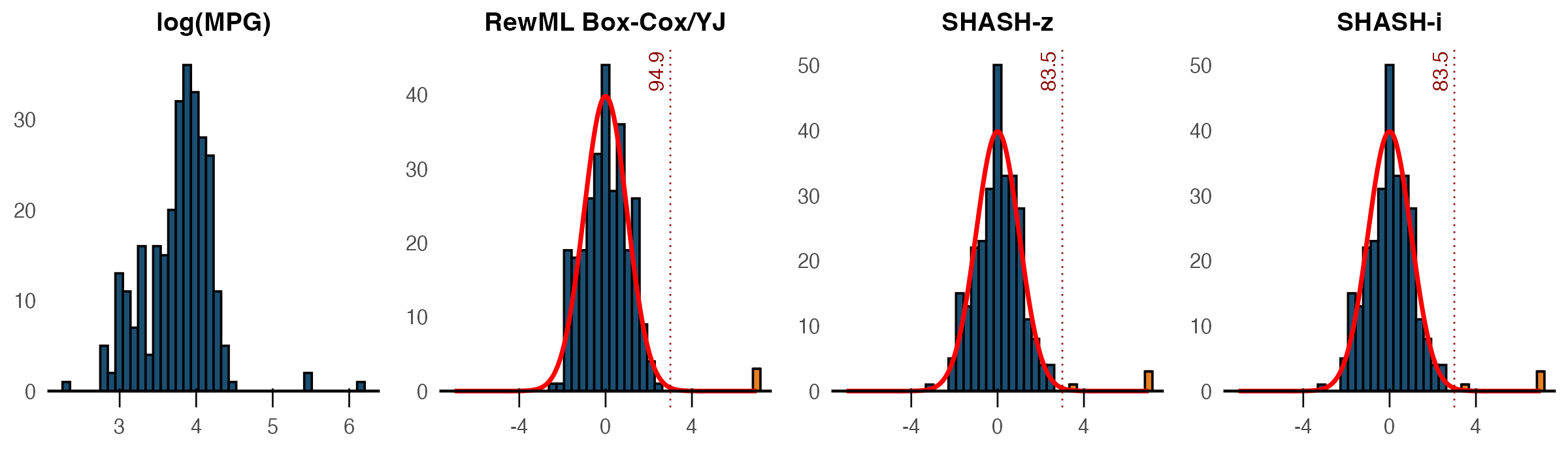} \\[6pt]
    \includegraphics[width=0.72\textwidth,
                 height=1\textheight,
                 keepaspectratio]{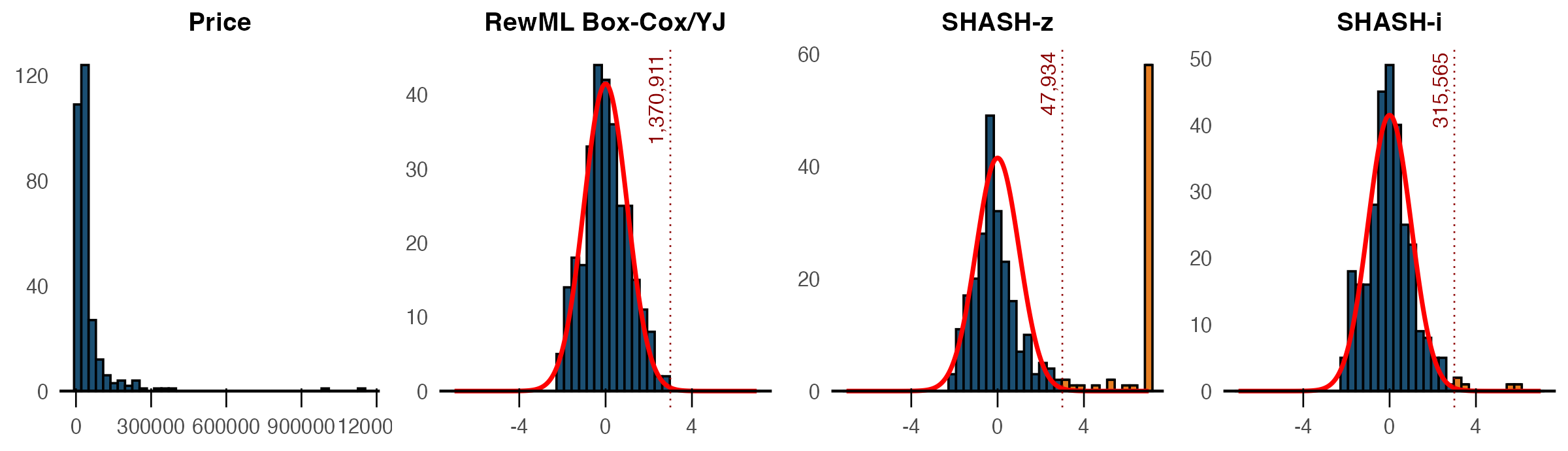} \\[6pt]
    \includegraphics[width=0.72\textwidth,
                 height=1\textheight,
                 keepaspectratio]{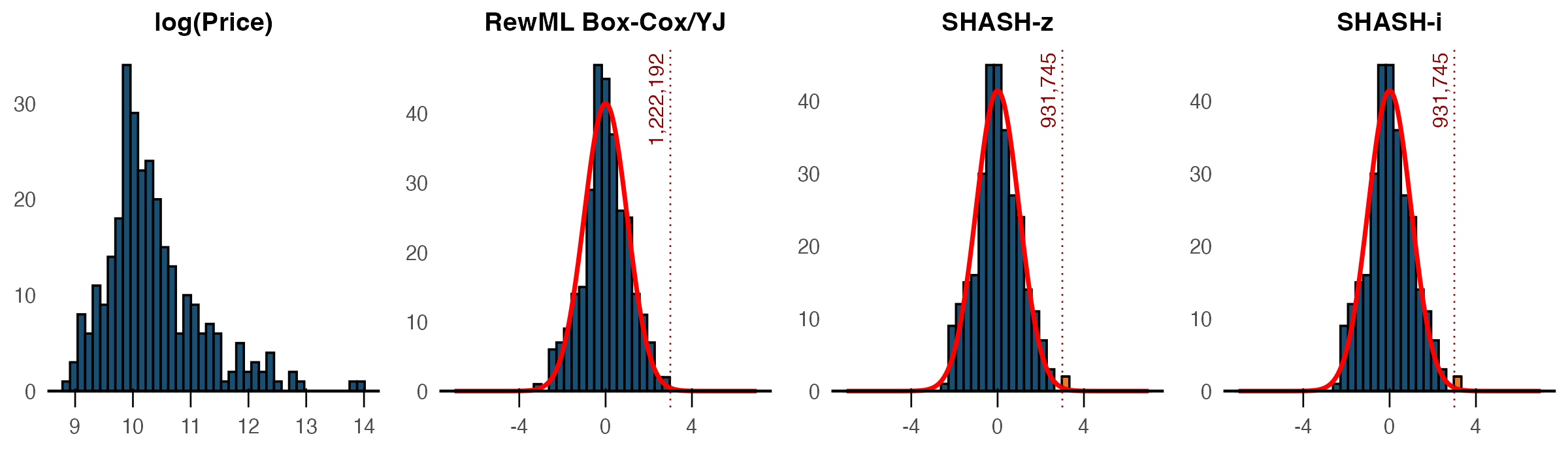} \\
    \includegraphics[width=5cm]{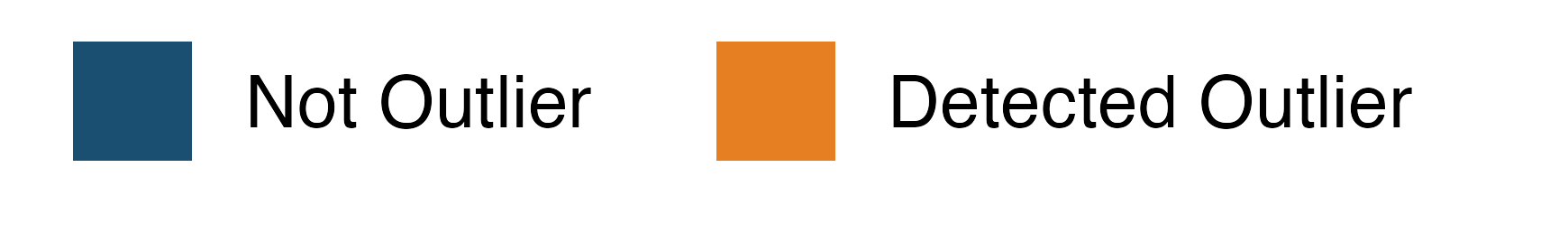}
    \caption{\small\textbf{\review{Application of outlier detection methods to the
Top Gear dataset.}} \review{Rows show Weight ($n = 264$), MPG and log(MPG)
($n = 285$), and Price and log(Price) ($n = 297$); vehicles with missing values
are excluded from each variable's analysis.} True outlier identities are
unknown; color indicates outlier flags based on each method. Values beyond
$\pm 7$ are truncated for visualization purposes. For each transformation, the
outlier detection threshold on the original scale of the data is annotated on
the plot\review{; for log(MPG) and log(Price), thresholds are back-transformed
to MPG and pounds}.}
    \label{fig:topgear}
\end{figure}

\review{\textbf{Figure~\ref{fig:topgear}} displays the results for the Top Gear variables Weight, MPG and Price, with MPG and Price also shown after applying log transformation as a preprocessing step to mitigate extreme right skew. For Weight, all three methods flag the same 5 lightweight vehicles, despite slightly different thresholds. For MPG, all three methods flag the most extreme vehicles, but both SHASH-z and SHASH-i yield lower thresholds than RewML Box-Cox/YJ, resulting in $1$ and $4$ additional detected outliers, respectively. Whether these additional vehicles should be regarded as outliers is context-dependent and can be controlled by selecting a more or less conservative post-transformation threshold. After log transformation, the methods are in closer agreement: SHASH-z and SHASH-i both flag $4$ vehicles, while RewML Box-Cox/YJ flags the same $3$ extreme outliers.}

\review{Price is the most challenging variable due to extreme right skew, with most vehicles concentrated at below \pounds100,000, a smaller subset between \pounds100,000 and \pounds400,000, and a long right tail that includes the two most expensive vehicles, the Bugatti Veyron (\pounds1,139,985) and the Pagani Huayra (\pounds990,000). RewML Box-Cox/YJ gives a threshold of \pounds1,370,911 and flags no vehicles, not even the two most expensive ones, suggesting masking due to the influence of the right tail on the transformation. SHASH-i gives a threshold of \pounds315,565 and flags the $5$ most expensive vehicles. The SHASH-z fit to Price is degenerate, resulting in $67$ outlier flags. With log transformation, the SHASH-z fit is no longer degenerate, and SHASH-z and SHASH-i are in perfect agreement, both giving a threshold of \pounds931,745 and flagging the $2$ most expensive vehicles. This illustrates that a log transformation can serve as a useful preprocessing step to facilitate SHASH transformation to normality on strictly positive, highly skewed data. RewML Box-Cox/YJ again flags no vehicles as price outliers, giving a threshold of \pounds1,222,192, again suggesting the influence of outliers on the transformation and their consequent masking.}

\begin{figure}
    \centering
    \includegraphics[width=0.9\textwidth,height=1\textheight,
         keepaspectratio]{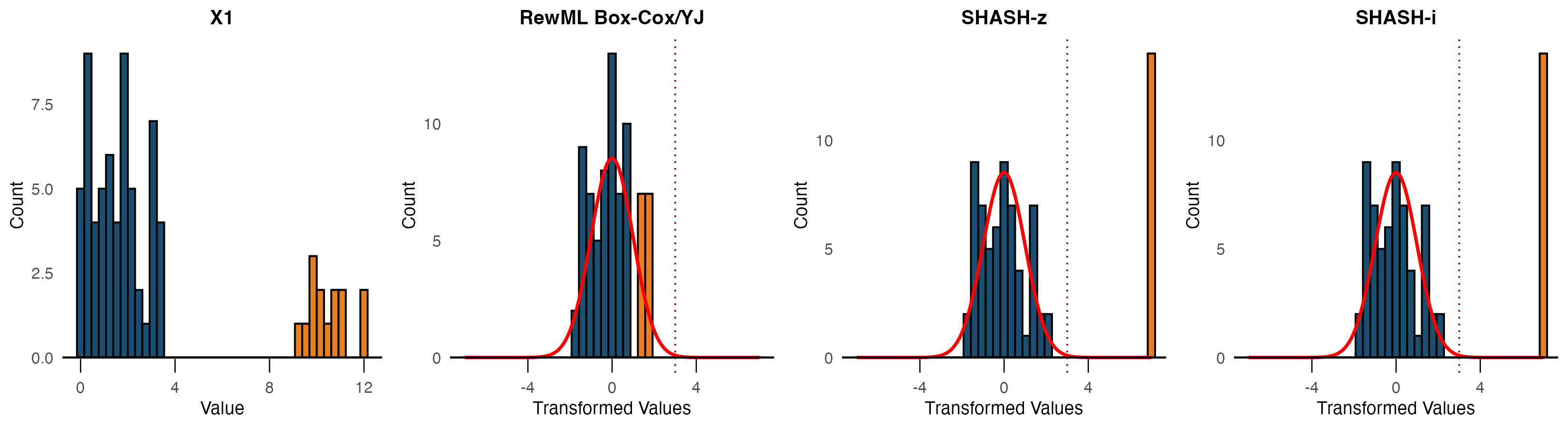} \\[12pt]
    \includegraphics[width=0.9\textwidth, height=1\textheight,
         keepaspectratio]{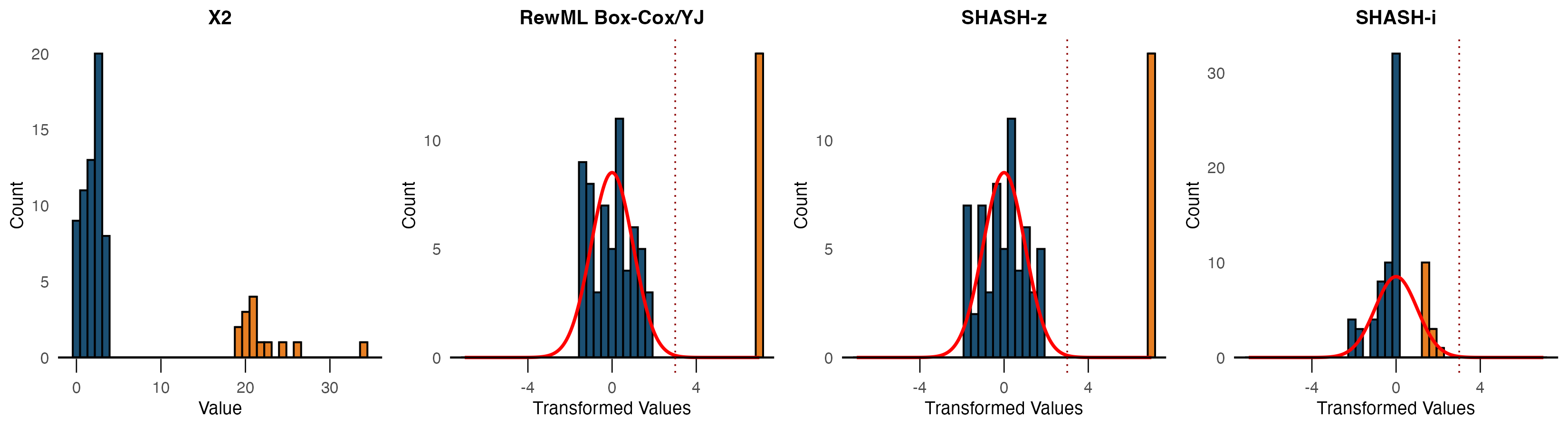} \\[12pt]
    \includegraphics[width=0.9\textwidth, height=1\textheight,
         keepaspectratio]{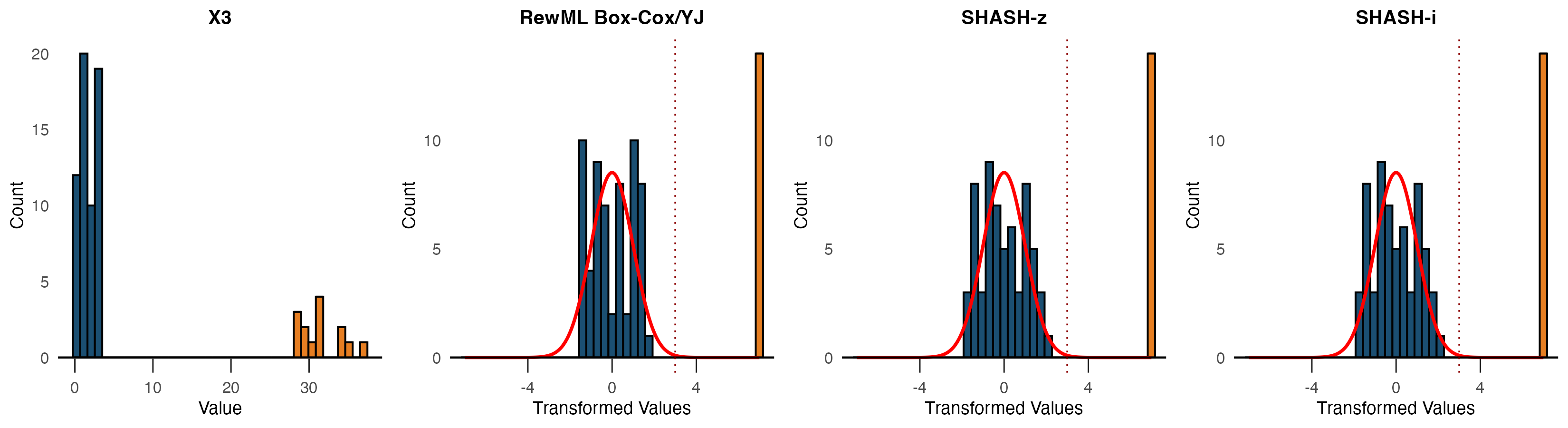} \\[12pt]
    \includegraphics[width=0.9\textwidth, height=1\textheight,
         keepaspectratio]{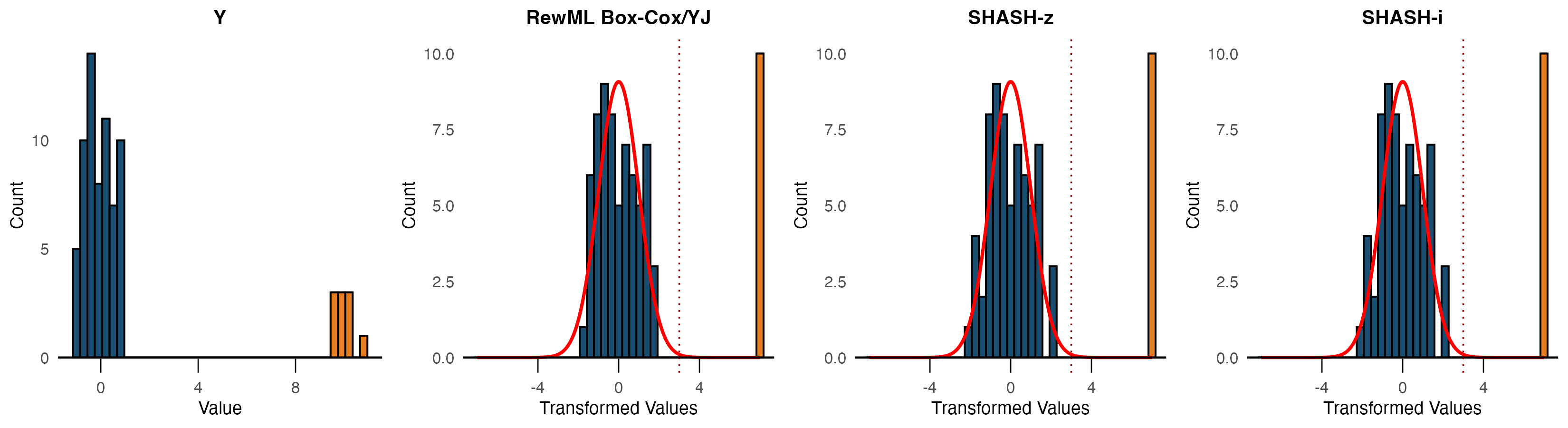} \\
    \includegraphics[width=4cm]{figures/toydatasets/status_legend_1X2.png}
 \caption{\small\textbf{\review{Application of outlier detection methods to four variables in the HBK dataset ($n = 75$).}} True outlier identities are indicated with color\review{; observations 1--14 are outliers in X1--X3, and observations 1--10 are outliers in Y}. Transformed values are truncated at $7$ for visualization purposes. Histograms with a large spike at $7$ reflect degenerate fits.}
    \label{fig:hbktoydatacomparison}
\end{figure}

\review{\textbf{Figure~\ref{fig:hbktoydatacomparison}} displays the results for the four variables in the HBK dataset, for which the true outliers are known: observations 1--14 in X1--X3 and observations 1--10 in Y. On the original scale, the outliers form a distinct cluster in the right tail of each variable, and given their number they have strong potential to influence the transformation. SHASH-z separates the true outliers from the non-outlying observations in all four variables. RewML Box-Cox/YJ and SHASH-i each succeed in three of the four variables, but mask all of the true outliers in the remaining variable: X1 for RewML Box-Cox/YJ and X2 for SHASH-i. For RewML Box-Cox/YJ, this is consistent with the limited robustness of the Huber estimators used during the algorithm to exclude outliers (see Section~\ref{sec:results_SHASH}). For SHASH-i, the failure reflects incomplete initialization of outliers via iForest+. Specifically, the smallest outlier with iForest anomaly score over 0.6 has a value of 23; the densely clustered group of true outliers below 23 are not flagged and are allowed to influence the transformation. 
iForest is known to be vulnerable to masking when anomalies form dense clusters, since it assumes that anomalies can be isolated
from other observations \citep{liu2008isolation, liu2012isolation}. 
SHASH-i inherits this limitation of iForest and may struggle in the presence of tightly clustered outliers. Note that large transformed values
are truncated at 7 for visualization purposes but may be much larger; these extremes reflect degenerate fits, which are common for extremely light tailed distributions---the case for all four HBK variables. Indeed, SHASH transformation yields transformed inlier distributions with lighter-than-Gaussian tails, even though it successfully segregates outliers and brings the inliers closer to standard normality. This illustrates that robust SHASH transformation can be useful even when it does not fully achieve normality.}

\begin{figure}
    \centering
    \includegraphics[width=0.9\textwidth, height=1\textheight, 
    keepaspectratio]{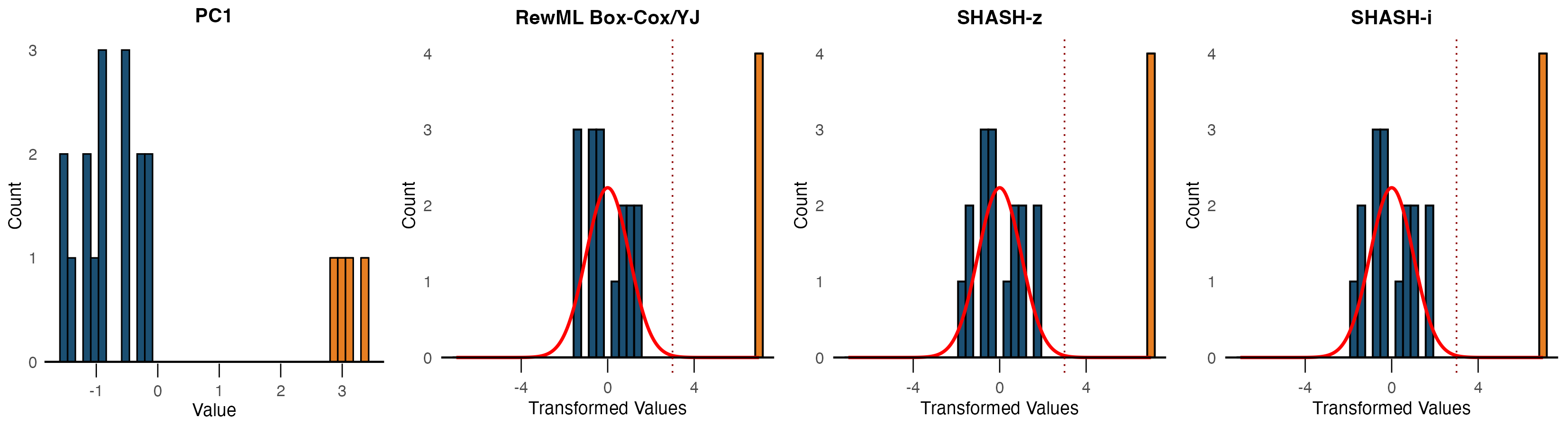} \\
    \includegraphics[width=4cm]{figures/toydatasets/status_legend_1X2.png}
 \caption{\small\textbf{\review{Application of outlier detection methods to the
Wood dataset ($n = 20$).}} The first robust principal component is shown.
True outlier identities are indicated with color. Transformed values are
truncated at $7$ for visualization purposes. Histograms with a large spike at $7$ reflect degenerate fits.}
    \label{fig:woodtoydatacomparison}
\end{figure}

Lastly, \review{\textbf{Figure~\ref{fig:woodtoydatacomparison}} displays the results for the first robust principal component of the Wood dataset, in which the 4 true outliers are multivariate rather than univariate outliers (\textbf{Figure~\ref{fig:wood_pairs}}). On the first robust principal component, the outliers are well separated from the bulk of the data, and RewML Box-Cox/YJ, SHASH-z and SHASH-i all place them beyond the outlier detection threshold without flagging any non-outlying observations. Thus, when the outliers are well separated from the main body of the data, the three approaches yield similar conclusions.}

\section{Application to functional neuroimaging}
\label{sec:neuroimaging}

Here, we illustrate the application of robust SHASH transformation in the context of multivariate outlier detection for denoising functional neuroimaging data, specifically functional magnetic resonance imaging (fMRI).  Functional MRI data is highly sensitive to contamination from head movements, scanner instabilities, physiological signals, and other sources. The conventional approach to detecting fMRI time points or \textit{volumes} contaminated by artifacts is using participant head motion \citep{power2012spurious}. However, although head motion is a major source of artifacts, relying on it alone has been shown to be suboptimal in terms of sensitivity and specificity \citep{pham2023less}. There is growing recognition of the need for statistically principled outlier detection techniques for automatic detection of contaminated fMRI volumes \citep{afyouni2018insight, mejia2017pca, parlak2023robust}. \review{In our previous work, we adopted robust multivariate outlier detection techniques, specifically robust distances based on minimum covariance determinant (MCD) estimates of location and scatter, to identify volumes exhibiting statistically significant abnormalities \citep{mejia2017pca, parlak2023robust}. Determining an appropriate MCD robust distance threshold is facilitated by previous work showing that those distances approximately follow a scaled $F$ distribution \citep{hardin2005distribution}, assuming multivariate normality of the underlying variables. In previous work, we found that MCD distances in this context deviate markedly from this reference distribution due to heavy tails and other deviations from normality, leading to overly aggressive outlier detection \citep{parlak2023robust}. Thus, here we consider robust SHASH transformation as a preprocessing step to satisfy the Gaussianity assumption, before computing robust distances and determining a parametric outlier detection threshold.}

\review{We analyze fMRI data from two participants in the Autism Brain Imaging Data Exchange (ABIDE) initiative \citep{di2014ABIDE}. The first participant's data contains $T=193$ volumes and exhibits substantial artifact-related contamination on visual inspection; the second participant's data contains $T=145$ volumes and is comparatively cleaner. For visualization and analysis purposes, we consider a single near-midline sagittal slice containing $V=4{,}675$ and $V=4{,}679$ in-mask voxels for participants 1 and 2, respectively.}

\review{Since artifacts typically affect many voxels simultaneously, we first project the $T \times V$ data onto a lower-dimensional space using independent component analysis (ICA), a well-established technique for separating signal and noise sources in fMRI \citep{mckeown1998analysis, griffanti2014, pruim2015ica, pham2023less}. ICA is performed using \texttt{fMRIscrub::pscrub} (v0.14.5) \citep{pham2023less}, which identifies the number of components automatically via penalized semi-integrated likelihood (PESEL; \citep{sobczyk2017bayesian}) and retains only high-kurtosis components, yielding $K=29$ and $K=24$ components for Subjects 1 and 2, respectively. This produces a $T \times K$ matrix $\mathbf{M}$ of component time courses. While the spatial maps are constructed to be approximately independent, the time courses are not, and artifacts may be expressed simultaneously across several components. We therefore analyze them jointly using a multivariate outlier detection framework. }

\review{We apply SHASH-i to each component time course to robustly transform it to normality, 
yielding the transformed matrix $\widetilde{\mathbf{M}}$. We then apply robust multivariate outlier detection separately for $\mathbf{M}$ and $\widetilde{\mathbf{M}}$ via MCD robust distances using \texttt{robustbase::covMcd} with subset fraction $\alpha_{\mathrm{MCD}}=0.75$. Robust distances are computed using the final reweighted estimates of location and scatter. Appropriate calibration of the cutoff is important for controlling the size of MCD-based outlier tests \citep{cerioli2009controlling}. We obtain the cutoff using the scaled-$F$ approximation of \citet{hardin2005distribution}, with the extension of \citet{green2014} to estimate the degrees of freedom for MCD fractions other than the maximum-breakdown setting. This calculation is implemented using \texttt{hr05CutoffMvnormal()} from the R package \texttt{CerioliOutlierDetection}, with \texttt{method = "GM14"}, selecting the returned \texttt{cutoff.pred} value. We choose the nominal upper-tail probability to match that of the univariate $\pm 3$ rule: for $Z \sim N(0,1)$, $\Pr(|Z|>3) \approx 0.0027$. Accordingly, a volume is flagged as an outlier when its squared robust distance exceeds the $0.9973$ quantile of the scaled-$F$ approximation.}

\begin{figure}[p]
  \centering
  \begin{minipage}{\textwidth}
    \centering
    {Subject 1}\par\vspace{0.3em}
    \includegraphics[width=\textwidth, trim=0 1cm 0 1.4cm, clip]{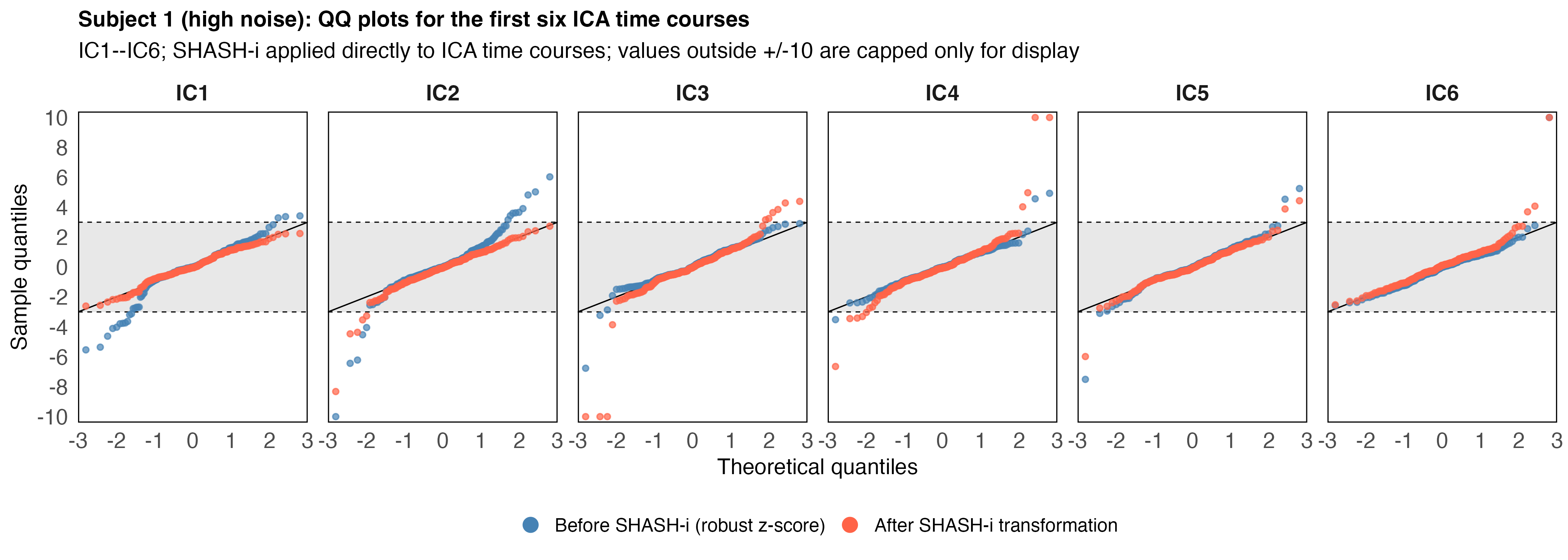}
  \end{minipage}
  \par\vspace{4mm}
  \begin{minipage}{\textwidth}
    \centering
    {Subject 2}\par\vspace{0.3em}
    \includegraphics[width=\textwidth, trim=0 1cm 0 1.4cm, clip]{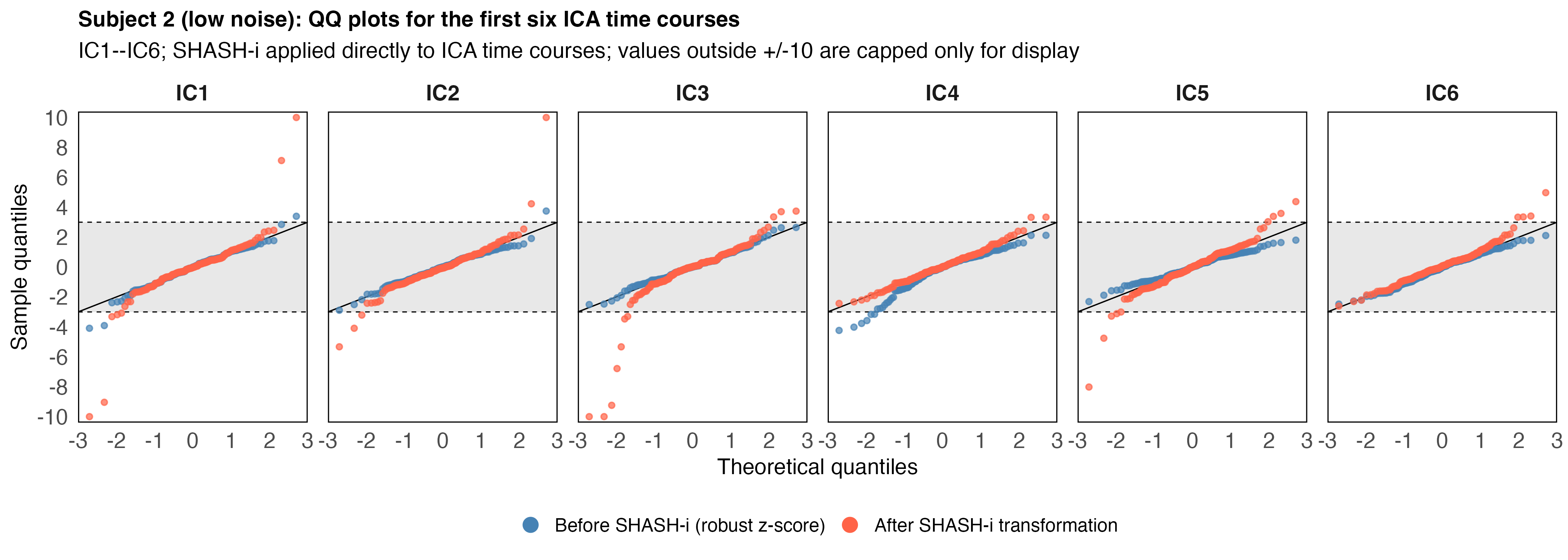}
  \end{minipage}
  \par\vspace{0.4em}
  \par \includegraphics[width=0.60\textwidth, trim = 0 12cm 0 10cm , clip]{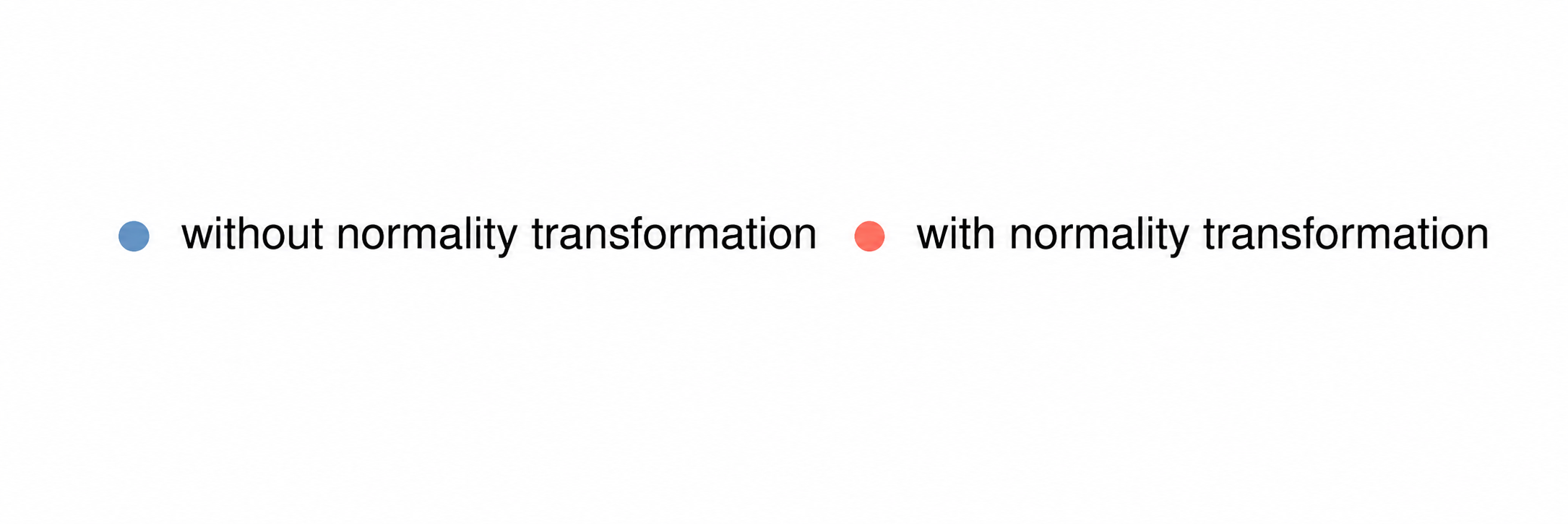}
  \vspace{0.3em}
  \caption{\review{\textbf{Standard \review{n}ormal QQ plots for the first six ICA time courses in ABIDE dataset} for subject 1 (high noise; ID:~0050048) and subject 2 (low noise; ID:~0051479). For each independent component (IC), the non-transformed values are robustly centered and scaled using the median and MAD; SHASH-i is used for robust normality transformation. The solid black line denotes the standard \review{n}ormal QQ reference line, and the dashed horizontal lines at $-3$ and $3$ indicate the outlier-detection thresholds. The gray region highlights the interval $[-3,3]$. Values outside the displayed range $[-10,10]$ are capped for visualization purposes.}}
  \label{fig:QQ_ABIDE}
\end{figure}

\review{The analysis is computationally fast: on a MacBook Pro work station with M2 processor and 8 GB of RAM, ICA required $1.29$ and $0.75$ seconds for each subject, respectively; SHASH-i transformation of all component time courses required $0.99$ and $1.24$ seconds; and MCD estimation required $0.76$ and $0.37$ seconds.
Since SHASH-i operates on the $T \times K$ component time courses, its cost depends on the number of volumes and components rather than the number of voxels, and fits are independent across components and subjects, so they can be readily parallelized.}

\review{\textbf{Figure~\ref{fig:QQ_ABIDE}} shows the effect of SHASH-i on the marginal distributions of the first six component time courses for each subject.  Prior to transformation, several components exhibit deviations from Gaussianity, including IC1 and IC2 of subject 1 (heavy tails), IC1, IC2 and IC4 of subject 2 (left skew), and IC5 of subject 2 (light tails). These non-Gaussian features are generally mitigated by robust SHASH transformation. As observed in the simulations studies above, we also see some examples of heavier-than-Gaussian tails, such as IC1 and IC3 of subject 1, which can occur due to the exclusion of true inliers from transformation parameter estimation.}


\begin{figure}[H]
    \centering
    \includegraphics[ width=\textwidth, trim= 0 0 0 1.5cm, clip]{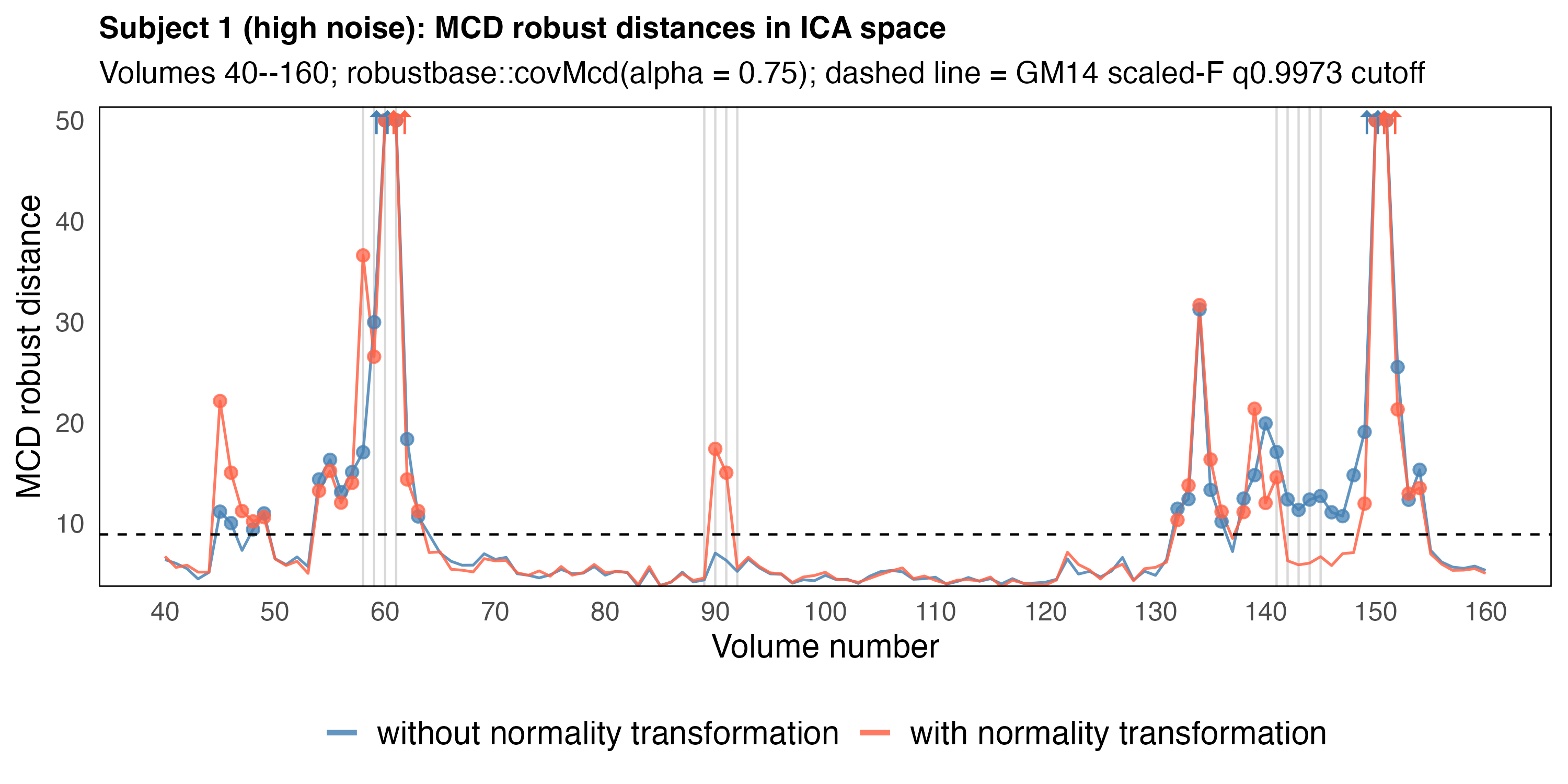}
    \caption{\review{
    \textbf{MCD robust distances and outlier detection before and after robust SHASH-i normality transformation in ABIDE dataset.} A subset of Subject 1's data is shown for illustration purposes. Points denote volumes classified as outliers. The dashed horizontal line indicates the outlier detection threshold, which equals the square root of the $0.9973$ quantile of the scaled-$F$ distribution for squared MCD robust distances.  Upward arrows mark values truncated for visualization purposes. Vertical gray lines indicate selected representative segments for image-based visualization in \textbf{Figure~\ref{fig:MANGO}}: the first segment illustrates agreement with and without normality transformation; the second segment illustrates outlier detection only after normality transformation; and the final segment illustrates a drop in outlier detection after normality transformation.}
    }
    \label{fig:ABIDE1_MCD_40_160}
\end{figure}

\begin{figure}[p]
    \centering

    \includegraphics[
        width=\textwidth, trim = 0.3cm 0 5mm 0, clip
    ]{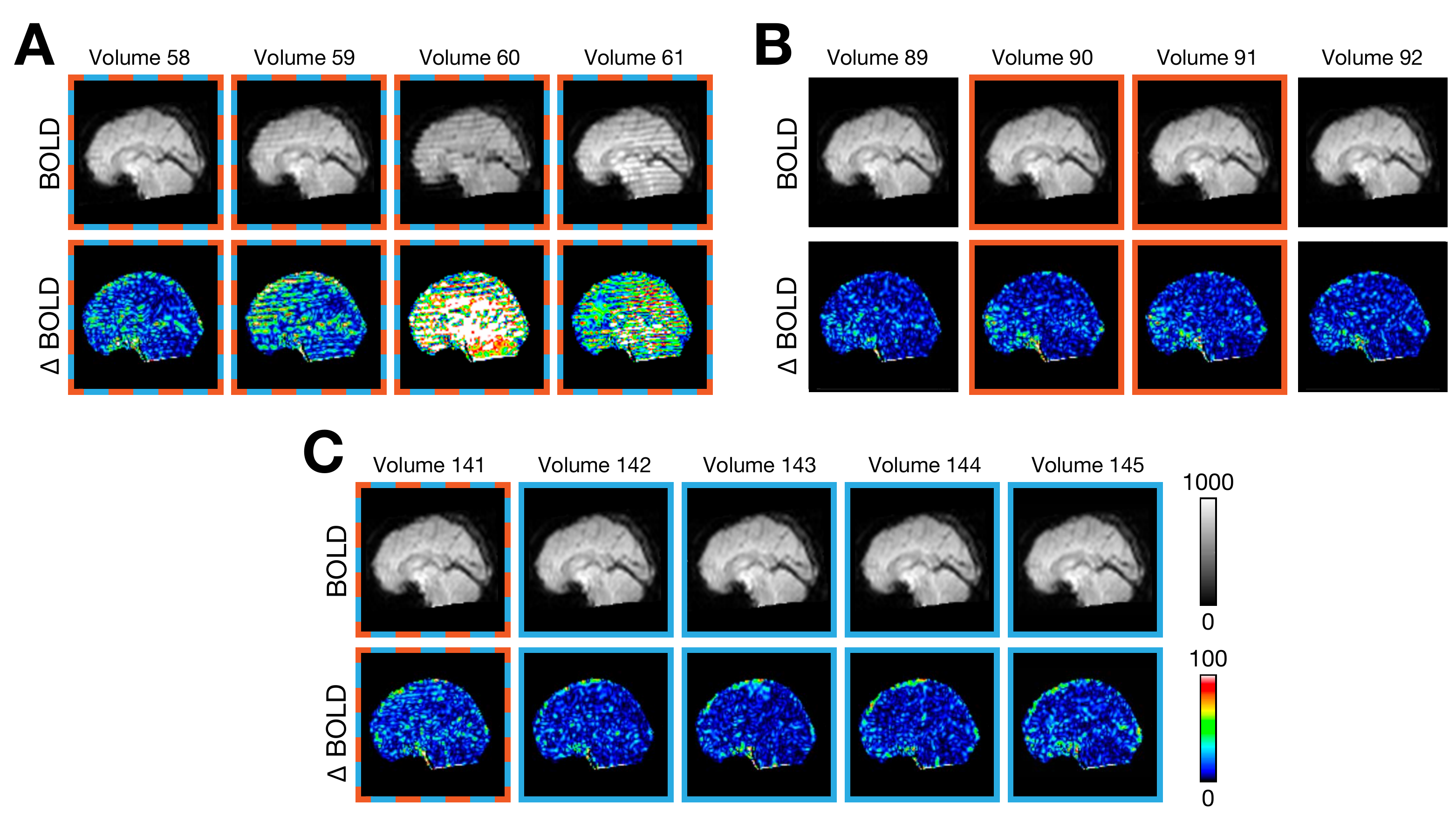}
    \par {\includegraphics[width=0.75\textwidth, trim = -8cm 10cm 0 8cm , clip]{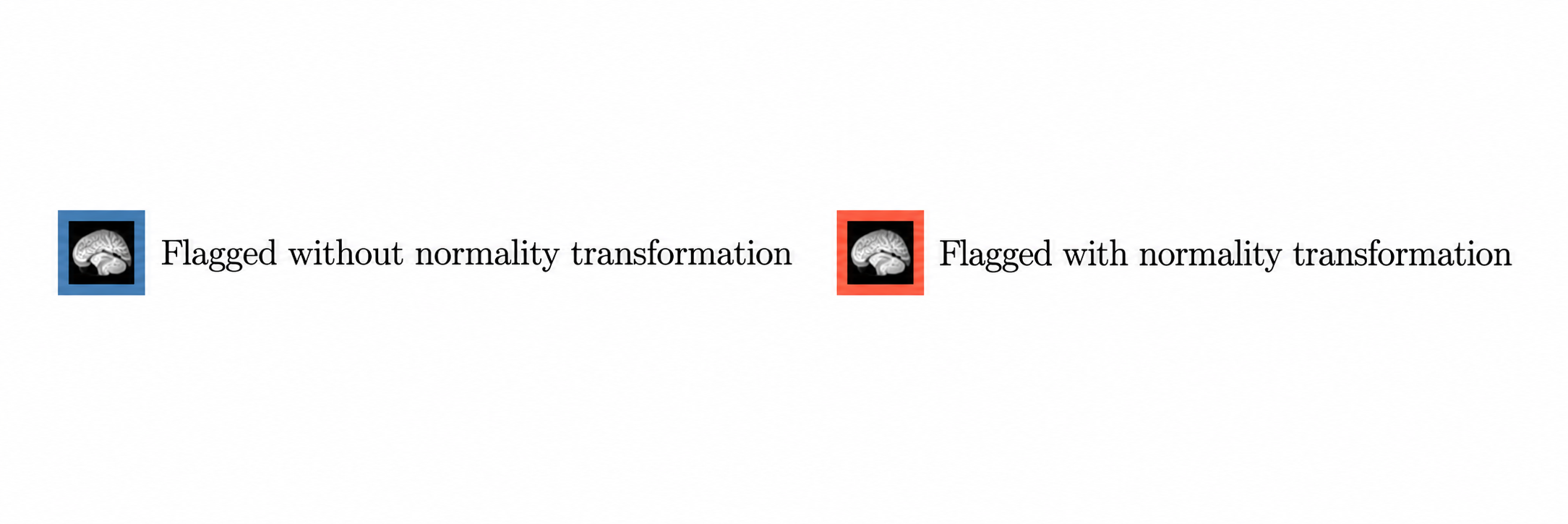}}
    \caption{\textbf{\review{Representative volumes illustrating effects of normality transformation on MCD-based multivariate outlier detection for ABIDE subject 1.}} \review{Robust, flexible normality transformation is performed on each ICA time course using SHASH-i. For each selected volume $t$, the top row shows the original BOLD image $X_t$, while the bottom row shows $\Delta\mathrm{BOLD}$, defined as the absolute difference between that volume and the median image, $ \Delta\mathrm{BOLD}_t = \left|X_t-\operatorname{med}_{s}\{X_s\}\right|.$  The difference image is shown to better display abnormal voxel intensities. (A) Example of concordance before and after normality transformation: an intense banding artifact is apparent in volumes 58-61.  (B) Example of outlier detection only after normality transformation, suggestive of \textit{masking}: artifacts in frontal regions are apparent on $\Delta\mathrm{BOLD}$ in volumes 90-91. (C) Example of volumes no longer identified as outliers after normality transformation, suggestive of \textit{swamping}: a banding artifact is only visually apparent in volume 141 but not volumes 142-145. However, note that in the absence of ground truth, these are subjective interpretations. 
    }}
    \label{fig:MANGO}
\end{figure}

\review{\textbf{Figure~\ref{fig:ABIDE1_MCD_40_160}} compares the robust distances for Subject 1 before and after robust SHASH transformation, with flagged outliers indicated by points. We display a subset of the scan (volumes 40--160) for illustration purposes. Several prominent peaks exceed the outlier detection threshold with or without normality transformation, and there is broad agreement between the two methods for most volumes. However, some volumes near the threshold change classification after transformation. To understand these changes, \textbf{Figure~\ref{fig:MANGO}} highlights segments for which the methods agree (panel A), for which volumes are flagged only after transformation (panel B), and for which volumes are no longer flagged after transformation (panel C). Since there is no ground truth, we rely on visual inspection of both the original BOLD image and its absolute deviation from the voxel-wise temporal median image, which more clearly highlights abnormal intensities. }

\review{As seen in \textbf{Figure~\ref{fig:MANGO}}, volumes 58--61 are flagged both before and after transformation and exhibit an intense banding artifact (panel A). Volumes 90 and 91 are flagged only after transformation and show subtle artifacts in frontal regions, which are more visible in the absolute deviation images. This suggests masking of these more subtle outliers without normality transformation (panel B). Conversely, volumes 142--145 are flagged only before transformation and show no obvious artifacts, unlike the neighboring volume 141, which shows a banding artifact and is flagged both before and after transformation; this suggests possible swamping of true inliers prior to transformation (panel C). These examples illustrate that robust, flexible SHASH normality transformation has the potential to improve both sensitivity and specificity of MCD-based robust multivariate outlier detection, with applicability to artifact detection in functional neuroimaging and other high-dimensional settings.}

\section{Discussion} \label{sec:discussion}
In this work, we have developed a novel robust \review{normality transformation method} based on the flexible sinh-arcsinh (SHASH) family of distributions\review{, with applications to univariate and multivariate outlier detection}. We developed two variations, which differ in their method of initializing outliers, a crucial step to avoid the undue influence of outliers on the transformation. SHASH-z uses a traditional robust z-score approach for initialization, while SHASH-i uses a novel version of the anomaly detection technique iForest. Through extensive simulation studies and a rich set of real datasets, we benchmarked the performance of \review{robust} SHASH transformation against \review{non-robust SHASH transformation, RewML Box-Cox and Yeo-Johnson power transformations, and robust z-scoring}. In both simulated and real data settings, \review{robust} SHASH transformation consistently exhibited strong performance\review{, both in transforming non-outlying observations toward normality and in identifying outliers}. \review{Beyond univariate outlier detection, robust normality transformation can serve as a preprocessing step for multivariate outlier detection, by improving conformity with the Gaussian assumptions underlying the distribution of robust distances. In an application to volume censoring in functional MRI, transforming ICA component time courses toward normality altered the classification of volumes near the outlier detection threshold in ways suggestive of reduced masking and swamping of artifact-contaminated volumes.}

\review{Our first simulation study highlights the importance of robustly estimating the transformation. When transformation parameters are estimated without accounting for contamination, the transformation accommodates the outliers, distorting the distribution of non-outlying observations in ways very likely to cause masking in downstream outlier detection. Robust SHASH transformation instead transforms non-outlying observations towards normality across distributional shapes, under contamination levels up to 20\% (\textbf{Figures~\ref{fig:density_n500}}, \textbf{\ref{fig:cvm_n500}} and \textbf{\ref{fig:density_sample_sizes}}). These simulations also reveal that after robust SHASH transformation, the distributions tend to have slightly heavier-than-Gaussian tails. This is due to the exclusion of some non-outlying observations from parameter estimation, which results in false positive rates between 1-2\%, above the nominal level of 0.3\% (\textbf{Figure~\ref{fig:FPR_TPR_plot}}).}

\review{Our second simulation study and real data analyses show that across distributions and contamination levels, robust SHASH transformation, and SHASH-i in particular, successfully transforms distributions towards normality under contamination. Downstream outlier detection shows high sensitivity to outliers, while keeping the false positive rate consistent around 1-2\% (\textbf{Figure~\ref{fig:FPR_TPR_plot}}). Competing methods each break down in characteristic ways. Robust z-scoring, which does not account for non-Gaussianity, maintains high sensitivity in heavy-tailed and skewed distributions but at the cost of many false positives. RewML power transformations mask outliers in skewed distributions, even at low levels of contamination.} This was surprising, since the Box-Cox and YJ transformations are designed to deal with skew. The culprit is not the transformation itself, but rather \review{its reliance on robust z-scores based on Huber estimators of location and scale, which are much less robust to outliers than the median and MAD (\textbf{Figure~\ref{fig:CenterScaleSideBySide}}), to identify outliers during iterative reweighting.} If outliers are allowed to influence the transformation, this is almost sure to result in masking of outliers after transformation, as we saw in both simulated and real data. \review{SHASH-z, which relies on robust z-scoring for initialization, likewise lost sensitivity at high contamination levels in Gaussian data. The main caveats to the strong performance of SHASH-i are consistently, if mildly, inflated false positive rates due to somewhat heavier-than-Gaussian tails, and vulnerability to dense clusters of outliers, which iForest+ may fail to initialize, as in the HBK data.} These findings suggest that combining the outlier initialization techniques underlying SHASH-z and SHASH-i may be an effective strategy to further avoid the influence of outliers on the transformation and consequent masking of outliers.

Our results also underscore the risk of over-reliance on robust z-scoring to initialize or detect outliers.  In simulated Gaussian data, robust z-scoring \review{might be expected to} exhibit ideal performance with good estimators of location and scale. Instead, its performance deteriorates \review{steadily} as outlier contamination increases. The reasons for this are readily apparent by examining the sampling distributions of the median and MAD (\textbf{Figure \ref{fig:MedianMADNormalAndGamma}})\review{, which reflect the fact that many robust measures of location and scale, including the median and MAD, are inconsistent under contamination \citep{martin1993bias, klooster2025inconsistency}}. The MAD exhibits positive bias \review{under contamination, as large outliers replace randomly occurring observations} (\textbf{Figure \ref{fig:CenterScaleSideBySide}}).  Additionally, \review{if outliers occur disproportionately in one tail,} the median \review{also} exhibits positive bias. The bias is larger for Huber estimators (location and scale), especially at high levels of outlier contamination.

These limitations of robust z-scoring underscore the need for alternative methods for outlier initialization. Our novel variant on iForest, which is tailored for univariate settings where outliers are considered to be defined based on magnitude, is one such alternative. While this approach has high sensitivity to outliers---our main goal in outlier initialization---its specificity is often suboptimal\review{: used alone as an outlier detection method, iForest+ flags substantially more non-outlying observations than SHASH-i (\textbf{Figure~\ref{fig:R_iforest}}), illustrating the value of the subsequent robust transformation and iterative updating. iForest is also known to be vulnerable to masking when anomalies form dense clusters \citep{liu2008isolation, liu2012isolation}, as we observed in the HBK data. Conversely, overly aggressive initialization can sharply truncate the tail of the non-outlying distribution, resulting in a degenerate transformation. Such fits were rare in our simulations and can often be remedied by a less aggressive initialization or, for strictly positive and highly skewed data, a log transformation prior to fitting.} Future work is needed to refine this approach, including handling densely clustered outliers and automatically detecting and remedying degenerate transformations.  
\review{Another possible modification of our robust SHASH transformation would be to progressively relax the outlier detection threshold after initial transformation to incorporate more non-outlying observations into the transformation and thus improve specificity. Alternatively, it may be possible to quantify and correct for the heavy-tail bias of robust SHASH transformation.}

This study has several limitations. First, \review{most of our simulation studies} consider a fixed sample size of $n=500$. Sample size will surely have an influence on the performance of all the methods considered here, given that they all depend on the estimation of transformation parameter(s).  Still, \review{our first simulation study shows that robust SHASH transformation is fairly stable across sample sizes from $100$ to $5,000$ (\textbf{Figure~\ref{fig:density_sample_sizes}}). Additionally,} our analysis of a rich variety of real datasets contain a wide range of sample sizes from small ($n=20$) to large ($n=297$). Another limitation is the use of \review{Huber} robust location and scale estimators for outlier initialization in the benchmark \review{RewML power} transformation method, which somewhat hinders fair comparison of that transformation to SHASH transformation. However, since that method was originally designed to use Huber estimators, and is implemented as such in the corresponding software, we chose to maintain that choice in order to provide a realistic assessment of its performance. In addition, here we only considered the SHASH transformation, and did not consider other transformations that can handle skew as well as tail weight \citep{Tsai2017HP, goerg2015LambertW}. Future research is needed to provide a comprehensive comparison of different transformations and outlier initialization techniques.

\review{Observations flagged as outliers may sometimes form structured groups that reflect meaningful heterogeneity or latent subpopulations rather than errors or artifacts. Robust SHASH transformation estimates the shape of a reference distribution despite such observations, but does not determine whether they are spurious or scientifically meaningful; when multiple subpopulations are of substantive interest, mixture modeling or group-specific analysis may be more appropriate.}

\review{Finally, we illustrated the use of robust SHASH transformation as a preprocessing step for multivariate outlier detection, with an application to volume censoring in functional MRI data. MCD-based robust distances approximately follow a scaled-$F$ distribution under multivariate normality \citep{hardin2005distribution}, but fMRI component time courses often depart markedly from Gaussianity, which in previous work led to overly aggressive outlier detection \citep{parlak2023robust}. Transforming each component time course toward normality changed the classification of volumes near the outlier detection threshold in ways suggestive of reduced masking and swamping. However, several limitations remain. Normality of each component does not imply multivariate normality, and fMRI volumes are temporally dependent, so the scaled-$F$ reference distribution remains an approximation. In addition, in the absence of ground truth, our evaluation relied on visual inspection of volumes from two subjects. Future work should evaluate this approach in larger cohorts, for example based on its effect on the reliability of downstream analyses such as functional connectivity \citep{pham2023less}. The approach is computationally well suited to such settings: SHASH transformation of all component time courses required about one second per subject, and SHASH fits are independent across components and subjects, so they can be readily parallelized.} \review{More broadly, robust and flexible normality transformation provides a principled basis for automatic multivariate outlier detection in high-dimensional settings where visual inspection is infeasible, such as large neuroimaging cohorts.}


\section*{Acknowledgments}
The authors gratefully acknowledge Peter Rousseeuw and Brian Caffo for their valuable comments and constructive feedback during the preparation of this manuscript. This work is supported by National Institute of Aging (NIA) grant R01AG083919 to A.F.M.  

\section*{\review{Data Availability Statement}}
\review{The proposed methods are implemented in the open-source \texttt{rrobot} R package (version~0.2), available on GitHub (\url{https://github.com/mandymejia/rrobot}) and CRAN. All code needed to reproduce the simulation studies, real data analyses, and figures in this paper, including random seeds, is available at \url{https://github.com/statMINDlab/SHASH-paper}. The benchmark datasets analyzed in Section~\ref{sec:toydatasets} are publicly available in R packages on CRAN: the Top Gear dataset in the \texttt{robustHD} package (version~0.6.1 or later) \citep{alfons2019robustHD}, and the HBK and Wood datasets in the \texttt{robustbase} package (version~0.99-6 or later) \citep{robustbase}. The fMRI data analyzed in Section~\ref{sec:neuroimaging} were obtained from the Autism Brain Imaging Data Exchange (ABIDE I) \citep{di2014ABIDE}, which is openly available through the International Neuroimaging Data-Sharing Initiative (\url{http://fcon_1000.projects.nitrc.org/indi/abide/}); the two participants analyzed have ABIDE IDs 0050048 and 0051479.}

\section*{\review{Declaration of Generative AI}}
\review{The authors used ChatGPT(OpenAI; 6.1 Sol) and  Claude (Anthropic; Opus 5.5) to assist with generating code from pseudocode, edit and assemble figures, and revise the manuscript text. This included condensing sections and checking the manuscript for internal consistency, typographical errors, and LaTeX formatting. The tool was not used to design the methods, conduct analyses, generate data, produce plots, or draft the manuscript. After using this tool, the authors reviewed and edited all content as needed and take full responsibility for the content of the publication.}

{\singlespacing
\bibliography{mybib}

@article{yuan2000robust,
  author  = {Yuan, Ke-Hai and Chan, Wai and Bentler, Peter M.},
  title   = {Robust transformation with applications to structural equation modelling},
  journal = {British Journal of Mathematical and Statistical Psychology},
  volume  = {53},
  pages   = {31--50},
  year    = {2000},
  doi     = {10.1348/000711000159169}
}

@article{martin1993bias,
  author  = {Martin, R. Douglas and Zamar, Ruben H.},
  title   = {Bias Robust Estimation of Scale},
  journal = {The Annals of Statistics},
  volume  = {21},
  number  = {2},
  pages   = {991--1017},
  year    = {1993},
  doi     = {10.1214/aos/1176349161}
}

@misc{klooster2025inconsistency,
  author        = {Klooster, Jens and Nielsen, Bent},
  title         = {On (in)consistency of {M}-estimators under contamination},
  year          = {2025},
  eprint        = {2502.09145},
  archivePrefix = {arXiv},
  primaryClass  = {math.ST},
  url           = {https://arxiv.org/abs/2502.09145},
  note          = {arXiv preprint, version 1, 13 February 2025}
}

@article{carroll1987diagnostics,

  title={Diagnostics and Robust Estimation When Transforming the Regression Model and the Response},

  author={Carroll, Raymond J. and Ruppert, David},

  journal={Technometrics},

  volume={29},

  number={3},

  pages={287--299},

  year={1987},

  doi={10.1080/00401706.1987.10488239}
  }

@article{carroll1980robust,
  title={A Robust Method for Testing Transformations to Achieve Approximate Normality},
  author={Carroll, Raymond J.},
  journal={Journal of the Royal Statistical Society: Series B (Methodological)},
  volume={42},
  number={1},
  pages={71--78},
  year={1980}
}

@article{pham2023less,
  title        = {Less is more: balancing noise reduction and data retention in fMRI with data-driven scrubbing},
  author       = {Pham, Damon {\DJ} and McDonald, Daniel J and Ding, Lei and Nebel, Mary Beth and Mejia, Amanda F},
  journal      = {NeuroImage},
  volume       = {270},
  pages        = {119972},
  year         = {2023},
  publisher    = {Academic Press},
  doi          = {10.1016/j.neuroimage.2023.119972}
}

@article{rousseeuw1993alternatives,
  title={Alternatives to the median absolute deviation},
  author={Rousseeuw, Peter J and Croux, Christophe},
  journal={Journal of the American Statistical association},
  volume={88},
  number={424},
  pages={1273--1283},
  year={1993},
  publisher={Taylor \& Francis}
}

@article{hardin2005distribution,
  title={The distribution of robust distances},
  author={Hardin, Johanna and Rocke, David M},
  journal={Journal of Computational and Graphical Statistics},
  volume={14},
  number={4},
  pages={928--946},
  year={2005},
  publisher={Taylor \& Francis}
}

@article{rousseeuw2011robust,
  title={Robust statistics for outlier detection},
  author={Rousseeuw, Peter J and Hubert, Mia},
  journal={Wiley interdisciplinary reviews: Data mining and knowledge discovery},
  volume={1},
  number={1},
  pages={73--79},
  year={2011},
  publisher={Wiley Online Library}
}

@article{raymaekers2021transforming,
  title={Transforming variables to central normality},
  author={Raymaekers, Jakob and Rousseeuw, Peter J},
  journal={Machine Learning},
  pages={1--23},
  year={2021},
  publisher={Springer}
}

@article{box1964analysis,
  title={An analysis of transformations},
  author={Box, George EP and Cox, David R},
  journal={Journal of the Royal Statistical Society: Series B (Methodological)},
  volume={26},
  number={2},
  pages={211--243},
  year={1964},
  publisher={Wiley Online Library}
}

@article{yeo2000new,
  title={A new family of power transformations to improve normality or symmetry},
  author={Yeo, In-Kwon and Johnson, Richard A},
  journal={Biometrika},
  volume={87},
  number={4},
  pages={954--959},
  year={2000},
  publisher={Oxford University Press}
}

@article{mejia2017pca,
  title={PCA leverage: outlier detection for high-dimensional functional magnetic resonance imaging data},
  author={Mejia, Amanda F and Nebel, Mary Beth and Eloyan, Ani and Caffo, Brian and Lindquist, Martin A},
  journal={Biostatistics},
  volume={18},
  number={3},
  pages={521--536},
  year={2017},
  publisher={Oxford University Press}
}

@article{power2012spurious,
  title={Spurious but systematic correlations in functional connectivity MRI networks arise from subject motion},
  author={Power, Jonathan D and Barnes, Kelly A and Snyder, Abraham Z and Schlaggar, Bradley L and Petersen, Steven E},
  journal={Neuroimage},
  volume={59},
  number={3},
  pages={2142--2154},
  year={2012},
  publisher={Elsevier}
}

@article{pruim2015ica,
  title={{ICA}-{AROMA}: A robust {ICA}-based strategy for removing motion artifacts from f{MRI} data},
  author={Pruim, Raimon HR and Mennes, Maarten and van Rooij, Daan and Llera, Alberto and Buitelaar, Jan K and Beckmann, Christian F},
  journal={Neuroimage},
  volume={112},
  pages={267--277},
  year={2015},
  publisher={Elsevier}
}

@article{jones2009sinh,
  title={Sinh-arcsinh distributions},
  author={Jones, M Chris and Pewsey, Arthur},
  journal={Biometrika},
  volume={96},
  number={4},
  pages={761--780},
  year={2009},
  publisher={Oxford University Press}
}

@book{hawkins1980identification,
  title={Identification of outliers},
  author={Hawkins, Douglas M},
  volume={11},
  year={1980},
  publisher={Springer}
}

@article{hawkins1984robustbase,
  title     = {Location of several outliers in multiple regression: Effects of leverage points},
  author    = {Hawkins, Douglas M. and Bradu, Dan and Kass, Gilbert V.},
  journal   = {Technometrics},
  volume    = {26},
  number    = {3},
  pages     = {197--208},
  year      = {1984},
  publisher = {Taylor \& Francis}
}

@article{tukey1957comparative,
  title={On the comparative anatomy of transformations},
  author={Tukey, John W},
  journal={The Annals of Mathematical Statistics},
  pages={602--632},
  year={1957},
  publisher={JSTOR}
}

@book{rigby2019distributions,
  title={Distributions for modeling location, scale, and shape: Using GAMLSS in R},
  author={Rigby, Robert A and Stasinopoulos, Mikis D and Heller, Gillian Z and De Bastiani, Fernanda},
  year={2019},
  publisher={CRC press}
}

@article{maronna2002robust,
  title={Robust estimates of location and dispersion for high-dimensional datasets},
  author={Maronna, Ricardo A and Zamar, Ruben H},
  journal={Technometrics},
  volume={44},
  number={4},
  pages={307--317},
  year={2002},
  publisher={Taylor \& Francis}
}

@article{maronna1976robust,
  title={Robust {M}-estimators of multivariate location and scatter},
  author={Maronna, Ricardo Antonio},
  journal={The annals of statistics},
  pages={51--67},
  year={1976},
  publisher={JSTOR}
}

@article{rousseeuw1984least,
  title={Least median of squares regression},
  author={Rousseeuw, Peter J},
  journal={Journal of the American statistical association},
  volume={79},
  number={388},
  pages={871--880},
  year={1984},
  publisher={Taylor \& Francis}
}

@article{griffanti2014,
	author = {Griffanti, Ludovica and Salimi-Khorshidi, Gholamreza and Beckmann, Christian F and Auerbach, Edward J and Douaud, Gwena{\"e}lle and Sexton, Claire E and Zsoldos, Enik{\H{o}} and Ebmeier, Klaus P and Filippini, Nicola and Mackay, Clare E and others},
	journal = {Neuroimage},
	pages = {232--247},
	publisher = {Elsevier},
	title = {{ICA}-based artefact removal and accelerated fMRI acquisition for improved resting state network imaging},
	volume = {95},
	year = {2014}}

@article{di2014ABIDE,
	author = {Di Martino, Adriana and Yan, Chao-Gan and Li, Qingyang and Denio, Erin and Castellanos, Francisco X and Alaerts, Kaat and Anderson, Jeffrey S and Assaf, Michal and Bookheimer, Susan Y and Dapretto, Mirella and others},
	journal = {Molecular psychiatry},
	number = {6},
	pages = {659--667},
	publisher = {Nature Publishing Group},
	title = {The autism brain imaging data exchange: towards a large-scale evaluation of the intrinsic brain architecture in autism},
	volume = {19},
	year = {2014}}

@unpublished{green2014,
  author = {Green, Christopher G. and Martin, R. Douglas},
  title = {An Extension of a Method of {Hardin} and {Rocke},
           with an Application to Multivariate Outlier Detection
           via the {IRMCD} Method of {Cerioli}},
  year = {2017},
  note = {Working paper, revised July 23, 2017},
  url = {https://christopherggreen.github.io/papers/hr05_extension.pdf}
}

@article{afyouni2018insight,
  title={Insight and inference for {DVARS}},
  author={Afyouni, Soroosh and Nichols, Thomas E},
  journal={Neuroimage},
  volume={172},
  pages={291--312},
  year={2018},
  publisher={Elsevier}
}

@article{mckeown1998analysis,
  title={Analysis of f{MRI} data by blind separation into independent spatial components},
  author={McKeown, Martin J and Makeig, Scott and Brown, Greg G and Jung, Tzyy-Ping and Kindermann, Sandra S and Bell, Anthony J and Sejnowski, Terrence J},
  journal={Human brain mapping},
  volume={6},
  number={3},
  pages={160--188},
  year={1998},
  publisher={Wiley Online Library}
}

@Manual{R,
     title = {R: {A} {L}anguage and {E}nvironment for {S}tatistical {C}omputing},
     author = {{R Core Team}},
     organization = {R Foundation for Statistical Computing},
     address = {Vienna, Austria},
     year = {2022},
     url = {https://www.R-project.org/},
   }

@book{rousseeuw1987robust,
  title     = {Robust Regression and Outlier Detection},
  author    = {Rousseeuw, Peter J. and Leroy, Annick M.},
  year      = {1987},
  publisher = {John Wiley \& Sons}
}

@article{parlak2023robust,
  title={A robust multivariate, non-parametric outlier identification method for scrubbing in f{MRI}},
  author={Parlak, Fatma and Pham, Damon D and Mejia, Amanda F}, 
  year={2023}}

@inproceedings{liu2008isolation,
  title={Isolation forest},
  author={Liu, Fei Tony and Ting, Kai Ming and Zhou, Zhi-Hua},
  booktitle={2008 Eighth IEEE International Conference on Data Mining},
  pages={413--422},
  year={2008},
  organization={IEEE},
  doi={10.1109/ICDM.2008.17}
}

@article{liu2012isolation,
  author  = {Liu, Fei Tony and Ting, Kai Ming and Zhou, Zhi-Hua},
  title   = {Isolation-Based Anomaly Detection},
  journal = {ACM Transactions on Knowledge Discovery from Data},
  volume  = {6},
  number  = {1},
  pages   = {3:1--3:39},
  year    = {2012},
  doi     = {10.1145/2133360.2133363}
}

@book{ruppert2010statistics,
  title     = {Statistics and Data Analysis for Financial Engineering},
  author    = {Ruppert, David},
  year      = {2010},
  publisher = {Springer},
  isbn      = {9781441977878},
  note      = {p. 118. Retrieved 2015-08-27}
}

@inproceedings{eftekhari2025gaussianizing,
  author    = {Eftekhari, Daniel and Papyan, Vardan},
  title     = {On the Importance of Gaussianizing Representations},
  booktitle = {Proceedings of the 42nd International Conference on
               Machine Learning},
  series    = {Proceedings of Machine Learning Research},
  volume    = {267},
  pages     = {15018--15037},
  publisher = {PMLR},
  year      = {2025}
}

@inproceedings{dettmers2023qlora,
  author    = {Dettmers, Tim and Pagnoni, Artidoro and Holtzman, Ari and
               Zettlemoyer, Luke},
  title     = {{QLoRA}: Efficient Finetuning of Quantized {LLMs}},
  booktitle = {Advances in Neural Information Processing Systems},
  volume    = {36},
  year      = {2023}
}

@book{huber1981robust,
  title     = {Robust Statistics},
  author    = {Huber, Peter J.},
  year      = {1981},
  publisher = {John Wiley \& Sons},
}

@article{huber1964robust,
  title={Robust Estimation of a Location Parameter},
  author={Huber, Peter J.},
  journal={The Annals of Mathematical Statistics},
  volume={35},
  number={1},
  pages={73--101},
  year={1964},
  doi={10.1214/aoms/1177703732}
}

@misc{alfons2019robustHD,
  author       = {Alfons, Andreas},
  title        = {{robustHD}: Robust Methods for High-Dimensional Data},
  year         = {2019},
  howpublished = {\url{https://CRAN.R-project.org/package=robustHD}},
  note         = {R package version 0.6.1}
}

@misc{robustbase,
  title        = {robustbase: Basic Robust Statistics},
  author       = {Martin Maechler and Peter Rousseeuw and Christophe Croux and Valentin Todorov and Andreas Ruckstuhl and Matias Salibian-Barrera and Tobias Verbeke and Manuel Koller and Eduardo L. T. Conceicao and Maria Anna di Palma},
  year         = {2025},
  note         = {R package version 0.99-6},
  url          = {https://cran.r-project.org/web/packages/robustbase/index.html}
}

@article{peterson2020,
  title   = {Ordered quantile normalization: a semiparametric transformation built for the cross-validation era},
  author  = {Peterson, Ryan A. and Cavanaugh, Joseph E.},
  journal = {Journal of Applied Statistics},
  year    = {2020},
  volume  = {47},
  number  = {13-15},
  pages   = {2312--2327},
  publisher = {Taylor \& Francis},
  doi     = {10.1080/02664763.2019.1630372}
}

@article{goerg2015LambertW,
  author  = {Goerg, Georg M.},
  title   = {The Lambert Way to Gaussianize Heavy-Tailed Data with the Inverse of Tukey's h Transformation as a Special Case},
  journal = {The Scientific World Journal},
  year    = {2015},
  volume  = {2015},
  pages   = {909231},
  doi     = {10.1155/2015/909231}
}

@article{Tsai2017HP,
  author  = {Tsai, Arthur C. and Liou, Michelle and Simak, Maria and Cheng, Philip E.},
  title   = {On hyperbolic transformations to normality},
  journal = {Computational Statistics and Data Analysis},
  year    = {2017},
  volume  = {115},
  pages   = {250--266},
  doi     = {10.1016/j.csda.2017.06.001}
}

@article{hampel1974influence,
  title={The influence curve and its role in robust estimation},
  author={Hampel, Frank R},
  journal={Journal of the american statistical association},
  volume={69},
  number={346},
  pages={383--393},
  year={1974},
  publisher={Taylor \& Francis}
}

@article{donoho1982breakdown,
  title={Breakdown properties of multivariate location estimators. unpublished qualifying paper, {H}arvard {U}niversity},
  author={Donoho, DL},
  journal={Statistics Dept},
  year={1982}
}

@article{chabchoub2022,
  title={An In-Depth Study and Improvement of Isolation Forest},
  author={Chabchoub, Yosra and Akermi, Imen and Gargouri, Faiez},
  journal={IEEE Access},
  volume={10},
  pages={17802--17818},
  year={2022},
  publisher={IEEE}
}

@article{Hubert2005ROBPCA,
  title   = {ROBPCA: A New Approach to Robust Principal Component Analysis},
  author  = {Hubert, Mia and Rousseeuw, Peter J. and Vanden Branden, Karlien},
  journal = {Technometrics},
  volume  = {47},
  number  = {1},
  pages   = {64--79},
  year    = {2005},
  doi     = {10.1198/004017004000000563}
}

@article{cerioli2009controlling,
  title={Controlling the size of multivariate outlier tests with the MCD estimator of scatter},
  author={Cerioli, Andrea and Riani, Marco and Atkinson, Anthony C},
  journal={Statistics and Computing},
  volume={19},
  number={3},
  pages={341--353},
  year={2009},
  publisher={Springer}
}

@article{sobczyk2017bayesian,
  title={Bayesian dimensionality reduction with PCA using penalized semi-integrated likelihood},
  author={Sobczyk, Piotr and Bogdan, Ma{\l}gorzata and Josse, Julie},
  journal={Journal of Computational and Graphical Statistics},
  volume={26},
  number={4},
  pages={826--839},
  year={2017},
  publisher={Taylor \& Francis}
}
\bibliographystyle{plainnat}   
}

\newpage

\appendix 

\renewcommand\thesection{\Alph{section}}
\renewcommand\thesubsection{\thesection.\arabic{subsection}}
\renewcommand\thetable{S.\arabic{table}}
\renewcommand\thefigure{S.\arabic{figure}}
\setcounter{figure}{0}
\setcounter{table}{0}

\section{The SHASH density function} 
\label{app:derivation}

The sinh–arcsinh transformation is defined as
\[
 Z = S_{\nu, \tau}(X) = \sinh\big(\tau \, \sinh^{-1}(X) - \nu\big),
\]
 and the inverse sinh-arcsinh transformation is given by
\[
X = S_{\nu, \tau}^{-1}(Z) = \sinh\big(\frac{\sinh^{-1}(Z) + \nu}{\tau}\big)
\].

Hence, if \( Z \sim N(0,1) \), then \( X = S_{\nu, \tau}^{-1}(Z) \) follows a sinh-arcsinh (SHASH) distribution. When location and scale parameters are introduced, the transformation extends to the four-parameter SHASHo form:
\[ Z = \sinh\!\left[ \tau\sinh^{-1}\!\left(\frac{X-\mu}{s}\right)-\nu \right], \]
where $\mu \in \mathbb{R}$ is a location parameter and $s>0$ is a scale parameter.
Defining a new scale parameter $\sigma=s/\tau$, so that $s=\sigma\tau$, gives the SHASHo2 parameterization used throughout this manuscript:
\[ Z = S_{\mu,\sigma,\nu,\tau}(X) = \sinh\!\left[ \tau\sinh^{-1}\!\left(\frac{X-\mu}{\sigma\tau}\right)-\nu \right].\]

To derive the probability density function of the SHASHo2 distribution, we apply the change-of-variables rule. Let $Z \sim N(0,1)$ with standard normal density $\phi(Z) = \frac{1}{\sqrt{(2 \pi)}}\exp{(\frac{-z^2}{2})}$. Since the sinh-arcsinh transformation is $Z = S_{\nu, \tau}(X) = \sinh\big(\tau \, \sinh^{-1}(X) - \nu\big)$, the corresponding density of $X$ is obtained as  
\[
f_X(x) = \phi(Z(X)) \left|\frac{dz}{dx}\right|
\]
Define 
\[
u = \sinh^{-1}\!\left( \frac{X - \mu}{\sigma \tau}\right),
\quad g = \tau u-\nu,
\quad Z(X) = \sinh(g).
\]
Then, differentiating $Z(X)$ with respect to $X$ gives
\[
\frac{dZ}{dX}
= \cosh(g) \, \tau \, \frac{du}{dX} 
= \cosh(g) \, \tau \, \frac{1}{\sigma \tau \sqrt{1+ \left(\frac{X-\mu}{\sigma \tau}\right)^2}}
= \frac{\cosh(g)}{\sigma \sqrt{1+ \left(\frac{X-\mu}{\sigma \tau}\right)^2}}
\]

Substituting into the change-of-variables formula gives
\[
f_X(x | \mu, \sigma, \nu, \tau) = \frac{\cosh\!\left(\tau \, \sinh^{-1} \!\left(\frac{x - \mu}{\sigma \tau} \right) - \nu\right)}{\sigma \sqrt{1+ \left(\frac{x - \mu}{\sigma \tau}\right) ^2}} \phi\!\left(\sinh\!\left[\tau \, \sinh^{-1}\!\left(\frac{x- \mu}{\sigma \tau}\right) - \nu\right]\right).
\]
To express the pdf in exponential form,
let \(w = \frac{x - \mu}{\sigma \tau}\), and define 
\[
r(x) = \sinh\!\big(\tau \, \sinh^{-1}(w) -\nu\big),
\qquad
c(x) = \cosh\!\big(\tau\, \sinh^{-1}(w) -\nu\big).
\]
Using the following hyperbolic identities
\[
\text{sinh}(y) = \frac{e^{y} - e^{-y}}{2},
\qquad
\text{cosh}(y) = \frac{e^{y} + e^{-y}}{2}
\]
we can express \(r(x) \) and \( c(x) \) in exponential form as
\[
r(x) = \tfrac{1}{2}\!\left[\exp\!\big(\tau \, \sinh^{-1}(w) - \nu\big) - 
\exp\!\big(-(\tau \, \sinh^{-1}(w) - \nu)\big)\right],
\]
\[
c(x) = \tfrac{1}{2}\!\left[\exp\!\big(\tau \, \sinh^{-1}(w) - \nu\big) + 
\exp\!\big(-(\tau \, \sinh^{-1}(w) - \nu)\big)\right].
\]
Hence, the density of the SHASHo2 distribution can be written as
\[
f_X(x; \mu, \sigma, \nu, \tau) =
\frac{c(x)}{\sigma \sqrt{1 + w^2}} \,
\frac{1}{\sqrt{2\pi}} 
\exp\!\big[-\tfrac{1}{2}r(x)^2\big].
\]




\section{Supplemental Figures} \label{app:Plots}

\begin{figure}[H]
    \centering
    \begin{minipage}[t]{0.47\textwidth}
        \centering
        \hspace{5mm}Symmetric Distributions\\[4pt]
        \includegraphics[
            width=\linewidth
        ]{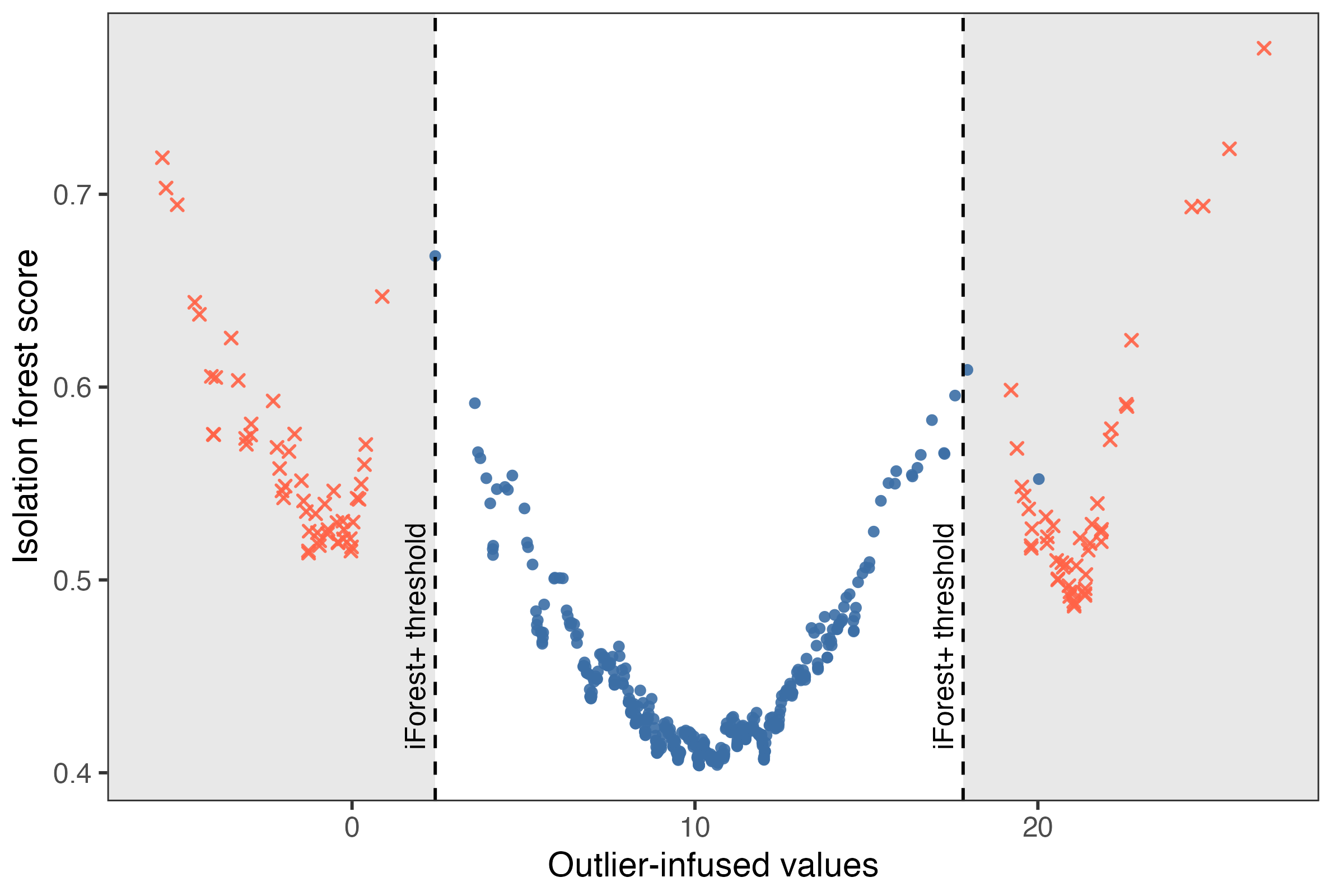}
    \end{minipage}%
    \hfill
    \begin{minipage}[t]{0.47\textwidth}
        \centering
         \hspace{5mm}Right-skewed Distributions\\[4pt]
        \includegraphics[
            width=\linewidth
        ]{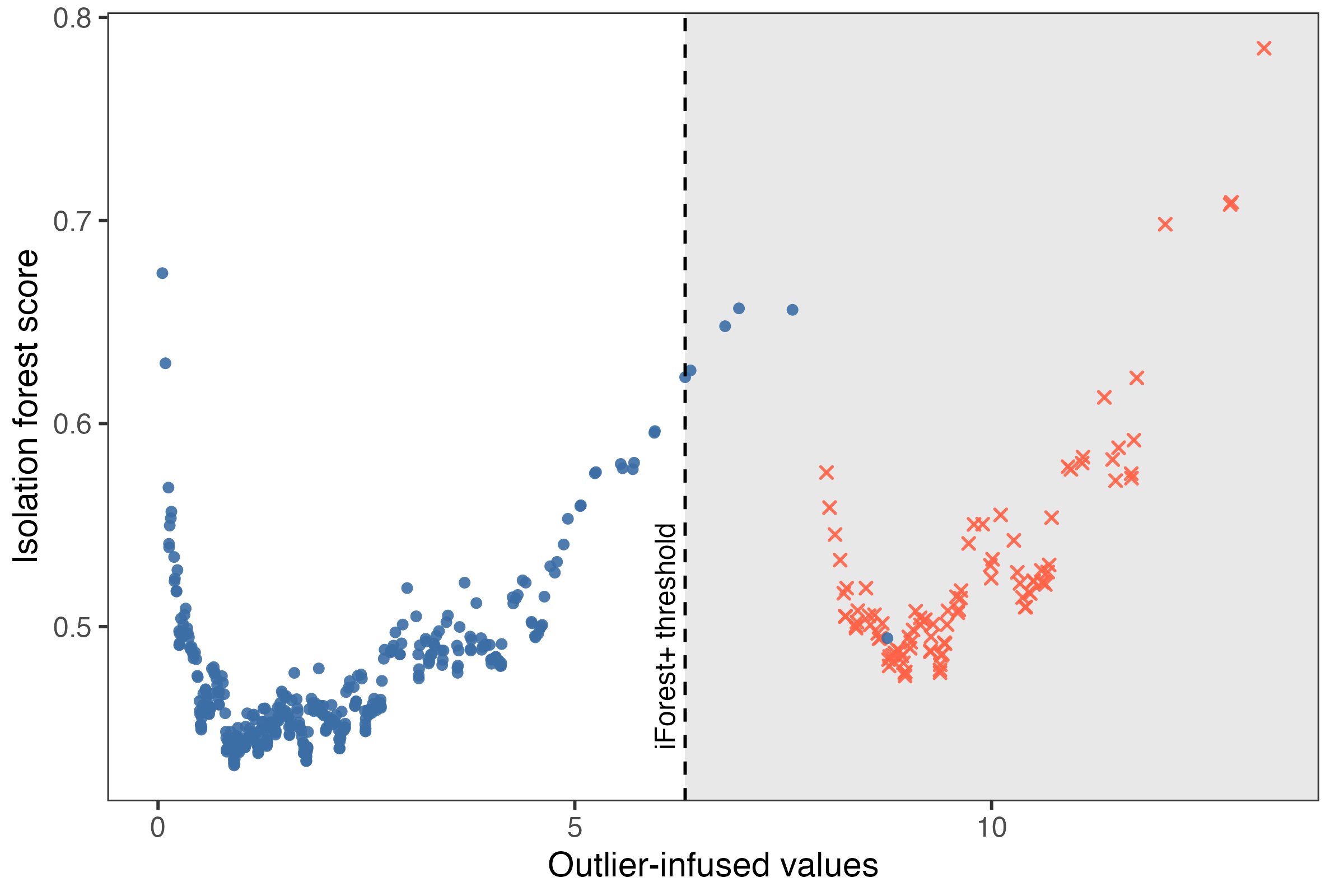}
    \end{minipage}
    \caption{\small Illustration of iForest+ initialization procedure, with $20\%$ outlier contamination, with outliers represented as red x's. Two distributions are shown for illustration purposes: a symmetric distribution, Normal$(10, 3)$; and a right-skewed distribution, Gamma$(2, 1)$. The iForest+ procedure \review{identifies the iForest anomaly (score above $0.6$) closest to the median and adopts it as the iForest+ threshold. For right-skewed distributions, only an upper threshold is set. For symmetric distributions, this is applied in both tails, and the anomaly closer to the median determines the threshold distance, so the upper and lower thresholds are equidistant from the median.}} \label{fig:iForestInit}
\end{figure}

\begin{figure}[H]
    \centering
    \includegraphics[width=\textwidth]{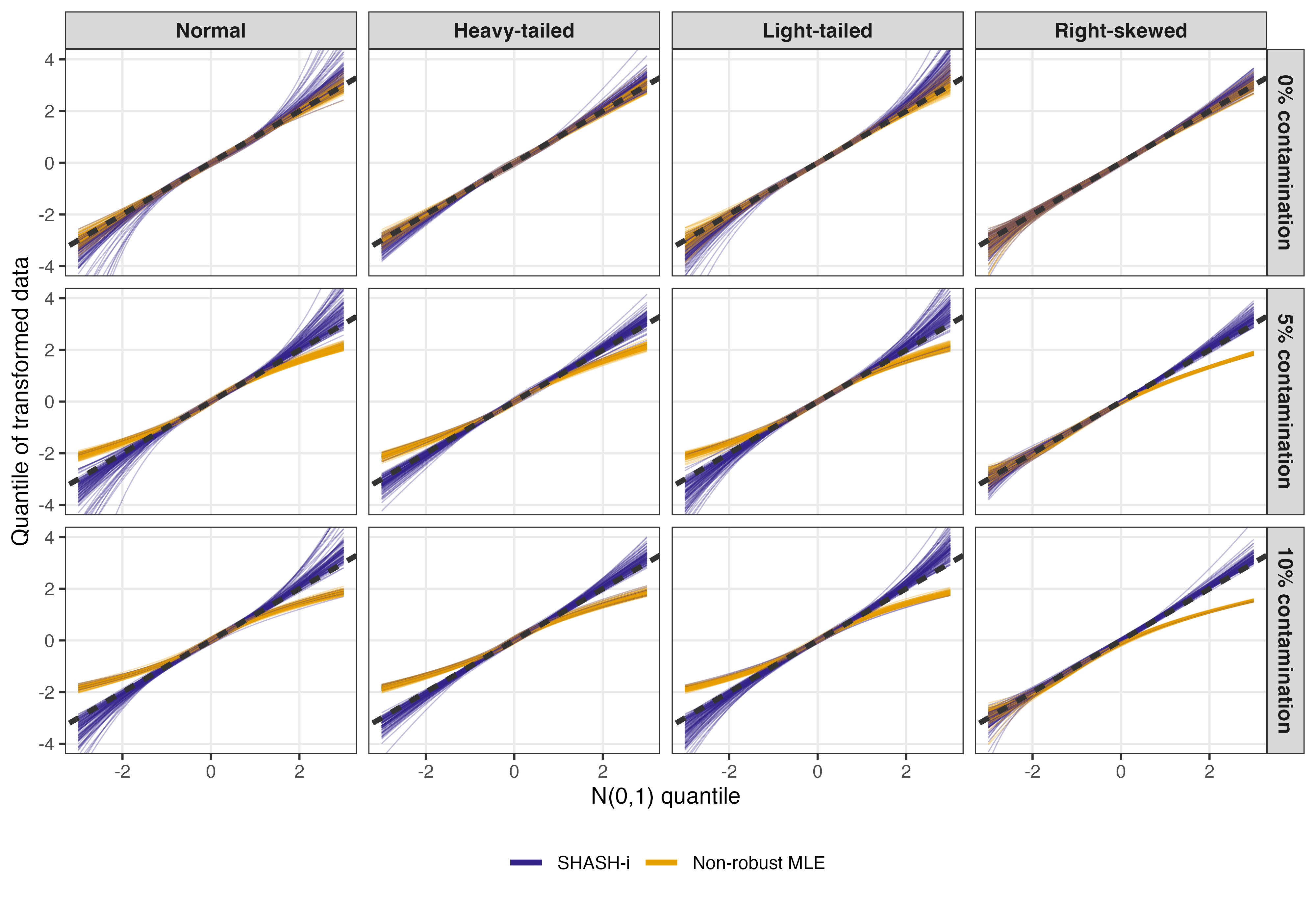}
    \caption{\review{\small\textbf{QQ plots after transformation.} Quantiles of the transformed non-outlying observations against $N(0,1)$ quantiles, at $n = 500$, for non-robust and robust SHASH transformation. For robust SHASH transformation, iForest+ outlier initialization was used.}}
    \label{fig:qq_n500}
\end{figure}

\begin{figure}[H]
    \centering
    \includegraphics[width=\textwidth, trim = 0 0 0 11mm, clip]{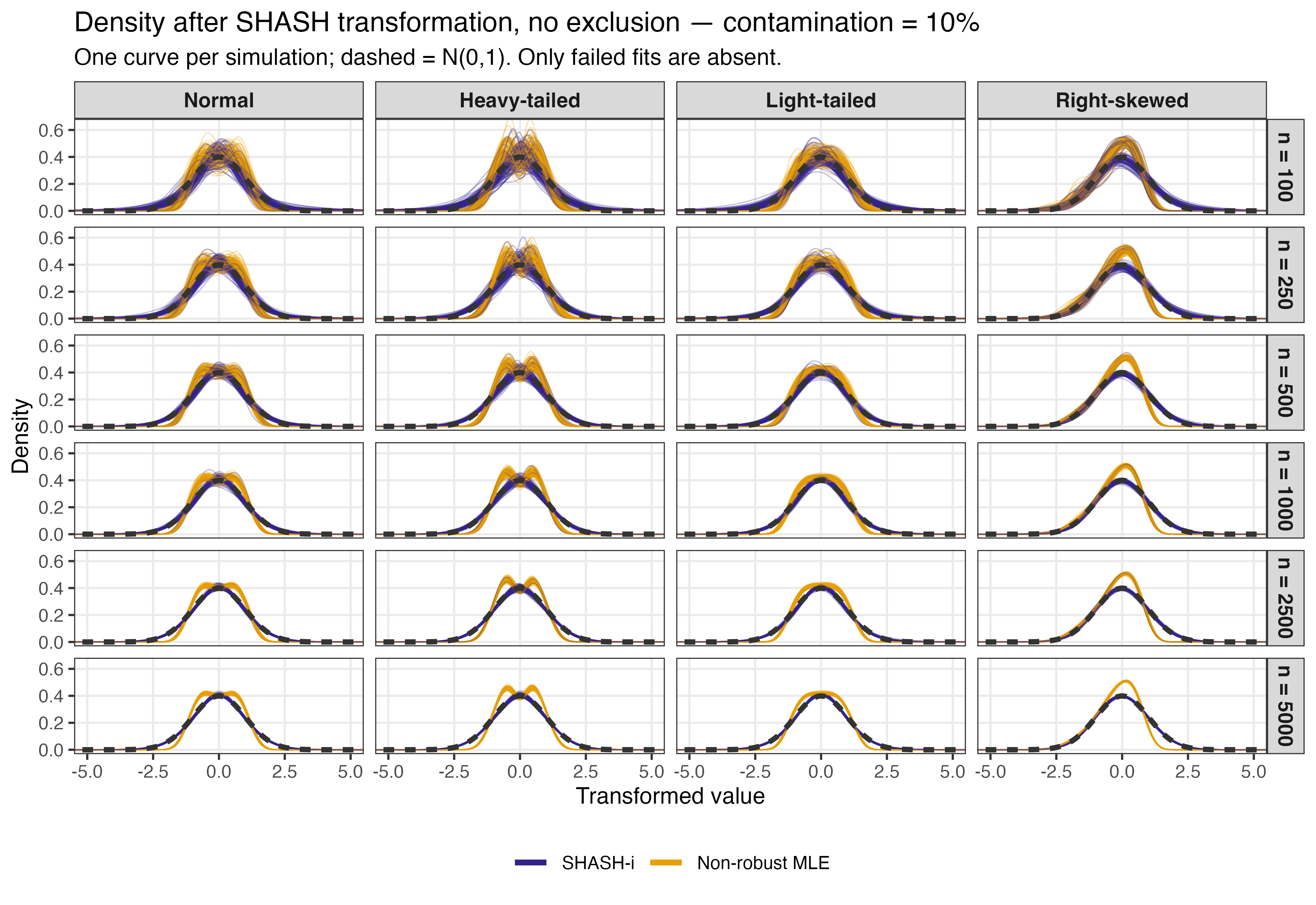}
    \caption{\review{\small\textbf{Density of true inliers after robust and non-robust SHASH transformation, at $10\%$ outlier contamination, varying sample size.} For each simulated
    dataset, SHASH parameters are estimated from the contaminated data ($10\%$ outliers) by robust SHASH transformation (SHASH-i) or
    non-robust maximum likelihood. Each panel and method displays $100$ curves, one for each iteration, and shows the distribution of true inliers after transformation with
    those parameter estimates. The dashed curve is the standard normal density, which is obtained exactly when the true parameters are used. As $n$ increases, the robust curves converge to close to standard normality, but with slightly heavier tails. The non-robust curves also become less variable but remain heavily biased due to the influence of outliers on the transformation parameter estimates.}}
    \label{fig:density_sample_sizes}
\end{figure}

\begin{figure}[H]
    \centering
    \begin{tabular}{|c|c|}
        \hline
        \begin{picture}(10,78)\put(0, 30){\rotatebox[origin=c]{90}{1\%}}\end{picture} &
        \includegraphics[width=0.8\textwidth,
                 height=1.7\textheight,
                 keepaspectratio]{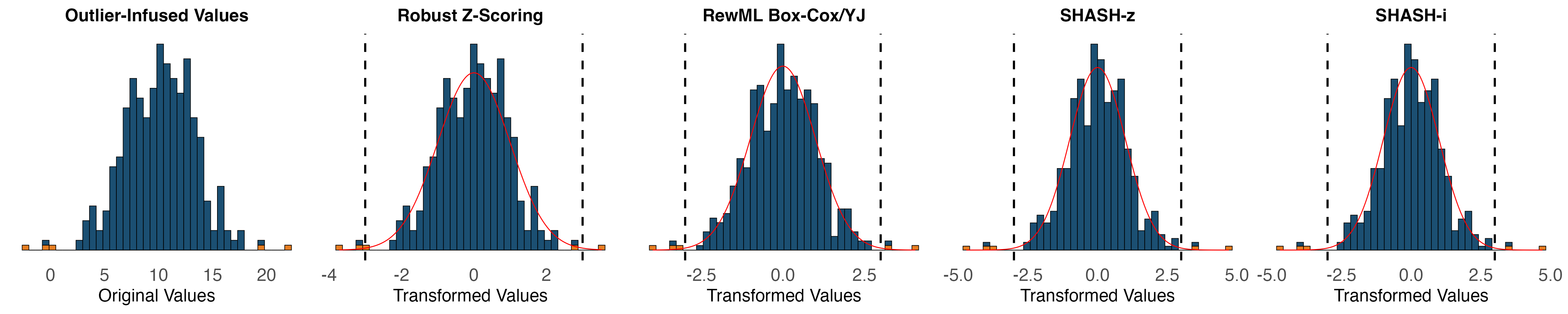} \\
        \hline
        \begin{picture}(10,78)\put(0, 30){\rotatebox[origin=c]{90}{5\%}}\end{picture} &
        \includegraphics[width=0.8\textwidth,
                 height=1.7\textheight,
                 keepaspectratio]{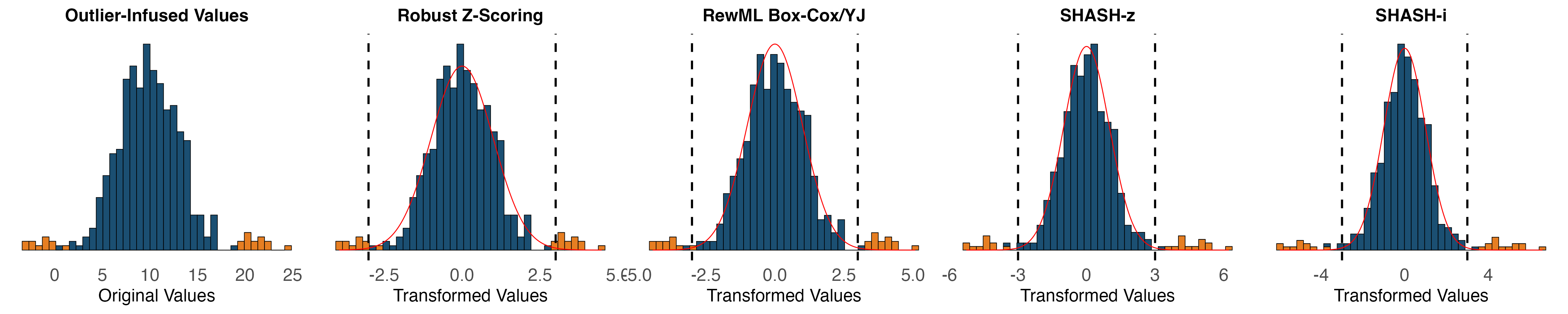} \\
        \hline
        \begin{picture}(10,78)\put(0,30){\rotatebox[origin=c]{90}{10\%}}\end{picture} &
        \includegraphics[width=0.8\textwidth,
                 height=1.7\textheight,
                 keepaspectratio]{figures/revision1/histograms/Normal-methods.png} \\
        \hline
        \begin{picture}(10,78)\put(0,30){\rotatebox[origin=c]{90}{20\%}}\end{picture} &
        \includegraphics[width=0.8\textwidth,
                 height=1.7\textheight,
                 keepaspectratio]{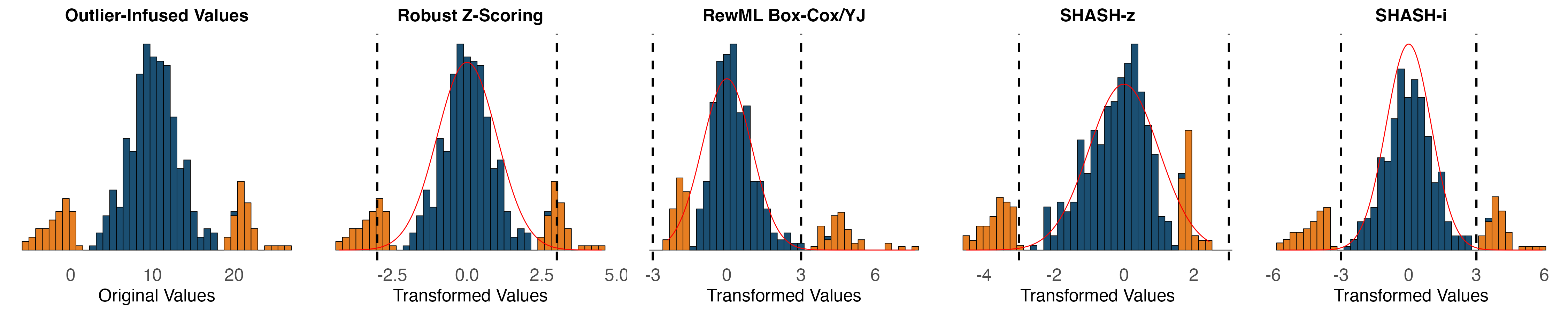} \\
        \hline
    \end{tabular}

        \includegraphics[width=4cm]{figures/toydatasets/status_legend_1X2.png}

     \caption{\review{\small\textbf{Histograms of simulated normally distributed data} and transformed values across contamination levels. Red curves show the standard normal density, and dashed lines indicate outlier detection thresholds based on standard normal quantiles.}}
    \label{fig:varyingpercentagenormal}
\end{figure}

\begin{figure}[H]
    \centering
    \begin{tabular}{|c|c|}
        \hline
        \begin{picture}(10,78)\put(0, 30){\rotatebox[origin=c]{90}{1\%}}\end{picture} &
        \includegraphics[width=0.8\textwidth,
                 height=1.7\textheight,
                 keepaspectratio]{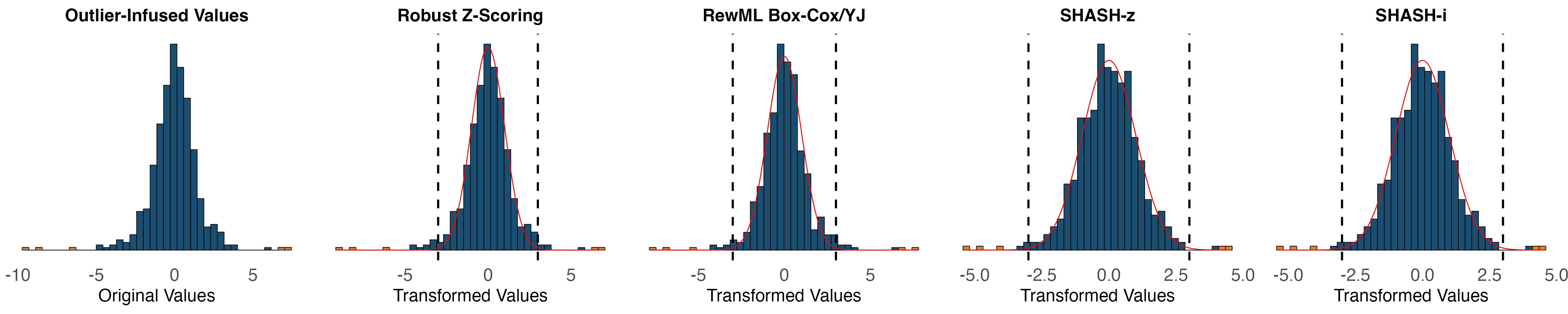} \\
        \hline
        \begin{picture}(10,78)\put(0, 30){\rotatebox[origin=c]{90}{5\%}}\end{picture} &
        \includegraphics[width=0.8\textwidth,
                 height=1.7\textheight,
                 keepaspectratio]{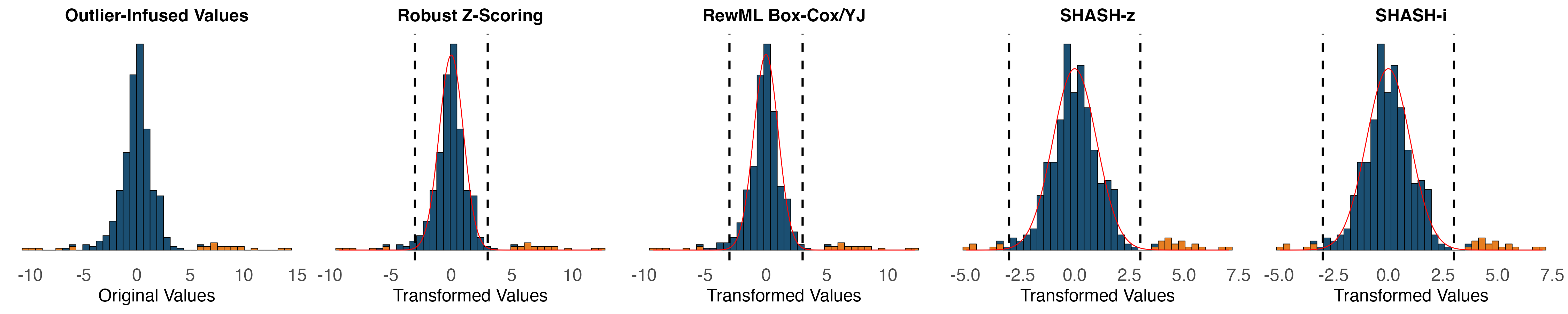} \\
        \hline
        \begin{picture}(10,78)\put(0,30){\rotatebox[origin=c]{90}{10\%}}\end{picture} &
        \includegraphics[width=0.8\textwidth,
                 height=1.7\textheight,
                 keepaspectratio]{figures/revision1/histograms/T-methods.png} \\
        \hline
        \begin{picture}(10,78)\put(0,30){\rotatebox[origin=c]{90}{20\%}}\end{picture} &
        \includegraphics[width=0.8\textwidth,
                 height=1.7\textheight,
                 keepaspectratio]{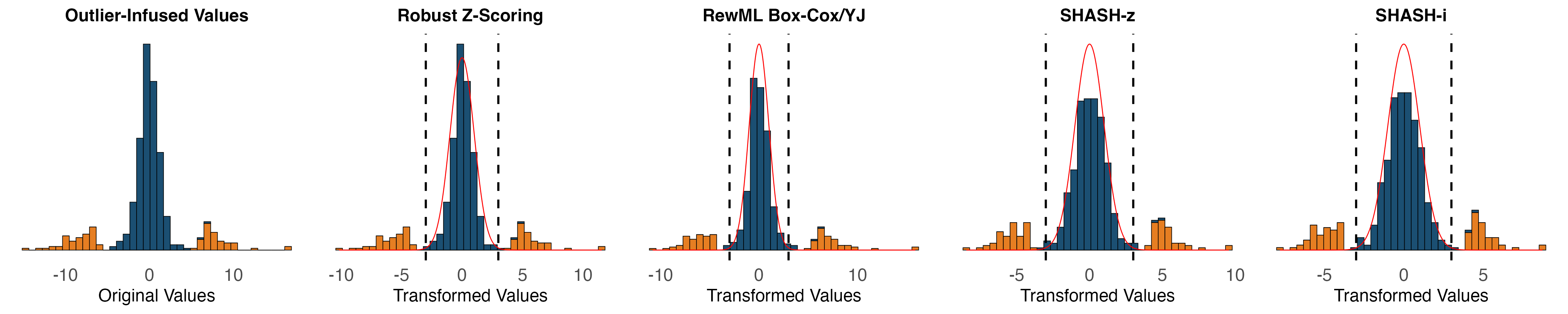} \\
        \hline
    \end{tabular}

        \includegraphics[width=4cm]{figures/toydatasets/status_legend_1X2.png}

     \caption{\small\textbf{Histograms of simulated $t(4)$ distributed data} and its transformation to \review{n}ormality using different methods, across varying outlier contamination levels. Red curves show the standard normal density, and dashed lines indicate outlier detection thresholds based on standard normal quantiles.}
    \label{fig:varyingpercentageT}
\end{figure}

\begin{figure}[H]
    \centering 
    \begin{tabular}{|c|c|}
        \hline
        \begin{picture}(10,78)\put(0, 30){\rotatebox[origin=c]{90}{1\%}}\end{picture} &
        \includegraphics[width=0.8\textwidth,
                 height=1.7\textheight,
                 keepaspectratio]{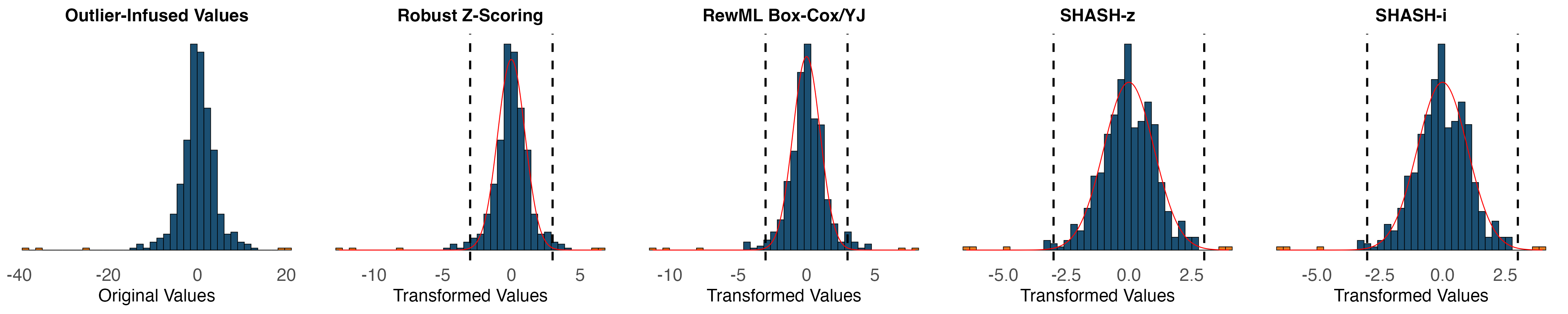} \\
        \hline
        \begin{picture}(10,78)\put(0, 30){\rotatebox[origin=c]{90}{5\%}}\end{picture} &
        \includegraphics[width=0.8\textwidth,
                 height=1.7\textheight,
                 keepaspectratio]{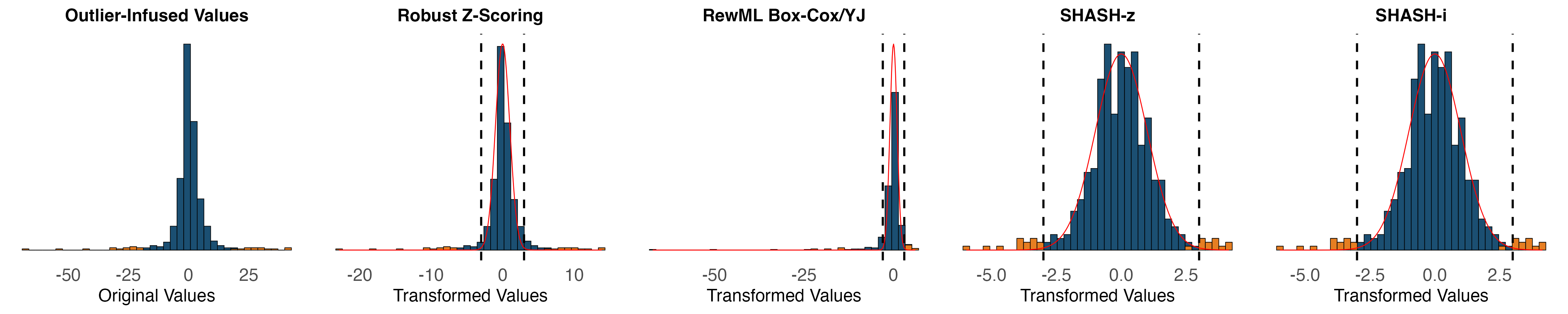} \\
        \hline
        \begin{picture}(10,78)\put(0,30){\rotatebox[origin=c]{90}{10\%}}\end{picture} &
        \includegraphics[width=0.8\textwidth,
                 height=1.7\textheight,
                 keepaspectratio]{figures/revision1/histograms/Laplace-methods.png} \\
        \hline
        \begin{picture}(10,78)\put(0,30){\rotatebox[origin=c]{90}{20\%}}\end{picture} &
        \includegraphics[width=0.8\textwidth,
                 height=1.7\textheight,
                 keepaspectratio]{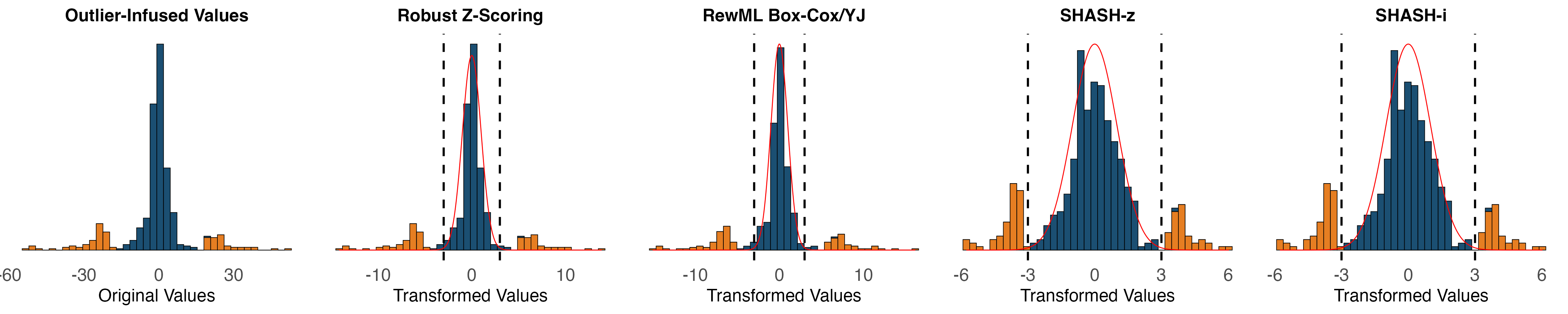} \\
        \hline
    \end{tabular}

        \includegraphics[width=4cm]{figures/toydatasets/status_legend_1X2.png}

     \caption{\small\textbf{Histograms of simulated Laplace distributed data} and its transformation to \review{n}ormality using different methods, across varying outlier contamination levels. Red curves show the standard normal density, and dashed lines indicate outlier detection thresholds based on standard normal quantiles.}
    \label{fig:varyingpercentagelaplace}
\end{figure}

\begin{figure}[H]
    \centering
    \begin{tabular}{|c|c|}
        \hline
        \begin{picture}(10,78)\put(0, 30){\rotatebox[origin=c]{90}{1\%}}\end{picture} &
        \includegraphics[width=0.8\textwidth,
                 height=1.7\textheight,
                 keepaspectratio]{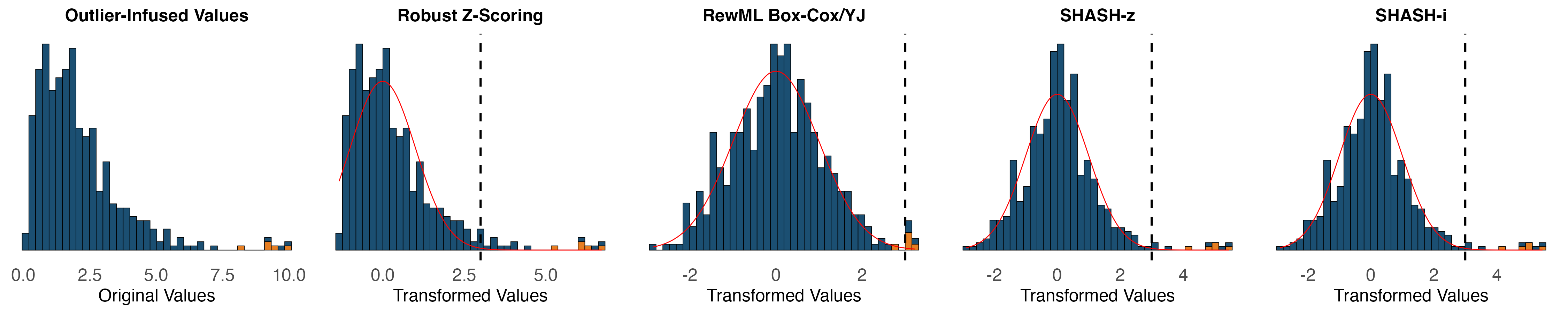} \\
        \hline
        \begin{picture}(10,78)\put(0, 30){\rotatebox[origin=c]{90}{5\%}}\end{picture} &
        \includegraphics[width=0.8\textwidth,
                 height=1.7\textheight,
                 keepaspectratio]{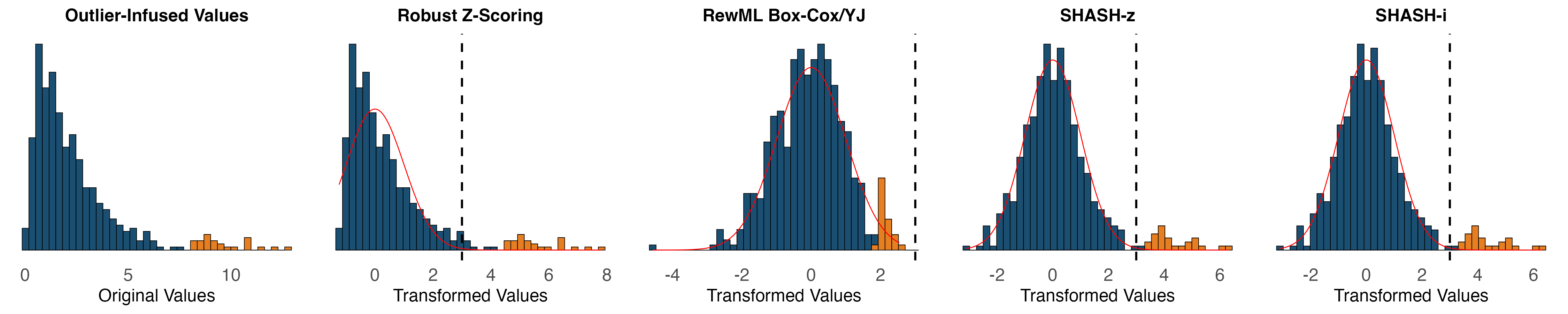} \\
        \hline
        \begin{picture}(10,78)\put(0,30){\rotatebox[origin=c]{90}{10\%}}\end{picture} &
        \includegraphics[width=0.8\textwidth,
                 height=1.7\textheight,
                 keepaspectratio]{figures/revision1/histograms/Gamma-methods.png} \\
        \hline
        \begin{picture}(10,78)\put(0,30){\rotatebox[origin=c]{90}{20\%}}\end{picture} &
        \includegraphics[width=0.8\textwidth,
                 height=1.7\textheight,
                 keepaspectratio]{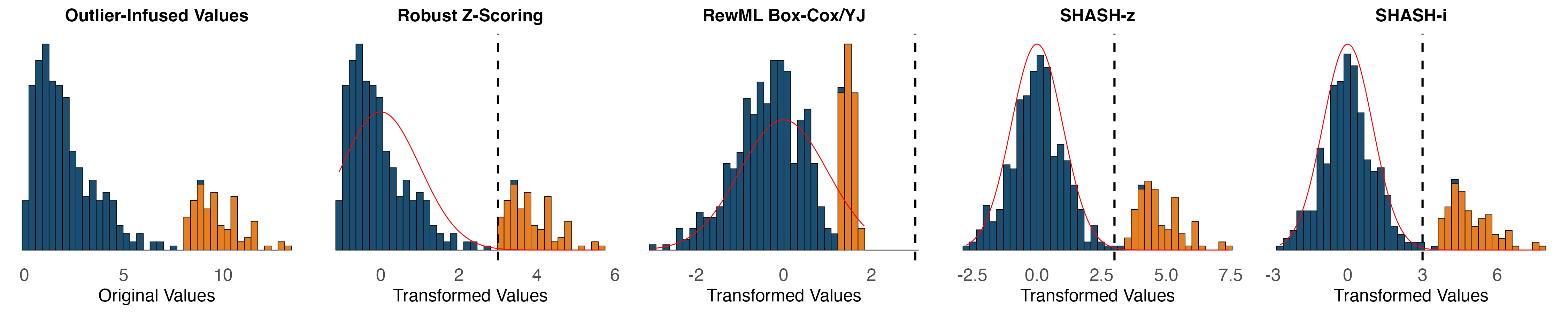} \\
        \hline
    \end{tabular}

       \includegraphics[width=4cm]{figures/toydatasets/status_legend_1X2.png}

     \caption{\small\textbf{Histograms of simulated Gamma distributed data} and its transformation to \review{n}ormality using different methods, across varying outlier contamination levels. Red curves show the standard normal density, and dashed lines indicate outlier detection thresholds based on standard normal quantiles.}
    \label{fig:varyingpercentagegamma}
\end{figure}

\begin{figure}[H]
    \centering
    \begin{tabular}{|c|c|}
        \hline
        \begin{picture}(10,78)\put(0, 30){\rotatebox[origin=c]{90}{1\%}}\end{picture} &
        \includegraphics[width=0.8\textwidth,
                 height=1.7\textheight,
                 keepaspectratio]{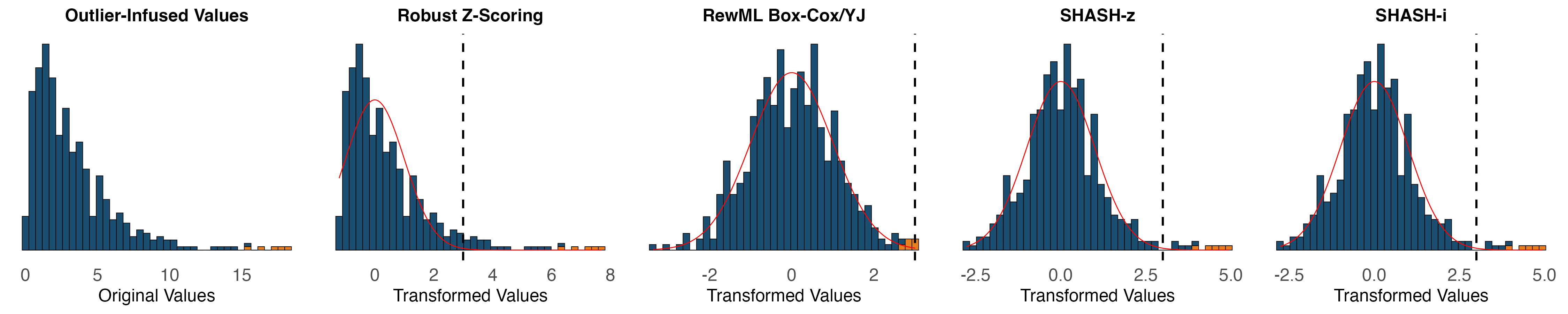} \\
        \hline
        \begin{picture}(10,78)\put(0, 30){\rotatebox[origin=c]{90}{5\%}}\end{picture} &
        \includegraphics[width=0.8\textwidth,
                 height=1.7\textheight,
                 keepaspectratio]{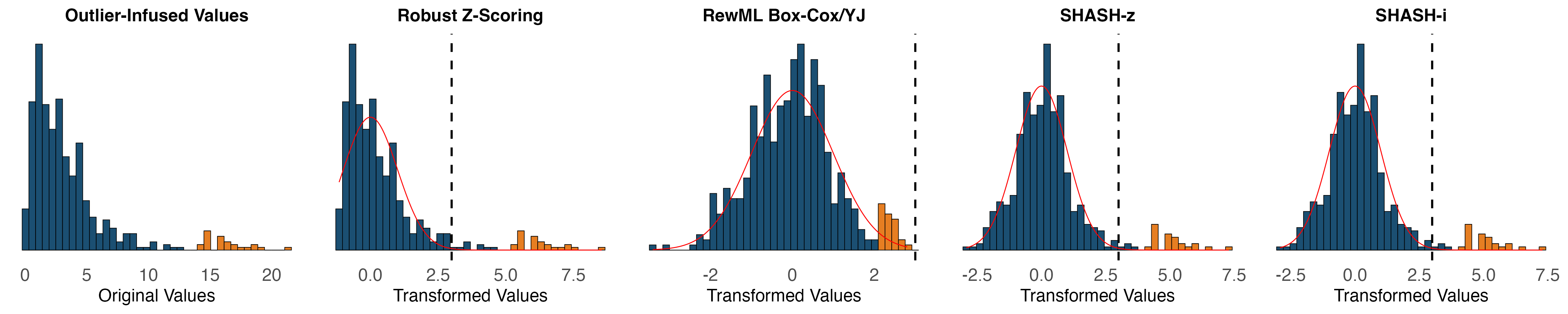} \\
        \hline
        \begin{picture}(10,78)\put(0,30){\rotatebox[origin=c]{90}{10\%}}\end{picture} &
        \includegraphics[width=0.8\textwidth,
                 height=1.7\textheight,
                 keepaspectratio]{figures/revision1/histograms/Chisquare-methods.png} \\
        \hline
        \begin{picture}(10,78)\put(0,30){\rotatebox[origin=c]{90}{20\%}}\end{picture} &
        \includegraphics[width=0.8\textwidth,
                 height=1.7\textheight,
                 keepaspectratio]{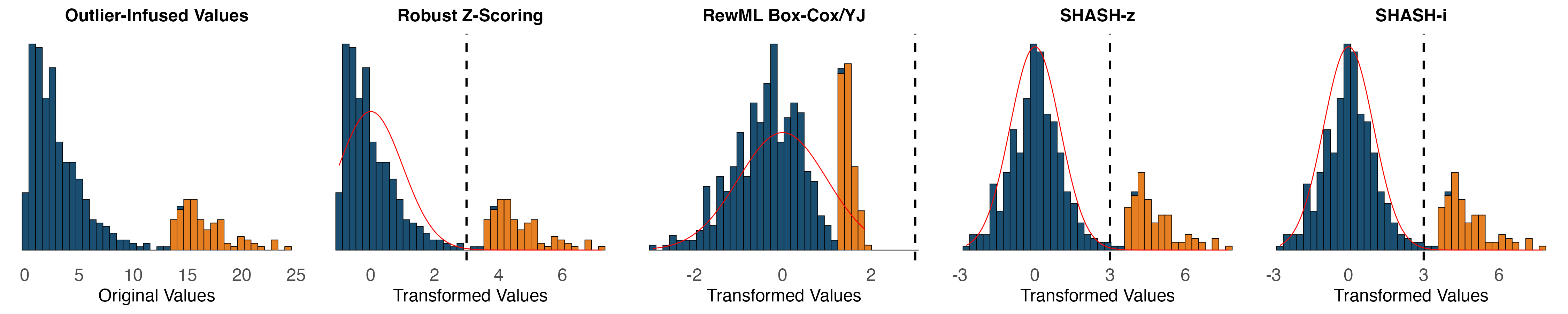} \\
        \hline
    \end{tabular}

        \includegraphics[width=4cm]{figures/toydatasets/status_legend_1X2.png}

     \caption{\small\textbf{Histograms of simulated Chi-Square distributed data} and its transformation to \review{n}ormality using different methods, across varying outlier contamination levels. Red curves show the standard normal density, and dashed lines indicate outlier detection thresholds based on standard normal quantiles.}
    \label{fig:varyingpercentagechisq}
\end{figure}

\begin{figure}[H]
    \centering
    \begin{tabular}{|c|c|}
        \hline
        \begin{picture}(10,78)\put(0, 30){\rotatebox[origin=c]{90}{1\%}}\end{picture} &
        \includegraphics[width=0.8\textwidth,
                 height=1.7\textheight,
                 keepaspectratio]{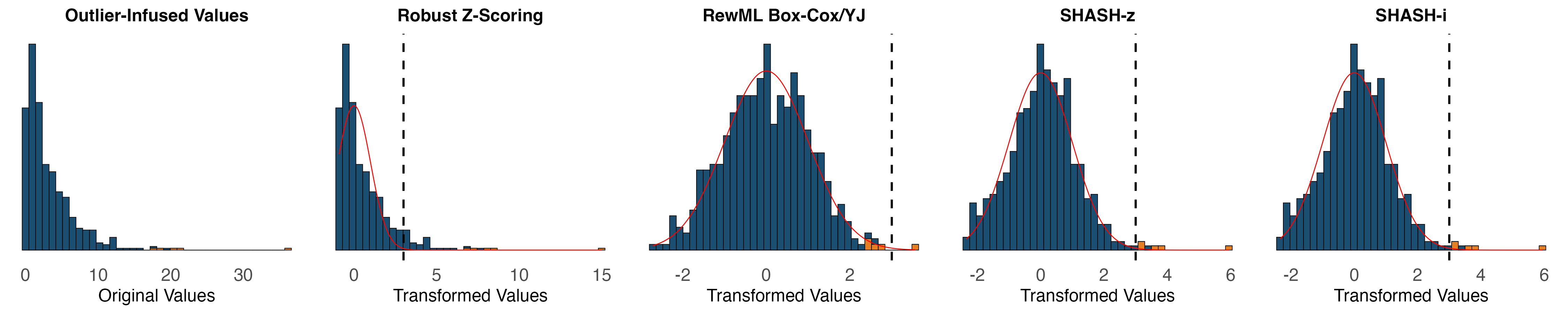} \\
        \hline
        \begin{picture}(10,78)\put(0, 30){\rotatebox[origin=c]{90}{5\%}}\end{picture} &
        \includegraphics[width=0.8\textwidth,
                 height=1.7\textheight,
                 keepaspectratio]{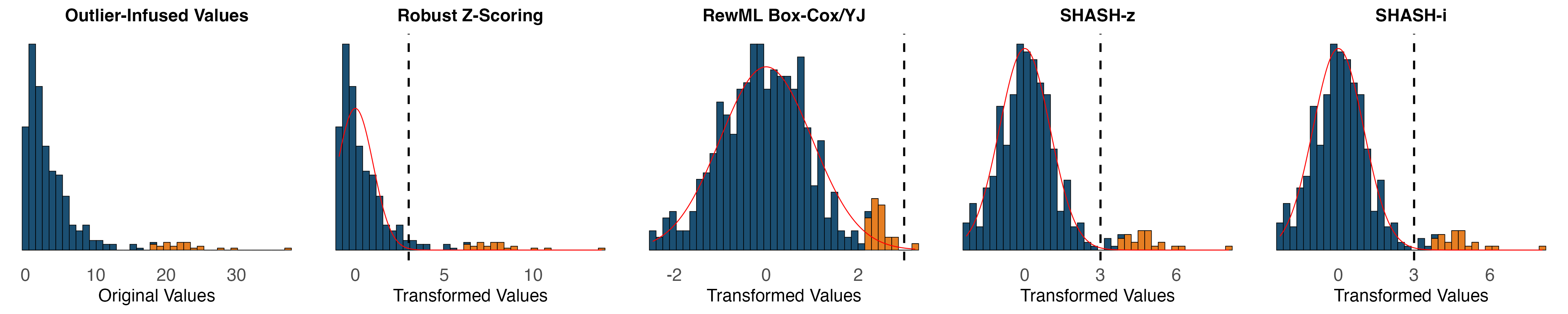} \\
        \hline
        \begin{picture}(10,78)\put(0,30){\rotatebox[origin=c]{90}{10\%}}\end{picture} &
        \includegraphics[width=0.8\textwidth,
                 height=1.7\textheight,
                 keepaspectratio]{figures/revision1/histograms/Weibull-methods.png} \\
        \hline
        \begin{picture}(10,78)\put(0,30){\rotatebox[origin=c]{90}{20\%}}\end{picture} &
        \includegraphics[width=0.8\textwidth,
                 height=1.7\textheight,
                 keepaspectratio]{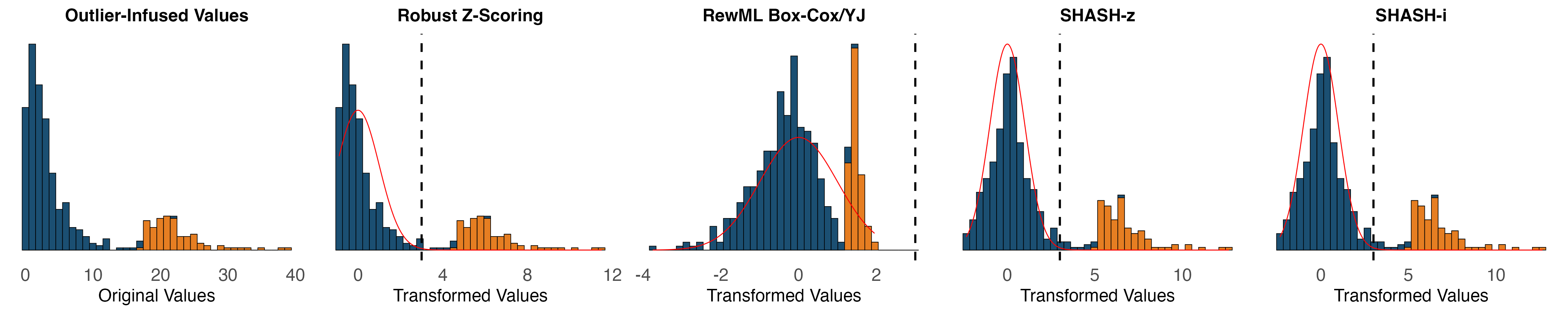} \\
        \hline
    \end{tabular}

        \includegraphics[width=4cm]{figures/toydatasets/status_legend_1X2.png}

     \caption{\small\textbf{Histograms of simulated Weibull distributed data} and its transformation to \review{n}ormality using different methods, across varying outlier contamination levels. Values over 50 are truncated for visualization purposes. Red curves show the standard normal density, and dashed lines indicate outlier detection thresholds based on standard normal quantiles.}
    \label{fig:varyingpercentageweibull}
\end{figure}

\begin{figure}
    \centering
    \includegraphics[width=\textwidth]{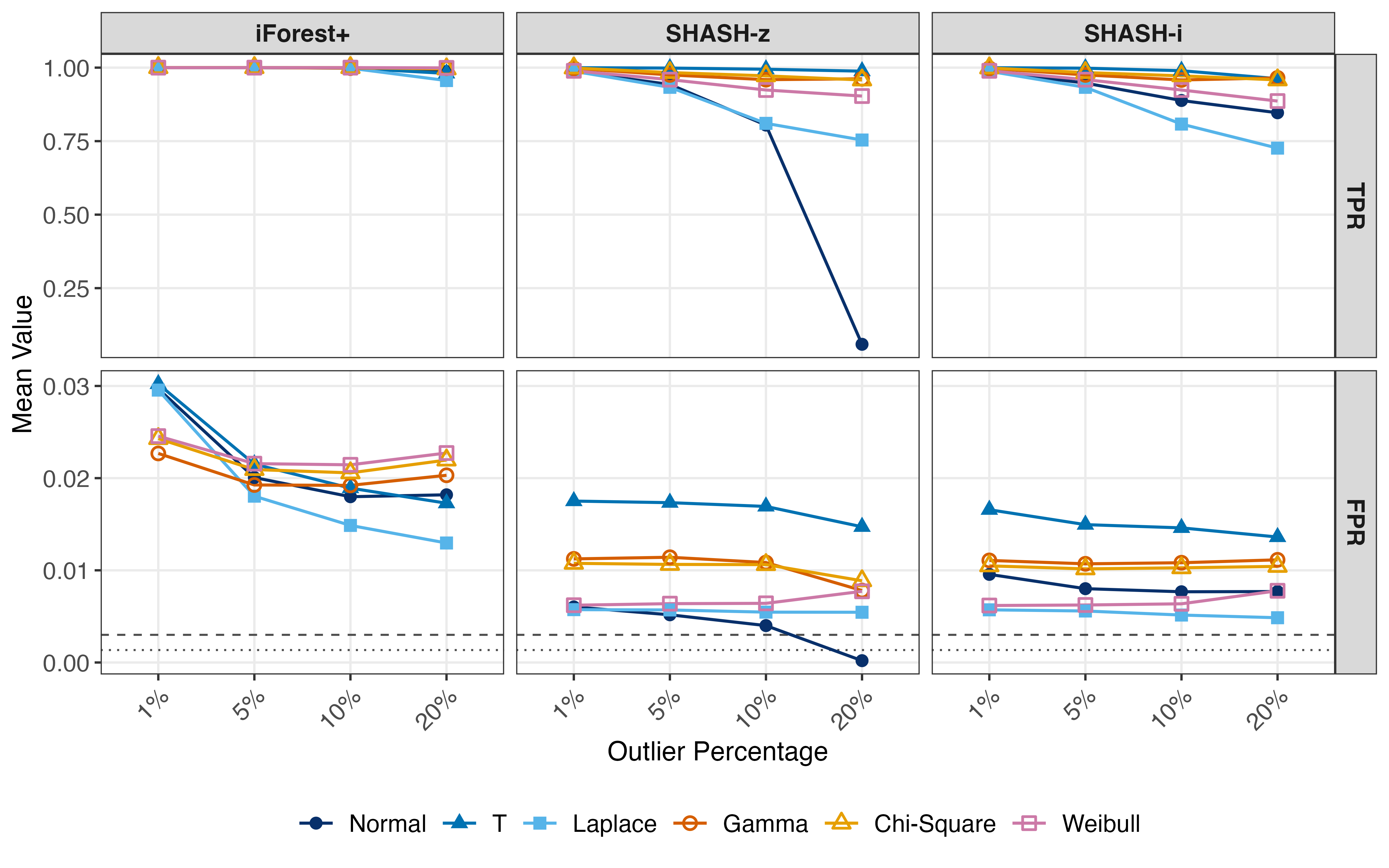}
    \caption{\review{\small\textbf{iForest+ used alone as an outlier detector, compared with SHASH-z and SHASH-i.} Mean TPR (top) and FPR (bottom) over 1{,}000 simulated datasets per distribution and contamination level. Reference lines mark the nominal FPR: dashed, 0.003 (two-sided, symmetric distributions); dotted, 0.00135 (one-sided, skewed distributions). Degenerate SHASH fits are excluded (see Figure~\ref{fig:degen_rate}).}}
    \label{fig:R_iforest}
\end{figure}

\begin{figure}[H]
    \centering
    Normal(10,3) \\
    \includegraphics[width=0.7\textwidth, trim = 0 0 0 0, clip]{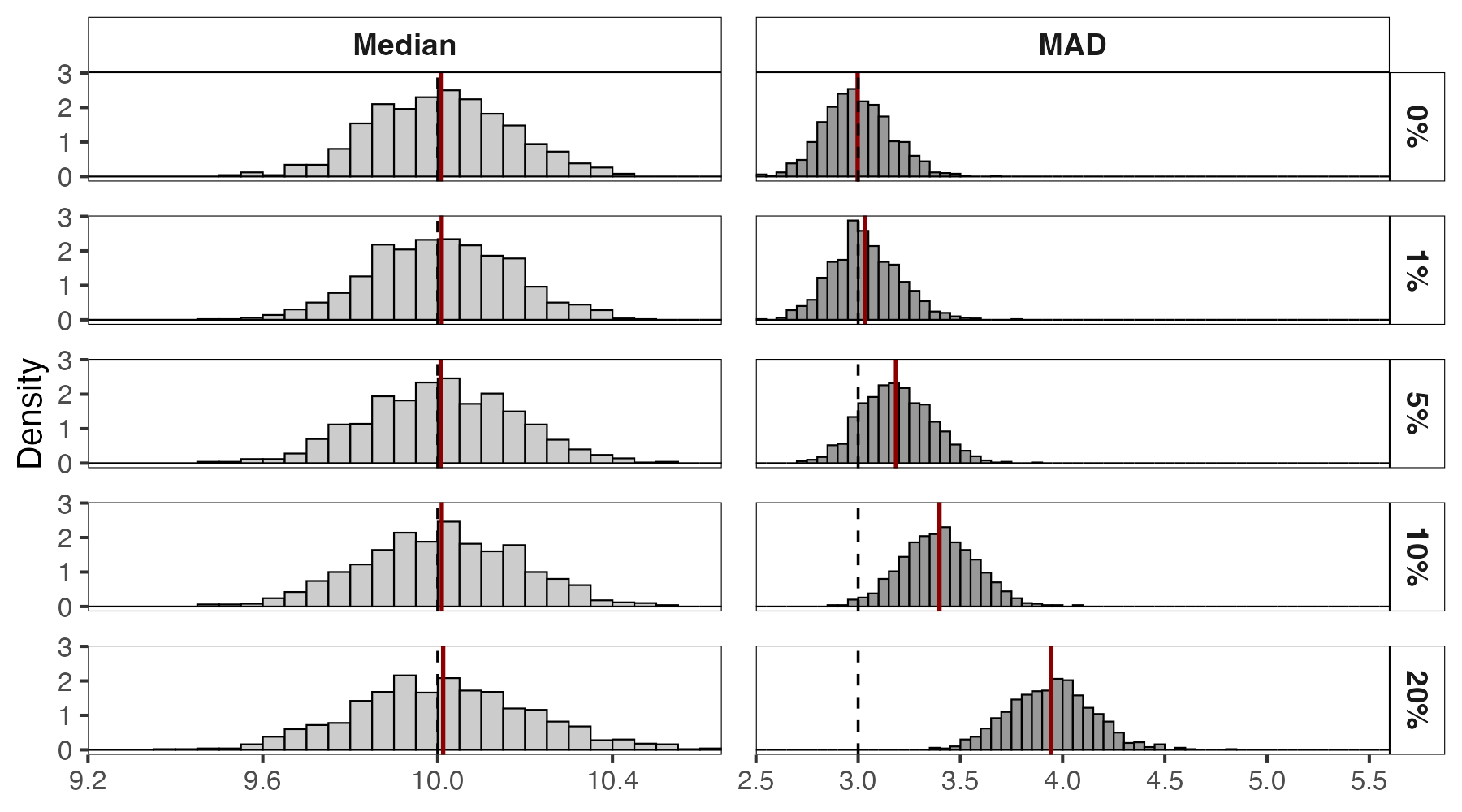} \\[10pt]
    \centering
    Gamma(2,1) \\   
    \includegraphics[width=0.7\textwidth, trim = 0 0 0 0, clip]{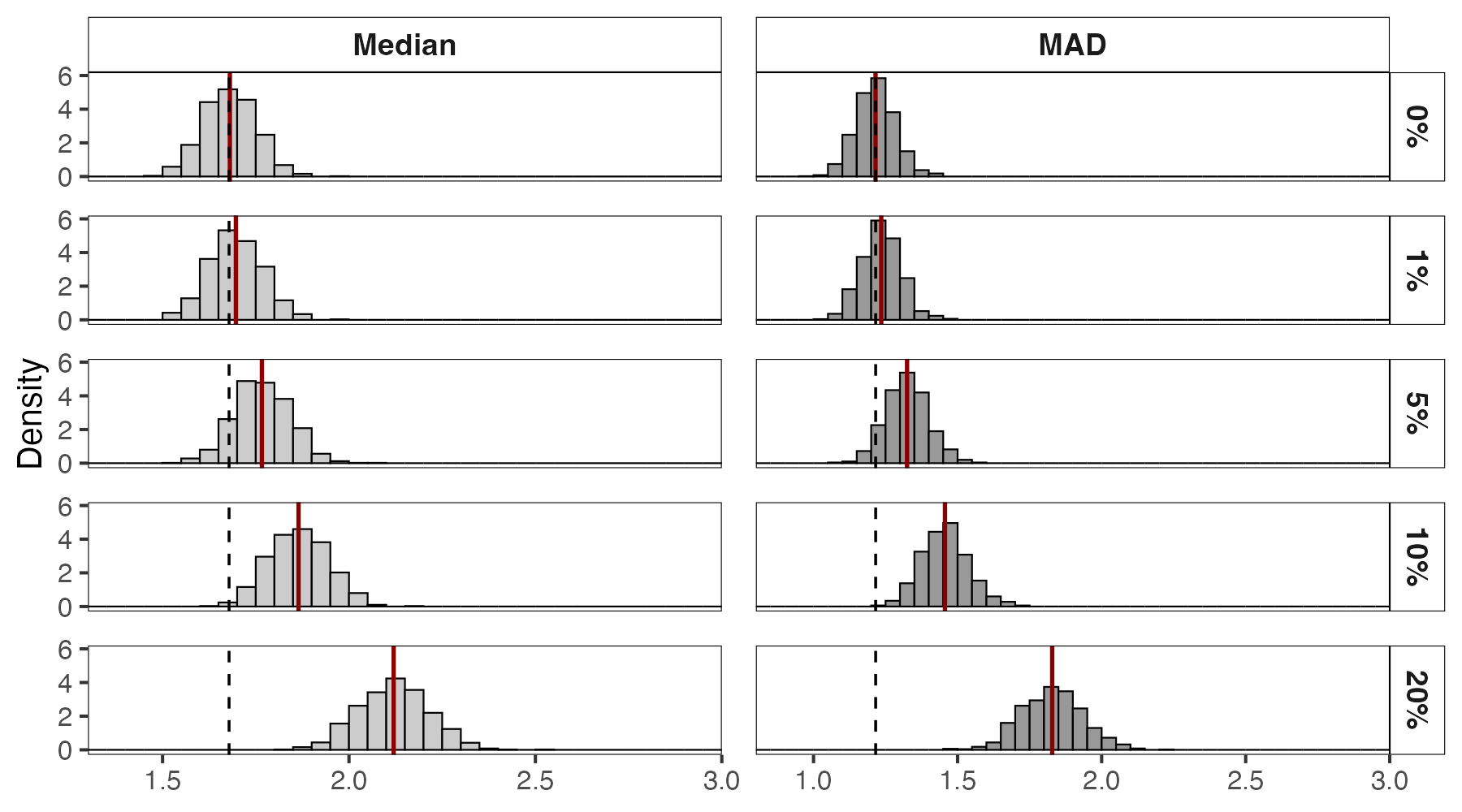}
    \caption{\textbf{Empirical sampling distributions of the median and MAD estimates
    by outlier contamination level \review{(0--20\%)}.} The Normal and Gamma
    distributions are shown to give an example of a symmetric and a skewed
    distribution. Dark Red dot dash lines represent the sampling distribution mean,
    while the black dashed lines indicate the true median and MAD values. In the
    symmetric case, the median remains unbiased, while it is positively biased in the
    right-skewed case due to the location of outliers in the upper tail. The MAD is
    upwardly biased in both cases, as regular observations become extreme values,
    shifting the location of the median deviation upwards.}

    \label{fig:MedianMADNormalAndGamma}
\end{figure}


\begin{figure}[H]
    \centering

    \begin{minipage}[t]{0.49\textwidth}
        \centering
        \includegraphics[width=\linewidth]%
            {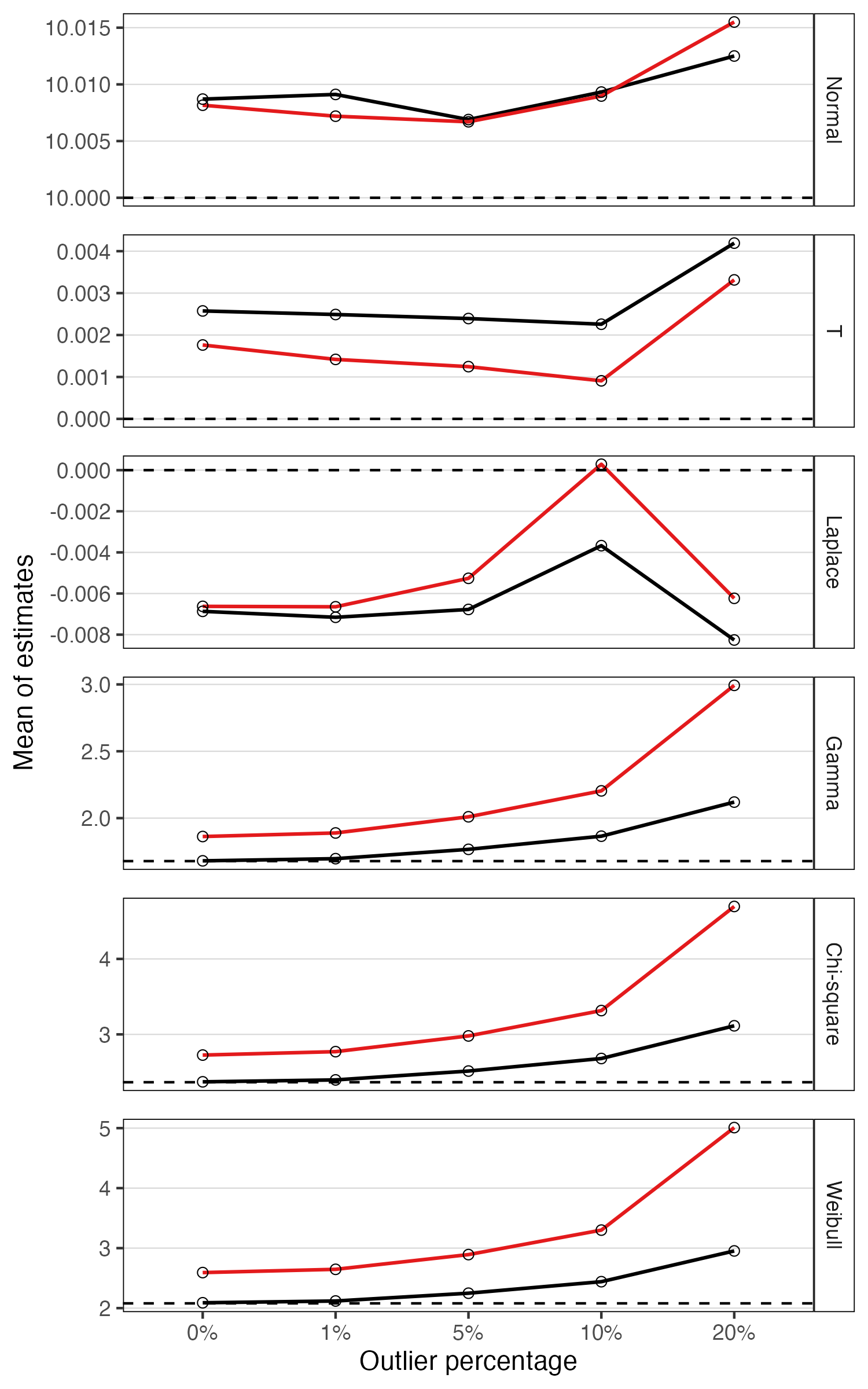}
        \vspace{0.4em}
        \includegraphics[width=1\linewidth]{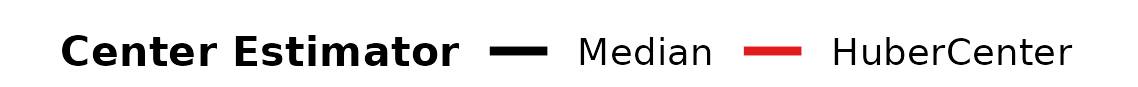}
        \vspace{-0.5em}
        \subcaption*{\textbf{A.} Center Estimators}
    \end{minipage}
    \hfill
    \begin{minipage}[t]{0.49\textwidth}
        \centering
        \includegraphics[width=\linewidth]%
            {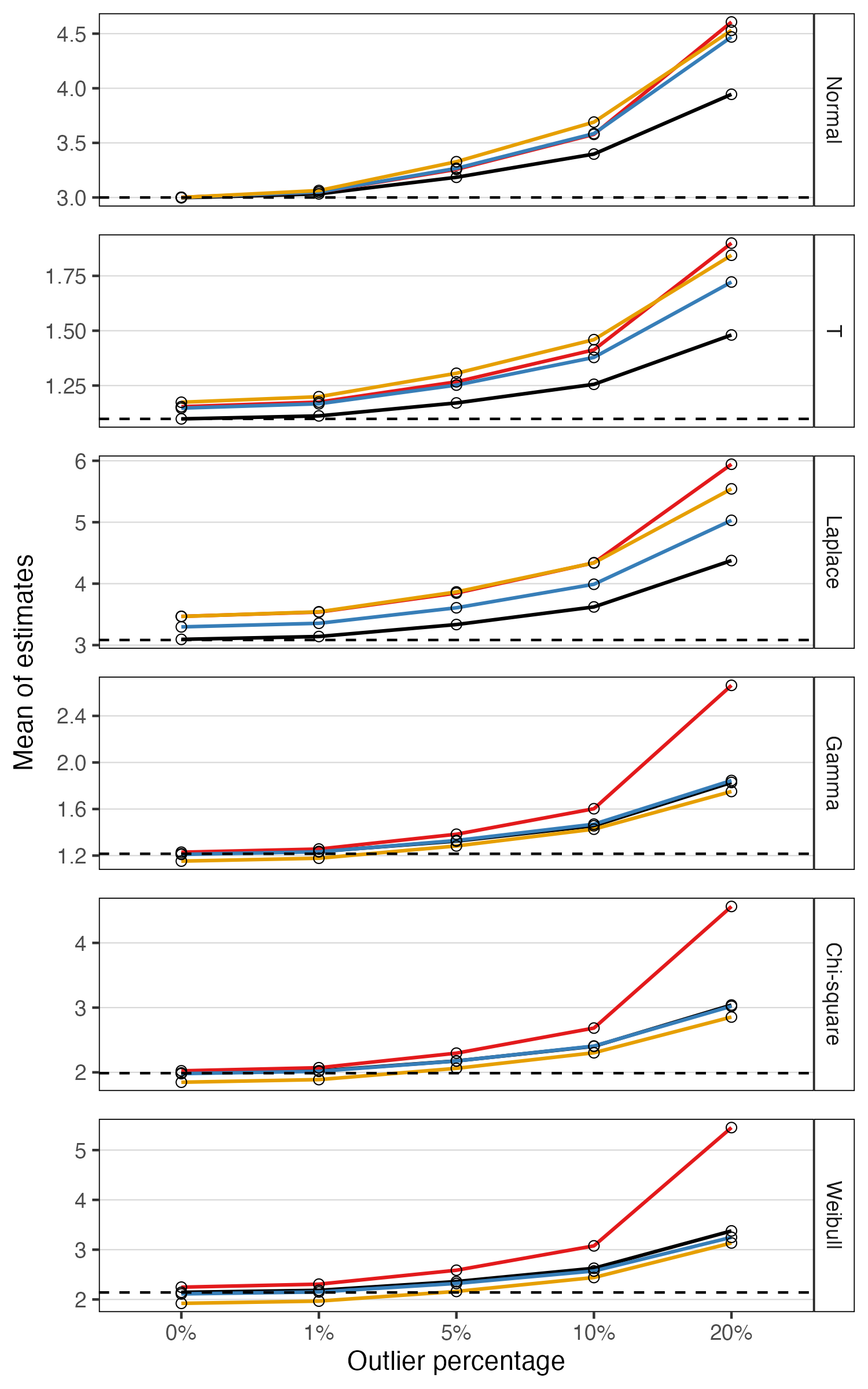}
        \vspace{0.4em}
        \includegraphics[width=1\linewidth]{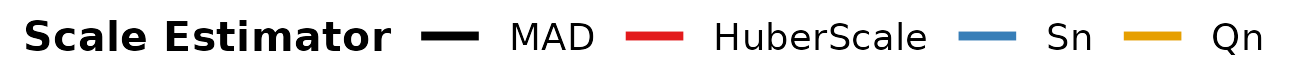}
        \vspace{-0.5em}
        \subcaption*{\textbf{B.} Scale Estimators}
    \end{minipage}

    \vspace{0.8em}

    \caption{\textbf{Robust estimator performance across contamination levels
    \review{(0--20\%)}.}
    Panel A shows average center estimates (Median, HuberCenter) with its legend below;
    Panel B shows average scale estimates (MAD, $S_n$, $Q_n$, HuberScale) with its own legend.
    \review{Black dashed lines indicate the true population median (A) and the true
    population scaled MAD (B).}}
    \label{fig:CenterScaleSideBySide}
\end{figure}


\begin{figure}[H]
    \centering
    \includegraphics[width=0.75\textwidth]{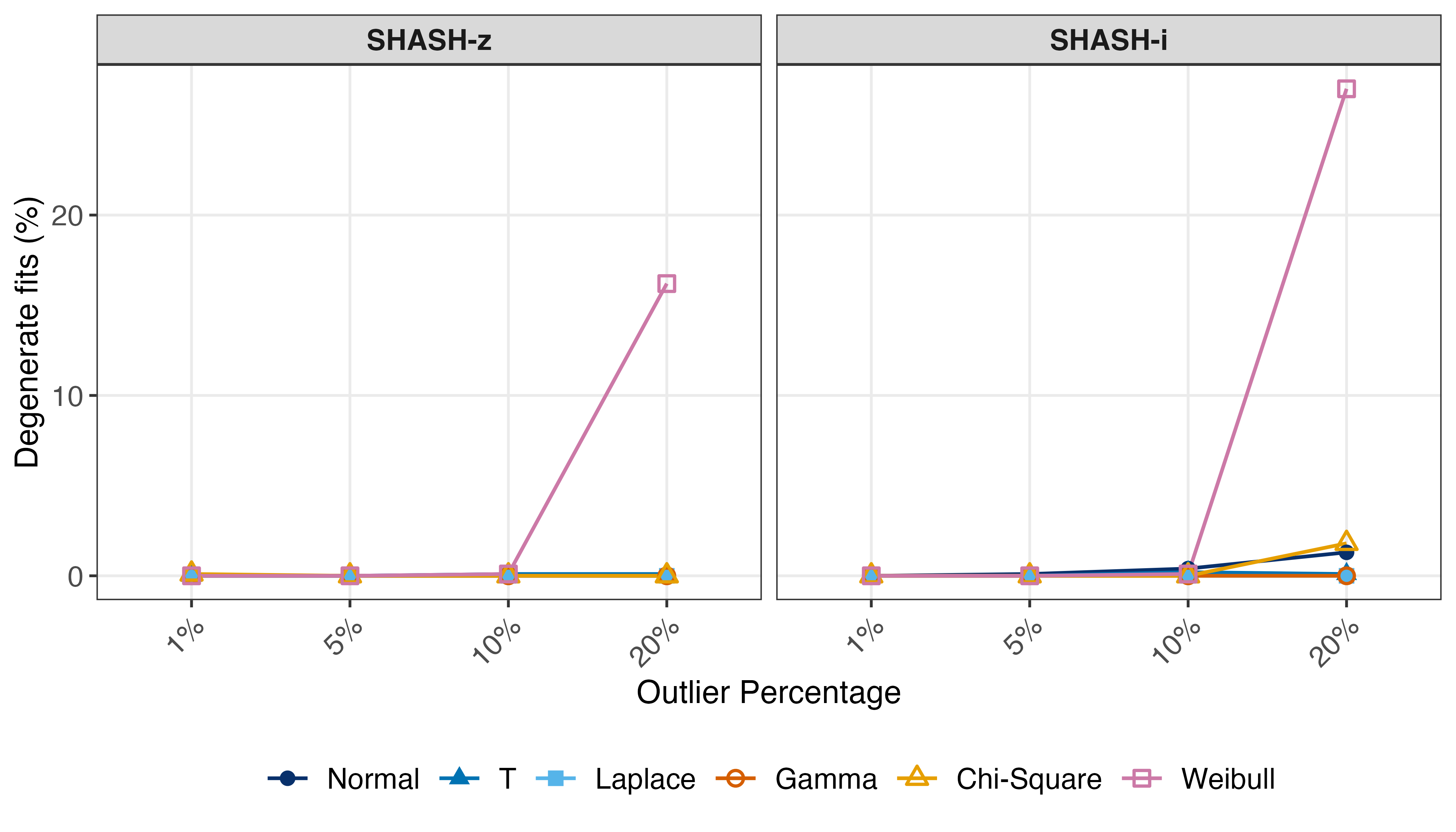}
    \caption{\review{\small\textbf{Degenerate fits by contamination level.} Percentage of the 1{,}000 simulated datasets per setting in which the SHASH fit was degenerate (more than 1\% of transformed values with $|z| > 20$), for SHASH-z and SHASH-i. Colours and symbols: generating distribution. Degenerate fits are rare except for Weibull at 20\% contamination (16\% for SHASH-z, 27\% for SHASH-i), and are excluded from Figure~\ref{fig:FPR_TPR_plot}.}}
    \label{fig:degen_rate}
\end{figure}


\begin{figure}[H]
    \centering
    \includegraphics[width=0.6\textwidth,
                    height=0.6\textheight,
                     keepaspectratio]{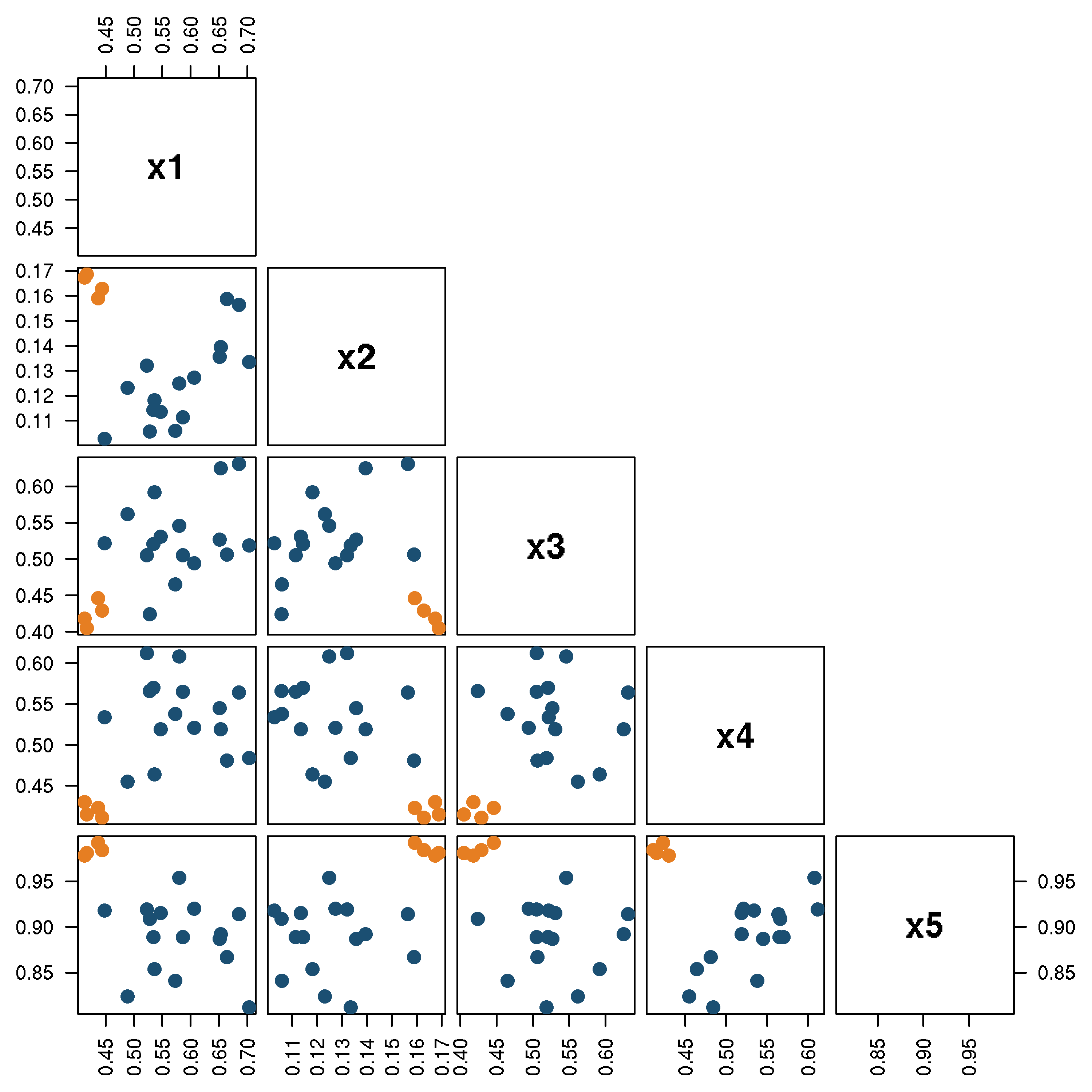}

    \caption{\small\textbf{Wood dataset – Pairwise scatterplot matrix.}
    Each panel shows pairwise relationships between predictors in the Wood dataset, 
    with true outliers (observations 4, 6, 8, and 19) highlighted in \review{orange}.}
    \label{fig:wood_pairs}
\end{figure}

\end{document}